# Day of the week submission effect for accepted papers in Physica A, PLOS ONE, Nature and Cell

**Catalin Emilian Boja[a], Claudiu Herțeliu[b,*], Marian Dârdală[a], Bogdan Vasile Ileanu[b]**

[a]Department of Economic Informatics and Cybernetics, Bucharest University of Economic Studies, Bucharest, Romania, [b]Department of Statistics and Econometrics, Bucharest University of Economic Studies, Bucharest, Romania

[*]Correspondence address. E-mail: hertz@csie.ase.ro, claudiu.herteliu@gmail.com, Phone: +4 0722 455 586

**Abstract:** The particular day of the week when an event occurs seems to have unexpected consequences. For example, the day of the week when a paper is submitted to a peer reviewed journal correlates with whether that paper is accepted. Using an econometric analysis (a mix of log-log and semi-log based on undated and panel structured data) we find that more papers are submitted to certain peer review journals on particular weekdays than others, with fewer papers being submitted on weekends. Seasonal effects, geographical information as well as potential changes over time are examined. This finding rests on a large (178 000) and reliable sample; the journals polled are broadly recognized (Nature, Cell, PLOS ONE and Physica A). Day of the week effect in the submission of accepted papers should be of interest to many researchers, editors and publishers, and perhaps also to managers and psychologists.

**Keywords:** day of the week effect, DWE, peer reviewed journals, seasonal effects, GIS

## 1. Introduction

In most of the world, weekends are set apart from weekdays. This delineation is a social convention and at first glance nothing seems to distinguish one day from another. However, Rossi and Rossi in (1977) found that positive moods are more likely to occur at weekends. French (1980) found that Monday stock returns tends to be negative while returns on Tuesday, Wednesday, Thursday and Friday tend to be positive. Chang *et al.*, 1993; Berument & Kiymaz, 2001; Bhattacharya *et al.*, 2003; Fidrmuc & Tena, 2015; Dhesi *et al.*, 2016 have also studied the DWE in the context of different financial settings. More recently DWE's have been identified in other areas such as ecology (Marr & Harley, 2002), medicine (Bell & Redelmeier, 2001), demography (Herteliu *et al.*, 2015), meteorology (Dessens *et al.*, 2001), risk calculations (Doherty *et al.*, 1998) and scientometrics (Cabanac & Hartley, 2013; Campos-Arceiz *et al.*, 2013; Hartley & Cabanac, 2017; Ausloos *et al.*, 2016; Ausloos *et al.*, 2017).

Recently Cabanac & Hartley (2013) studied acceptance of technical manuscripts to journals and found big differences between week-day and week-end submissions. They found that submissions are at a maximum on Mondays. Hartley & Cabanac (2017) confirmed this result and also noticed that submissions are at a minimum at weekends. Ausloos *et al.* (2016) found that papers submitted on either Tuesdays or Wednesdays were more likely to be accepted for publication than papers submitted on weekends. Other reports (Shalvi *et al.*, 2010; Bornmann & Daniel, 2011; Schreiber, 2012) observed a seasonal effect – the so-called "seasonal overloading of editorial desks" - for decisions made during peer review process.

Here using a large data set of papers published in academic journals, we confirm a DWE for the submission day of accepted papers, however seasonal or geographical signals are less noticeable. Papers submitted on Tuesdays are more likely to be accepted by Physica A and Nature whereas Wednesdays seem the most likely day to submit and secure acceptance to PLOS ONE. For Cell Mondays and Tuesdays seem the best submission days in case of accepted papers. Relative to previous researches, we introduce methodological improvements (meta-data scraper). Even where we confirm conclusions from previous studies, we rely on different econometric models (mix of log-log and semi-log based on undated and panel structured data) and visual elements (localization quotients).

## 2. Materials and Methods

### 2.1. Data collection process

The core data upon which the current paper relies is wholly public. Our approach is similar to that of Cabanac & Hartley (2013). Even journals that hide papers behind a paywall freely offer (i) abstracts, (ii) authors' information and (iii) a brief history of peer review process on their web sites. Using a tailor-made web scraper we extracted information from the following journals: Nature, Cell, PLOS ONE and Physica A: Statistical Mechanics and its Applications. The date when a manuscript is received by the journal's system is provided by metadata and is labelled in similar manner: Nature uses "date received" while Cell, PLOS ONE and Physica A use "received". If a first parsing of a paper's information failed to retrieve the required data, the algorithm tried to parse again. In this way, we generated a large number of inquiries. Sometimes a publisher's server banned our web scraper for several days. For this reason, the retrieved information is not exhaustive. Nevertheless, we recovered significant totals of the available data (Physica A: 83.4%;

PLOS ONE: 99.9%; Nature: 76.4% and Cell: 68.7%) in our target journals (as may be seen in the Table 1) and believe it is representative of the total. Moreover, since only electronic submissions offer precise information we collected data only for submissions after 2000.

Papers that apparently did not pass a peer review process (editors' lists, editorials, corrigenda, errata, etc.) signaled by no date for reception of either an initial or revised submission were removed from the data set.

**Table 1.** Parsed papers sample by journal

| **No.** | **Journal** | **Time span (publication)** | **Papers retrieved** | **Items in Web of Science (WoS)** | **Citable items in WoS** | **Share of retrieved paper (%)** |
|---|---|---|---|---|---|---|
| (0) | (1) | (2) | (3) | (4) | (5) | (6) =(3/5)*100 |
| 1. | Nature | 2010-2016 | 4 653 | 18 462 | 6 092 | 76.4 |
| 2. | Cell | 2002-2016 | 3 777 | 8 408 | 5 497 | 68.7 |
| 3. | PLOS ONE | 2006-2016 | 160 172 | 168 289 | 160 306 | 99.9 |
| 4. | Physica A | 2001-2016 | 9 825 | 11 998 | 11 779 | 83.4 |
| **Total** | | | **178 427** | **207 107** | **183 674** | **97.1** |

**Note:** We distinguish between the number of published papers/items according to WoS simple query (WWW1) and the citable items as so referred in Journal Citations Report JCR (WWW2). The Items per Citable Items Ratio (ICIR) varies: 1.02 (Physica A), 1.53 (Cell), 1.05 (PLOS ONE), 3.03 (Nature). Since JCR2016 is not released yet, in order to estimate the number of citable items in 2016, historical values for ICIR were used. The indexing process in WoS is not exactly synchronized with a calendar. E.g., on the 17$^{th}$ of May 2018, the Physica A most recent issue (506) is (will be!) published on 15$^{th}$ of September 2018 while for Nature the most recent issue (7705) was published on 17$^{th}$ of May 2018. Therefore, the share of retrieved items in our dataset could occur to be larger than 100% sometimes (which is actually the case of PLOS ONE in 2009: 103.3%).

We gathered dates of accepted papers' submission using a Web scraper (Kobayashi & Takeda, 2000) developed to automatically search and retrieve needed information on the journals' Web sites. The software application was developed by the authors using a Java platform. For parsing Web page content and extracting searched information we used the open-source Jsoup library (WWW3). Retrieved data was stored in a MySQL database (WWW4). The Web scraper executed recursively the following sequence of steps for each analyzed journal:

1. The process starts with the journal Web page that contains web archive information for each issue. Starting from this page the scraper was able to extract the web link for each issue. Each link was recorded, to avoid reaching the same link in future searches, and it was accessed, to process each issue's Web page;

2. From each issue's Web page, recorded in the previous step, we extracted information for each article: title, authors, pages (if available) and their Web page~~s~~ links for further processing. The sequence that searched for articles was implemented by a prior analysis Web page structure. For example, each online issue of the PLOS ONE journal defined the articles using a list of div HTML tags, each having the "article-block" class:

```
    <div class="articles-list cf" data-subst="article-list">
      <div class="article-block">
        <a class="article-url"
href="/plosone/article?id=10.1371%2Fjournal.pone.0156445">
        <img class="grayscale" src=" ">
        </a>
        <div class="details">
                <!--article page Web link -->
                <!--article title -->
                <!--article authors -->
        </div>
      </div>
      <!--next article block -->
```

3. Each article's Web page was accessed and analyzed in order to get submission date, accepted date and publication date. As for the journal issue Web page, a prior analysis of the article Web page structure has been done in order to write the code sequence that will find and retrieve the article's bibliographic information.

Given the restricted number of analyzed journals it was more efficient to define a search & retrieval pattern particular to each journal's website internal structure.

For all analyzed journals, the recorded information was retrieved from their public pages. Because a Web scraper is capable of requesting Web pages at a much higher rate than a human user, special care was taken not to disrupt or affect the Web server functionality, e.g., by pausing the scraper process for several seconds after parsing each issue page. The scraper script and its technicalities are available in Supplementary Information section.

The first journal where our scraper attempted to get the information was Physica A[1]. During 29-30 April 2016 period, we collected more than 9000 papers. On the second day of inquiries, Elsevier banned access to its computer which meant that all journals included in Science Direct (not just Physica A) were inaccessible. Starting with 29th April 2016 till 22nd May 2016, our scraper got data (almost 149 000 papers) from PLOS ONE journal[2]. Within first ten days of May 2016 our scraper got more than 4 500 papers from Nature[3]. We took into consideration only research papers (labeled as letters or articles). Because the format of Nature's website changed in 2010, no information was retrieved from Nature prior to 2010. Between 22nd May 2016 and 26th May 2016, our scraper investigated Cell website[4] and retrieved information for papers published between 2003 and 2016. Prior to 2003 there was no information about date when a manuscript was accepted/published. Thus, we included within our dataset more than 4 000 papers from Cell. Between January and April 2017, during the second data collection process, information regarding corresponding authors' affiliation country has been appended to the initial database.

[1] Scraper worked on the following address: http://www.sciencedirect.com/science/journal/03784371/

[2] Scraper worked on the following address: http://journals.plos.org/plosone/browse/ while for the second round of data collection an archive (available on: https://www.dropbox.com/s/n6ldppdrgq11xyo/All%20of%20PLOS.tgz?dl=0 ) with metadata information regarding PLOS ONE's papers was used.

[3] Scraper worked on the following address: http://www.nature.com/nature/archive/index.html

Even if the day of submission can be determined, there is a potential bias induced by the server location (e.g. a paper submitted from Europe at 1:00 GMT is registered on the previous day if the server location is in USA). We were unable to resolve this issue since the hour of submission is unavailable. In our opinion, this is not an important constraint on the findings, because we have so much data. Another potential bias could occur if the online manuscripts submission systems (portals) do not provide accurate meta data (due to internal procedures required) regarding the submission date. In line with Cabanac & Hartley (2013), the editorial offices of the four journals which our dataset rely on were approached by email with the following question:

> *[..] may I ask you if there is any delay/ bias from the moment which a co-author hits "submit" button and the one (calendar date) which is registered in your system as submitted date?*

All four editorial offices confirmed to us that there is no such bias in their processes. One could claim that since Nature employs full-time editors, whom typically do not work on weekends, their metadata is special. Therefore, we provide below the clarification message received from Nature's editorial office (Teresa Dudley) regarding this issue:

> *"At Nature, the submission date that appears on the published paper is the date of submission of the version that received a decision to revise and resubmit, in other words a positive decision. In some cases, authors whose paper was rejected without a request to revise after peer review may return to the journal much later with a substantially revised manuscript; if this manuscript is eventually published, the submission date on the published paper will be that of the later submission."*

## 2.2. Recorded variables

For each paper which passed a peer review process several variables were recorded: (i) journal; (ii) title; (iii) volume; (iv) author(s); (v) initial reception date; (vi) revised version reception date – if any; (vii) acceptance date – if any; (viii) online availability date – if any and (ix) number of pages – if available. Numerical values were assigned as follows: (x) number of pages – if available; (xi) number of authors; (xii) week-day of initial submission (1 for Monday, 2 for Tuesday, 3 for Wednesday, 4 for Thursday, 5 for Friday, 6 for Saturday and 7 for Sunday); (xiii) week-day of revised version (same codification like for variable xii); (xiv) week of initial submission (a number between 1 and 53); (xv) year of initial submission; (xvi) year of publication/ acceptance. In a second round of data collection, information regarding (xvii) corresponding authors' affiliation country is recorded. For few thousand papers, the process of choosing affiliation country has been performed manually. Around a few hundred papers for which affiliation country was not clear enough (e.g. multiple corresponding authors from multiple countries, international organizations with no clear affiliation country, not enough data etc.) have been withdrawn from the database.

[4] Scraper worked on the following address: http://www.cell.com/

### 2.3. The dependent variable

For each journal distribution of papers by day of the week was tested (via Chi Square) against uniformity. The critical value for a significance level 0.05 and 6 degrees of freedom is 12.59. The same approach was conducted for the consolidated dataset. In addition the same procedure was applied if considering only week-days. The critical value being in this case 9.49 (for a significance level of 0.05 and 4 degrees of freedom).

Furthermore, in order to test the sensitivity of deviations from uniformity, a regression model was designed. The time span for rolling over the calendar is the week. For each week and for each journal the following sequence was computed:

$$UD_{ij}=N_{ij}/7=\frac{\sum_{k=1}^{7} n_{kij}}{7}=\frac{1}{7}\sum_{k=1}^{7}\sum_{a=1}^{p} a_{kij} \quad (1)$$

where $UD_{ij}$ denotes the uniform distribution of papers per day submitted during week *i* to journal *j*, $N_{ij}$ is the total number of papers received during week *i* by journal *j* and $n_{kij}$ is the number of papers submitted in the k[th] day of the week starting from Monday (1) to Sunday (7), in the week, *i*, of the year, for a particular journal *j*.

Particularly *a* is the lowest unit recorded, the article and it is referred by the day of the week of submission (*k*), week of the year (*i*) and journal where it was submitted (*j*). In a given day, *k* from week *i*, journal *j* we have 0 to maximum *p* articles submitted.

In this context, ratio to uniform distribution (R*UD*) for a specific day *k*, of the week *i* for journal *j* is defined as follows:

$$RUD_{kij}=\frac{\frac{1}{N_{ij}}\sum_{k=1}^{7} n_{kij}}{UD_{ij}} \quad (2.1)$$

N.B. For “a” articles submitted in the same day of the week($k$), in the same week($i$) of the year we have the same RUD.

For the consolidated data set of the journals, we have:

$$RUD_{ki} = \frac{\frac{1}{N_i}\sum_{j=1}^{p}\sum_{k=1}^{7} n_{kij}}{UD_i}, \qquad (2.2)$$

where $p$=4, represents the number of journal taken into consideration

In order to be clearer how the dependent variable is computed, an extract from the database of 47 PLOS ONE papers received in 14 weeks is presented in table 2. The step by step computation details are also available in table 2. The reception flow is quite diverse: there are weeks with one, two, three or four received papers, but there is one with 26. The daily flows are diverse too; most of the days show only one incoming paper but there is an exceptional one (7th August, 2006) with 16.

**Table 2.** Database extract and step by step computation for the dependent variable $RUD_{kij}$

| ID article (a) | Received date | Weekday (k=1=Monday to 7=Sunday) | Received Week(i) | $UD_{ij}$ | | $RUD_{kij}$ | |
|---|---|---|---|---|---|---|---|
| 167044 | 09-12-05 | 5 | 50 | 1/7= | 0.142857 | 1/0.14= | 7 |
| 167045 | 13-02-06 | 1 | 7 | 4/7= | 0.571429 | 1/0.57= | 1.75 |
| 167046 | 14-02-06 | 2 | 7 | 4/7= | 0.571429 | 2/0.57= | 3.5 |
| 167047 | 14-02-06 | 2 | 7 | 4/7= | 0.571429 | 2/0.57= | 3.5 |
| 167043 | 17-02-06 | 5 | 7 | 4/7= | 0.571429 | 1/0.57= | 1.75 |
| 167042 | 08-03-06 | 3 | 10 | 1/7= | 0.142857 | 1/0.14= | 7 |
| 167041 | 16-03-06 | 4 | 11 | 1/7= | 0.142857 | 1/0.14= | 7 |
| 167040 | 27-03-06 | 1 | 13 | 3/7= | 0.428571 | 1/0.43= | 2.33333 |
| 167033 | 28-03-06 | 2 | 13 | 3/7= | 0.428571 | 1/0.43= | 2.33333 |
| 166855 | 29-03-06 | 3 | 13 | 3/7= | 0.428571 | 1/0.43= | 2.33333 |
| 167038 | 06-04-06 | 4 | 14 | 1/7= | 0.142857 | 1/0.14= | 7 |
| 166926 | 09-05-06 | 2 | 19 | 2/7= | 0.285714 | 1/0.29= | 3.5 |
| 166885 | 11-05-06 | 4 | 19 | 2/7= | 0.285714 | 1/0.29= | 3.5 |
| 167034 | 16-05-06 | 2 | 20 | 2/7= | 0.285714 | 1/0.29= | 3.5 |
| 166750 | 18-05-06 | 4 | 20 | 2/7= | 0.285714 | 1/0.29= | 3.5 |
| 167037 | 02-06-06 | 5 | 22 | 2/7= | 0.285714 | 2/0.29= | 7 |
| 167039 | 02-06-06 | 5 | 22 | 2/7= | 0.285714 | 2/0.29= | 7 |
| 167036 | 19-06-06 | 1 | 25 | 1/7= | 0.142857 | 1/0.14= | 7 |
| 167032 | 30-06-06 | 5 | 26 | 1/7= | 0.142857 | 1/0.14= | 7 |
| 166830 | 26-07-06 | 3 | 30 | 1/7= | 0.142857 | 1/0.14= | 7 |
| 167029 | 01-08-06 | 2 | 31 | 1/7= | 0.142857 | 1/0.14= | 7 |
| 138586 | 07-08-06 | 1 | 32 | 26/7= | 3.714286 | 16/3.71= | 4.30769 |
| 166695 | 07-08-06 | 1 | 32 | 26/7= | 3.714286 | 16/3.71= | 4.30769 |
| 166757 | 07-08-06 | 1 | 32 | 26/7= | 3.714286 | 16/3.71= | 4.30769 |
| 166870 | 07-08-06 | 1 | 32 | 26/7= | 3.714286 | 16/3.71= | 4.30769 |
| 166937 | 07-08-06 | 1 | 32 | 26/7= | 3.714286 | 16/3.71= | 4.30769 |
| 166954 | 07-08-06 | 1 | 32 | 26/7= | 3.714286 | 16/3.71= | 4.30769 |
| 166957 | 07-08-06 | 1 | 32 | 26/7= | 3.714286 | 16/3.71= | 4.30769 |
| 166963 | 07-08-06 | 1 | 32 | 26/7= | 3.714286 | 16/3.71= | 4.30769 |
| 166970 | 07-08-06 | 1 | 32 | 26/7= | 3.714286 | 16/3.71= | 4.30769 |
| 166988 | 07-08-06 | 1 | 32 | 26/7= | 3.714286 | 16/3.71= | 4.30769 |
| 166992 | 07-08-06 | 1 | 32 | 26/7= | 3.714286 | 16/3.71= | 4.30769 |
| 166996 | 07-08-06 | 1 | 32 | 26/7= | 3.714286 | 16/3.71= | 4.30769 |
| 167018 | 07-08-06 | 1 | 32 | 26/7= | 3.714286 | 16/3.71= | 4.30769 |
| 167022 | 07-08-06 | 1 | 32 | 26/7= | 3.714286 | 16/3.71= | 4.30769 |
| 167024 | 07-08-06 | 1 | 32 | 26/7= | 3.714286 | 16/3.71= | 4.30769 |
| 167030 | 07-08-06 | 1 | 32 | 26/7= | 3.714286 | 16/3.71= | 4.30769 |
| 166749 | 08-08-06 | 2 | 32 | 26/7= | 3.714286 | 5/3.71= | 1.34615 |
| 166922 | 08-08-06 | 2 | 32 | 26/7= | 3.714286 | 5/3.71= | 1.34615 |
| 167004 | 08-08-06 | 2 | 32 | 26/7= | 3.714286 | 5/3.71= | 1.34615 |
| 167009 | 08-08-06 | 2 | 32 | 26/7= | 3.714286 | 5/3.71= | 1.34615 |
| 167031 | 08-08-06 | 2 | 32 | 26/7= | 3.714286 | 5/3.71= | 1.34615 |
| 167012 | 09-08-06 | 3 | 32 | 26/7= | 3.714286 | 2/3.71= | 0.53846 |
| 167025 | 09-08-06 | 3 | 32 | 26/7= | 3.714286 | 2/3.71= | 0.53846 |
| 167003 | 11-08-06 | 5 | 32 | 26/7= | 3.714286 | 2/3.71= | 0.53846 |
| 167028 | 11-08-06 | 5 | 32 | 26/7= | 3.714286 | 2/3.71= | 0.53846 |
| 167023 | 12-08-06 | 6 | 32 | 26/7= | 3.714286 | 1/3.71= | 0.26923 |

Furthermore, a look on the dependent variable is important. This variable should be inserted as the dependent variable within the regression models. In order to do so, a preliminary inspection of its distribution (within each individual journal and also to the consolidated dataset) is necessarily. In order to use “Ordinarily Least Squares” (OLS) regression parameter estimations at least a normal distribution is expected for the dependent variable. A classical test – Jarque-Bera (JB)[5] – for normality is used (table 3.a.).

[5] For each dataset, there are two hypothesis: $H_0$: normal (Gaussian) distribution and $H_1$: the distribution is not a normal one. The statistic test is computed by: $JB = N\left(\frac{Skewness^2}{6} + \frac{Kurtosis^2}{24}\right)$ where N is number of cases and skewness and kurtosis are measured basically by Pearson’s moment coefficients. This test follows a Chi square distribution with two degrees of freedom (most common critical value is: $\chi^2_{0.05,2}$=5.99). A detailed presentation about formula used for Kurtosis and its components is available within supplementary information (SI_ Kurtosis.docx).

**Table 3.a.** Jarque-Bera normality test for RUD

| Dataset (sample) | Skewness | Kurtosis | Cases (N) | Jarque-Bera test |
|---|---|---|---|---|
| Physica A | 1.291 | 3.417 | 9 825 | 7 509.01 |
| PLOS ONE | 0.649 | 14.069 | 160 172 | 1 332 241.06 |
| Nature | 1.865 | 8.340 | 4 653 | 16 182.46 |
| Cell | 1.348 | 2.429 | 3 777 | 2 072.39 |
| Consolidated | 2.606 | 20.306 | 178 427 | 3 267 433.88 |

The potential dependent variable – RUD – does not follow normal distribution, no matter which dataset is considered. The most common way to adapt a non-Gaussian distribution for a potential use within the regression models is the log transformation (table 3.b.).

**Table 3.b.** Jarque-Bera normality test for $\log_{10}$RUD

| Dataset (sample) | Skewness | Kurtosis | Cases (N) | Jarque-Bera test |
|---|---|---|---|---|
| Physica A | -0.395 | 0.056 | 9 825 | 256.77 |
| PLOS ONE | -2.056 | 4.855 | 160 172 | 270 153.90 |
| Nature | -0.441 | 0.374 | 4 653 | 177.94 |
| Cell | -0.032 | -0.397 | 3 777 | 25.45 |
| Consolidated | -1.414 | 4.014 | 178 427 | 179 243.15 |

The log transformation helps and the computed values for JB test became three digits for Physica A and Nature or even better – two digits – for Cell while within the consolidated dataset and PLOS ONE there are six digits figures. Two steps classical transformation (Linnet, 1987) are used. First one aims to avoid skewness while the second one tries to avoid kurtosis. Various parameters are tested for each dataset and those which minimize the JB test are kept (table 3.c.).

**Table 3.c.** Two steps transformations and their subsequent Jarque-Bera normality test

| Dataset (sample) | First step transformation | Second step transformation | Skewness | Kurtosis | Cases (N) | Jarque-Bera test |
|---|---|---|---|---|---|---|
| Physica A | $y^*_{PA} = \log_{10} RUD - 0.06$ | $y^{**}_{PA} = {[(y^*_{PA} + 1)^{1.82} - 1]}/{1.82}$ | 0.056 | -0.135 | 9 825 | 12.60 |
| PLOS ONE | $y^*_{PO} = {[(RUD)^6 - 1]}/{6}$ | $y^{**}_{PO} = {[(y^*_{PO} + 1)^{-0.8} - 1]}/{-0.8}$ | 0.129 | -0.094 | 160 172 | 503.21 |
| Nature | $y^*_{NT} = \log_{10} RUD + 0.05$ | $y^{**}_{NT} = {[(y^*_{NT} + 1)^{0.8} - 1]}/{0.8}$ | -0.065 | 0.224 | 4 653 | 13.00 |
| Cell | $y^*_{CL} = \log_{10} RUD - 0.05$ | $y^{**}_{CL} = {[(y^*_{CL} + 1)^{1.15} - 1]}/{1.15}$ | 0 | -0.386 | 3 777 | 23.45 |
| Consolidated | $y^*_{CT} = {[(RUD)^4 - 1]}/{4}$ | $y^{**}_{CT} = {[(y^*_{CT} + 1)^{-0.954} - 1]}/{-0.954}$ | -0.067 | 0.302 | 178 427 | 811.55 |

After the transformations, the computed JB test for each sample looks much better even if the null hypothesis (Gaussian distribution) is still rejected. For Physica A, Nature and Cell the outcomes are small (two digits not far away from the critical value) while for PLOS ONE and the consolidated dataset there are three digits level records. A visual inspection (figure 1) of these distributions shows that there are plenty of outliers.

**Physica A**

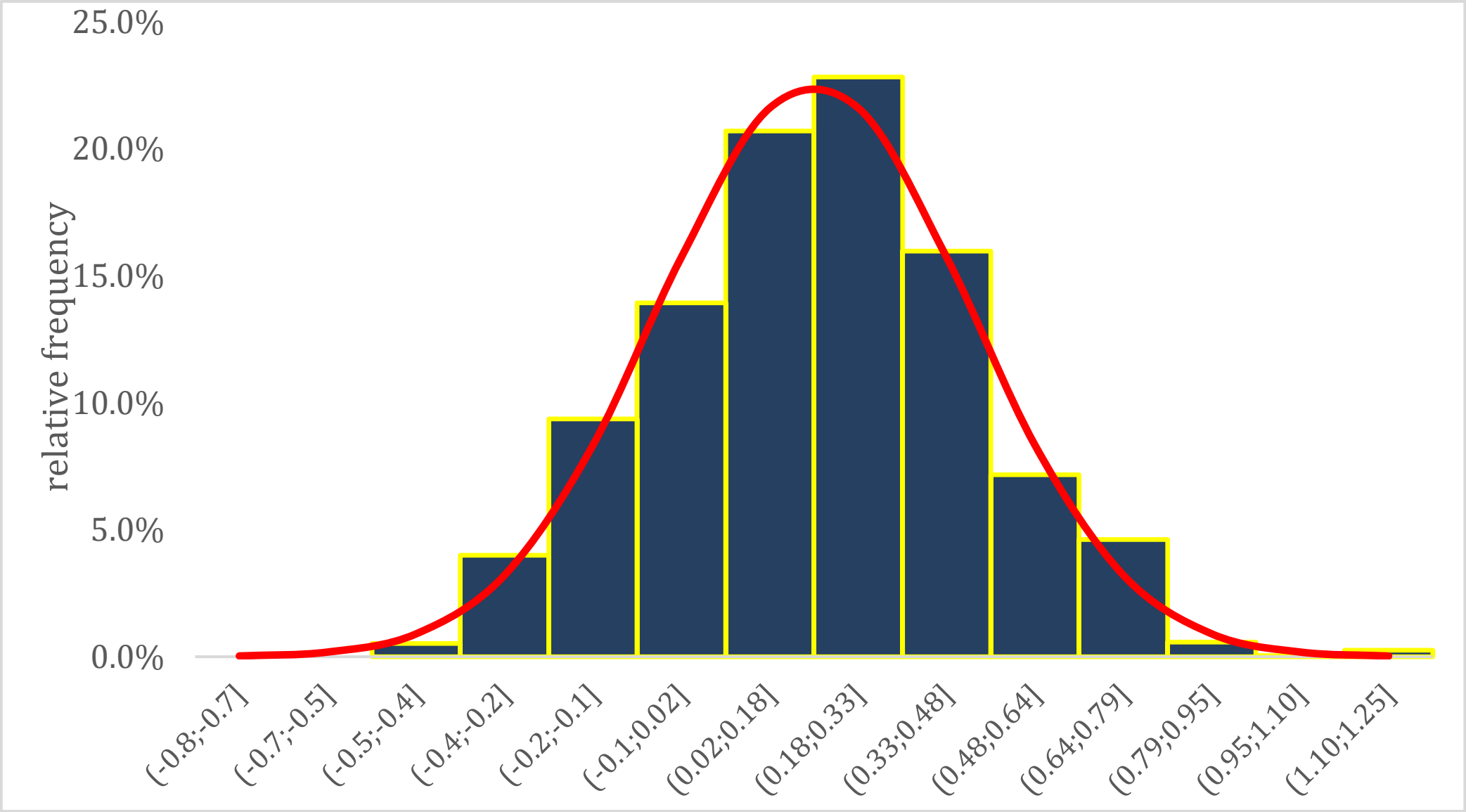

**PLOS ONE**

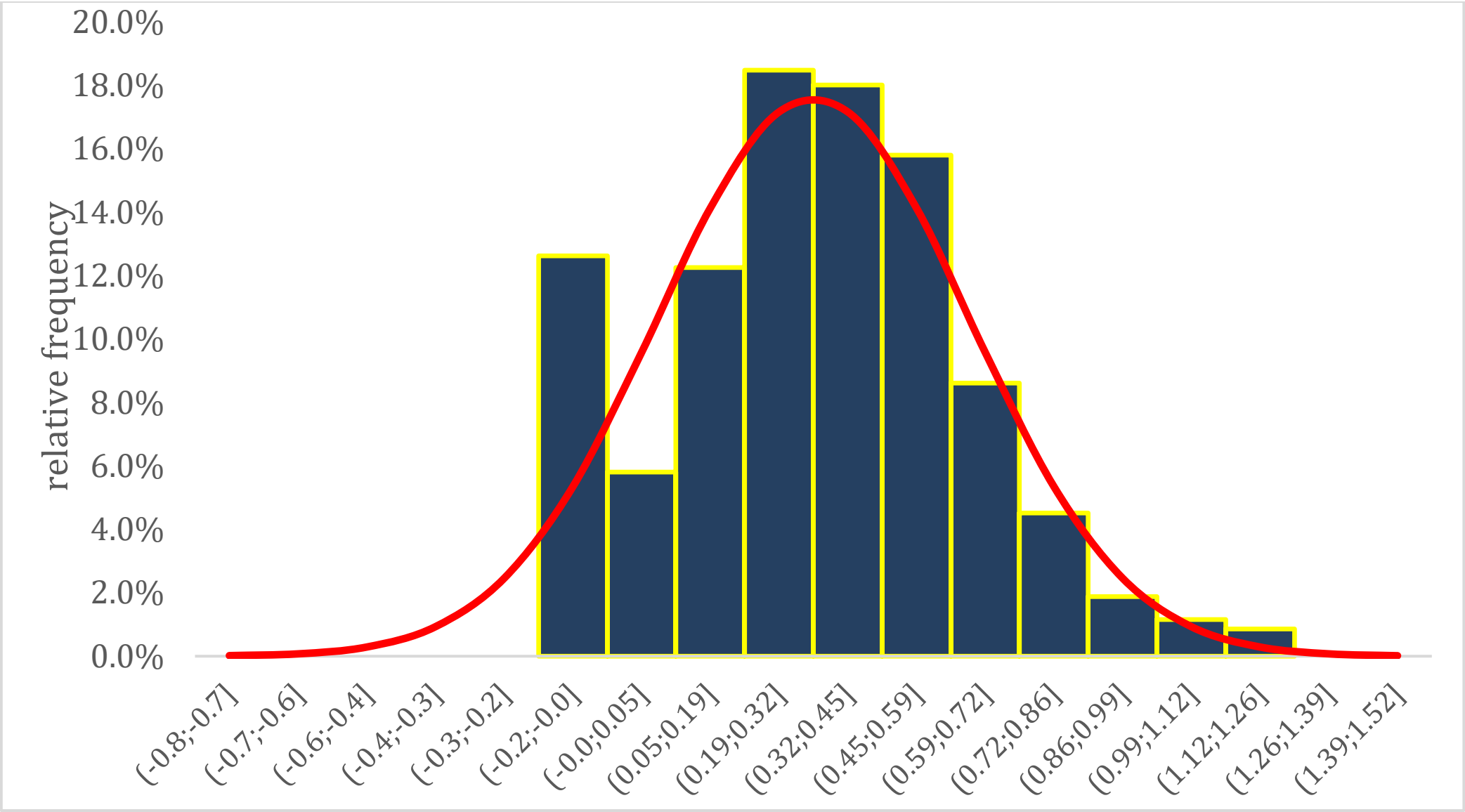

**Nature**

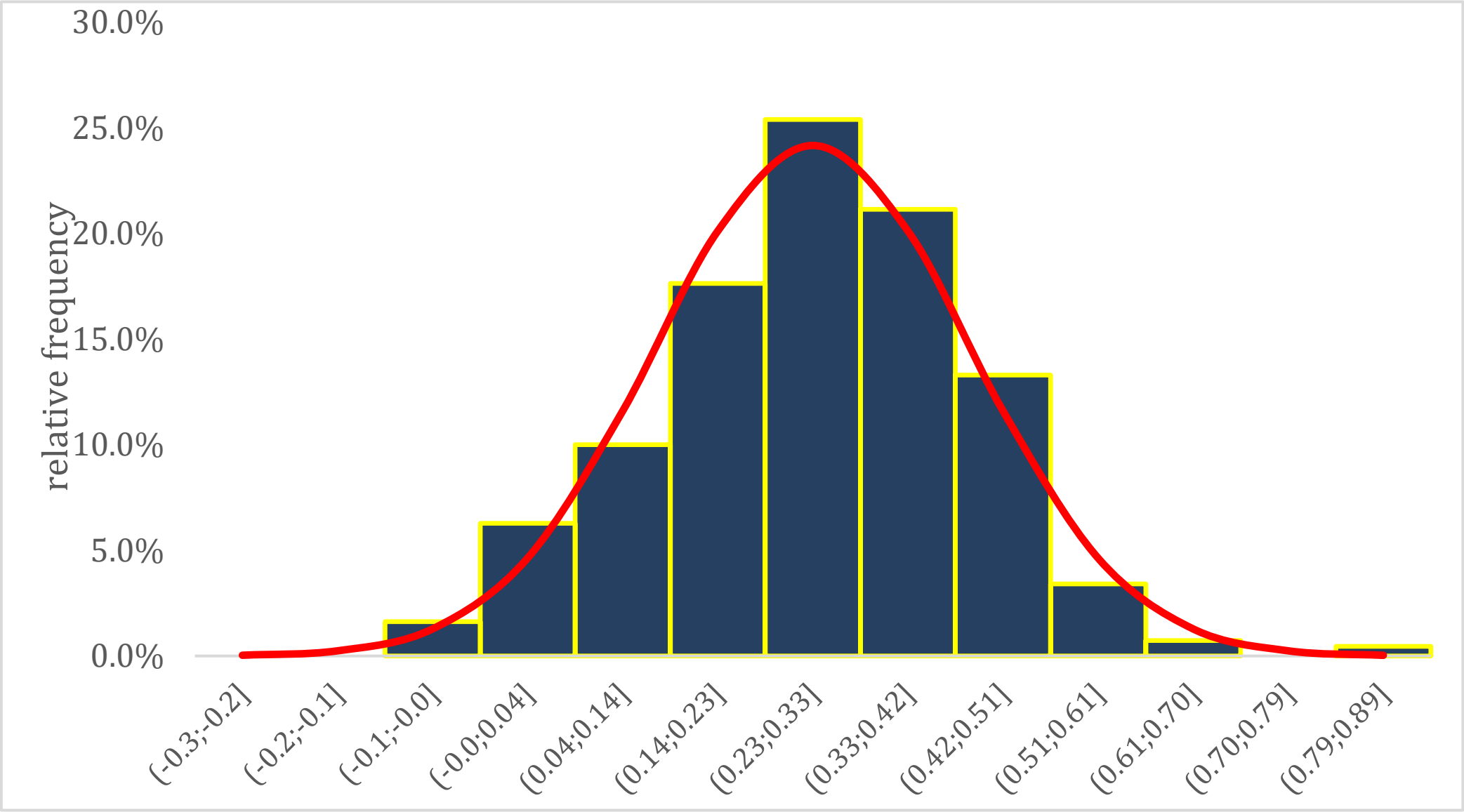

**Cell**

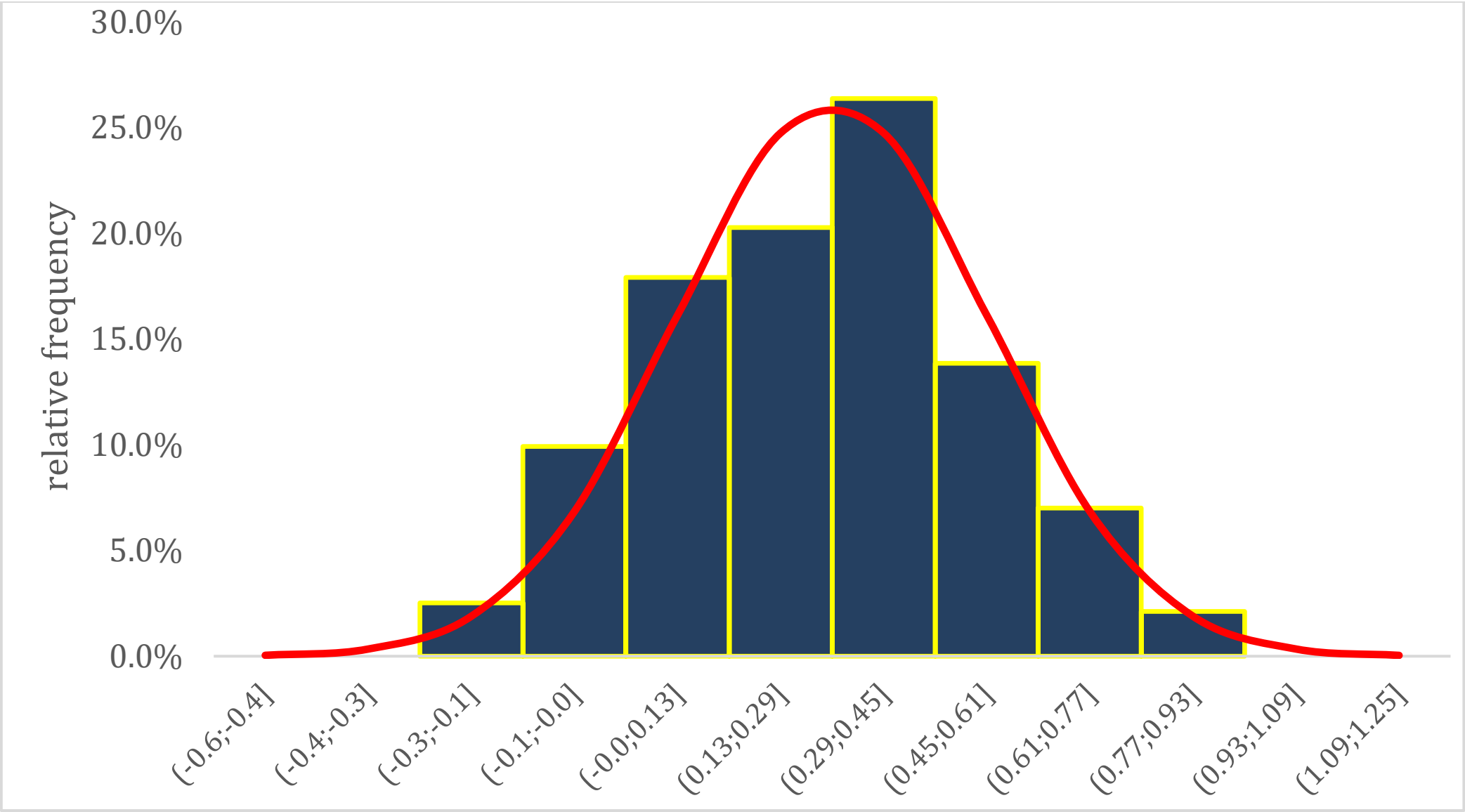

**Consolidated dataset**

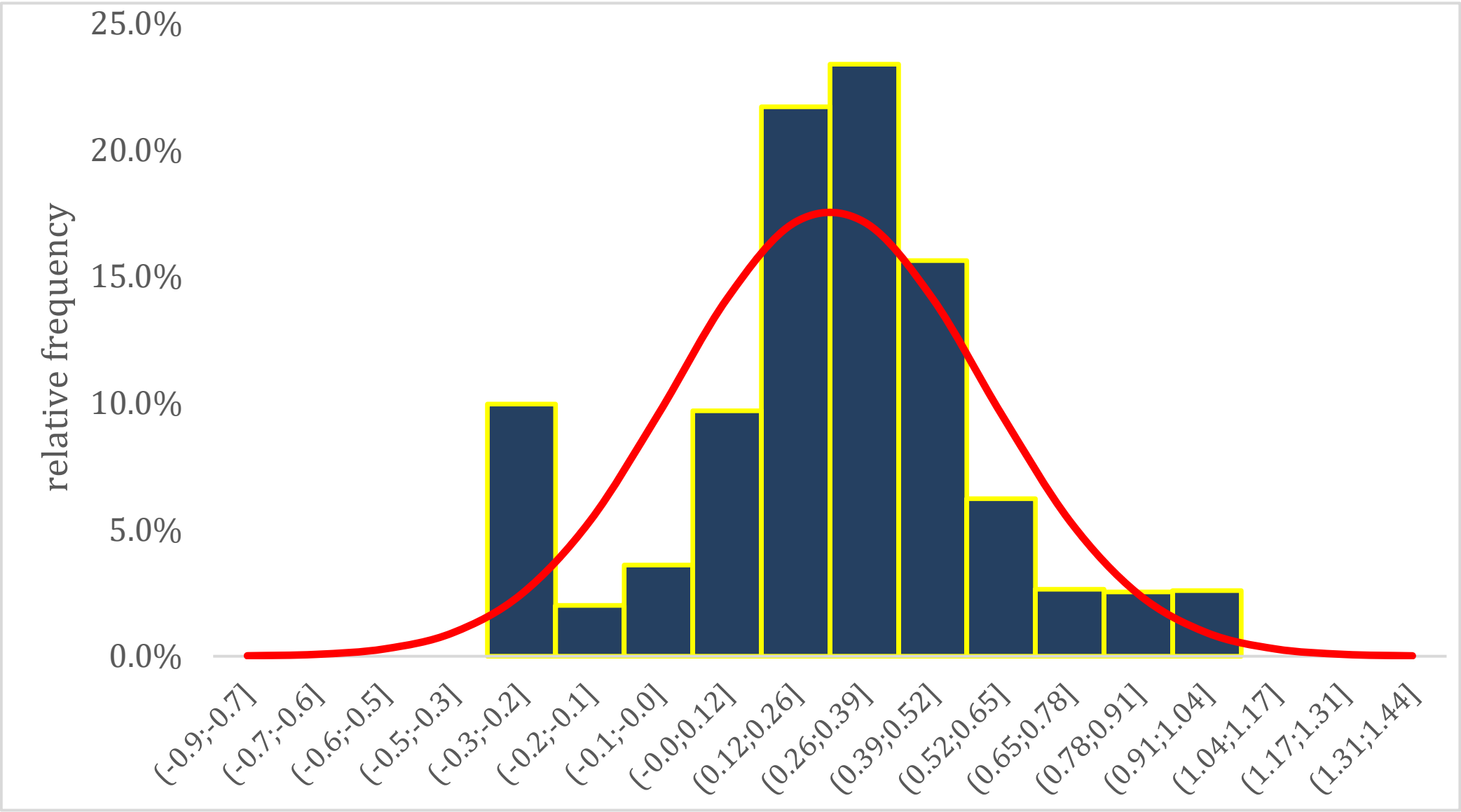

**Figure 1.** Distributions of dependent variable by intervals

## 2.4. The regression models

*a) Article unit model with week-day component*

This model is a linear one:

$$y_{akij}^{**} = \beta_0 + \beta_1 MON_{a1ij} + \beta_2 TUE_{a2ij} + \beta_3 WED_{a3ij} + \beta_4 THU_{a4ij} + \beta_6 \sum_{k=6}^{7} WEEKEND_{akij} + \varepsilon_{akij} \quad (3)$$

where $\beta_0$ to $\beta_6$ are regression parameters to be estimated. In this kind of model *a* is the index for a certain article received in the day *k*, week *i* for journal *j*, as defined above at (1).

It worth to underline that day of the week dummy from (3) will vary only by index *k*.

*MON, TUE, WED,* and *THU* are dummy variables (1 if paper was submitted in a Monday or Tuesday or Wednesday or Thursday and 0 otherwise). *WEEKEND* is also a dummy variable (generally, 1 if paper was submitted in a Saturday or in a Sunday and 0 otherwise). However, a subtle approach, introduced recently (Campos-Arceiz *et al.*, 2013), is used for manuscripts originating from almost two dozen of Countries having Special Weekends (SWC). It is about: Afghanistan, Algeria, Bahrain, Bangladesh, Brunei, Egypt, Hong-Kong, Iraq, Israel, Jordan, Kuwait, Libya, Mauritania, Nepal, Oman, Qatar, Saudi Arabia, Sudan, Syria, United Arab Emirates, and Yemen. Most common weekend type for this SWC is, currently, Friday and Saturday. It was also taken into account the changing of the weekend rules over time (2006-2009) as has been detailed by Campos-Arceiz *et al.* (2013). FRIDAY is considered to be the reference point. ε denoting the "perturbance" (here as well as in the next models), due to other factors, is assumed to be a white noise as usual in such regression model schemes.

*b) Article unit model with seasonal component*

In order to test other reported (Shalvi *et al.*, 2010; Bornmann & Daniel, 2011; Schreiber, 2012) seasonal effects, three dummy variables are used (1 if paper was submitted within a season and 0 otherwise). Classic definition of seasons, as Hartley (2011) suggests, is used: *SPRING* (March to May), S*UMMER* (June to August), *FALL* (September to November) and *WINTER* (December to February).

$$y_{akij}^{**} = \beta_0 + \beta_1 SPRING_{akij} + \beta_2 SUMMER_{akij} + \beta_3 FALL_{akij} + \varepsilon_{akij} \qquad (4)$$

WINTER is considered to be the reference point and, the indexes *a*, *k*, *i*, *j* have the same significance as in equation (3).

Here we note that season dummies from (4) will vary only by index *i*, according to the position of the week, *i* 1 to 52 in one season or another.

*c) Article unit model with geographic component*

Geographical data (corresponding authors' affiliation country) is included in this model using four dummy variables (1 if the paper is originated within a specific continent and 0 otherwise): *AFRICA*, *AMERICA*, *ASIA*, and *OCEANIA*.

$$y_{akij}^{**} = \beta_0 + \beta_1 AMERICA_{akij} + \beta_2 AFRICA_{akij} + \beta_3 ASIA_{akij} + \beta_4 OCEANIA_{akij} + \varepsilon_{akij} \qquad (5)$$

*EUROPE* is considered to be the reference point, the indexes *a*, *k*, *i*, *j* have the same significance as in equation (3).

Here the geographic component from (5) will vary only by index *a*, since this variable is associated with the author of a certain article: *a*.

*d) Control variables and subsequent models*

We are aware that there are plenty of factors which can influence our dependent variable. Previous research demonstrates that the scientific production, in general, can be affected by various factors: country size (Lippi & Mattiuzzi, 2017); economic level (Bernardes & Albuquerque, 2003; Lippi & Mattiuzzi, 2017); level of funding and/ or science policy (Henriques & Larédo, 2013; Crespi & Geuna, 2008; Ebadi & Schiffauerova, 2016); team size (Ebadi & Schiffauerova, 2016); non-economic factors (Inönü, 2003). We select from this list two factors which can affect scientific papers' production (and flows): *HDI* (the numeric figure for each country regarding Human Development Index[6]) as a proxy for the economic (lack) of development and *AUTHORS* which denotes the number of authors for the paper *j*. In case of the AUTHORS, it is expected that an increase in team size also increases the chance that a co-author who does not work (no matter what kind of activity: manuscript design, preparation for submission, agreement on the final form etc.) outside of the regular program (e.g. during the weekends). Subsequently, this control factor should be positively correlated to the dependent variable.

Furthermore, since the current paper is related in a great extent to the timing, we are identifying a very easy to measure important period of time which may interfere on leisure/ working time: C*HRISTMAS* is a dummy variable (1 if paper was submitted between 20th of December and 10th of January and 0 otherwise). When is about leisure/ working time balance, an interesting idea was coined by Hofstede *et al.* (2010:251). Here they explain what are the differences between Short Term (STO) and Long Term Oriented (LTO) countries. They state that for LTO countries "*leisure time is not important*" while for STO countries "*leisure time is important*". Therefore, in our case is expected that countries with a higher *LTO* index to work harder during the week-ends. Hence, we take exact figures about *LTO* for 93 countries (Hofstede *et al.* 2010:255-258) and insert this variable within the models.

Based on this the following general multiple regression models are designed:

$$y_{akij}^{**} = \beta_0 + \beta_1 MON_{akij} + \beta_2 TUE_{akij} + \beta_3 WED_{akij} + \beta_4 THU_{akij} + \\ + \beta_5 WEEKEND_{akij} + \beta_6 SPRING_{akij} + \beta_7 SUMMER_{akij} + \beta_8 FALL_{akij} + \varepsilon_{akij} \quad (6)$$

$$y_{akij}^{**} = \beta_0 + \beta_1 MON_{akij} + \beta_2 TUE_{akij} + \beta_3 WED_{akij} + \beta_4 THU_{akij} + \\ \beta_5 WEEKEND_{akij} +$$

[6] Data for HDI2016 is available on: http://hdr.undp.org/en/data . It is known that HDI is a geometric mean of three normalized indices (health, education, and income).

$$+\beta_6 SPRING_{akij}+\beta_7 SUMMER_{akij}+\beta_8 FALL_{akij}+\beta_9 AMERICA_{akij}+\beta_{10} AFRICA_{akij} + +\beta_{11} ASIA_{akij}+\beta_{12} OCEANIA_{akij} + \varepsilon_{akij} \quad (7)$$

$$y^{**}_{akij} = \beta_0 + \beta_1 MON_{akij} + \beta_2 TUE_{akij} + \beta_3 WED_{akij} + \beta_4 THU_{akij} + \beta_5 WEEKEND_{akij} + +\beta_6 SPRING_{akij}+\beta_7 SUMMER_{akij}+\beta_8 FALL_{akij}+\beta_9 AMERICA_{akij}+\beta_{10} AFRICA_{akij} + +\beta_{11} ASIA_{akij}+\beta_{12} OCEANIA_{akij}+\beta_{13} CHRISTMAS_{akij} + \varepsilon_{akij} \quad (8)$$

$$y^{**}_{akij} = \beta_0 + \beta_1 MON_{akij} + \beta_2 TUE_{akij} + \beta_3 WED_{akij} + \beta_4 THU_{akij}+\beta_5 WEEKEND_{akij} + \beta_6 SPRING_{akij}+\beta_7 SUMMER_{akij}+\beta_8 FALL_{akij}+\beta_9 AMERICA_{akij}+\beta_{10} AFRICA_{akij} + +\beta_{11} ASIA_{akij}+\beta_{12} OCEANIA_{akij}+\beta_{13} CHRISTMAS_{akij} + \beta_{14}\log_{10} AUTHORS_{akij} + \varepsilon_{akij} \quad (9)$$

$$y^{**}_{akij} = \beta_0 + \beta_1 MON_{akij} + \beta_2 TUE_{akij} + \beta_3 WED_{akij} + \beta_4 THU_{akij}+\beta_5 WEEKEND_{akij} + \beta_6 SPRING_{akij}+\beta_7 SUMMER_{akij}+\beta_8 FALL_{akij}+\beta_9 AMERICA_{akij}+\beta_{10} AFRICA_{akij} + +\beta_{11} ASIA_{akij}+\beta_{12} OCEANIA_{akij}+\beta_{13} CHRISTMAS_{akij} + \beta_{14}\log_{10} AUTHORS_{akij} + +\beta_{15}\log_{10} HDI_{akij} + \varepsilon_{akij} \quad (10)$$

$$y^{**}_{akij} = \beta_0 + \beta_1 MON_{akij} + \beta_2 TUE_{akij} + \beta_3 WED_{akij} + \beta_4 THU_{akij}+\beta_5 WEEKEND_{akij} + \beta_6 SPRING_{akij}+\beta_7 SUMMER_{akij}+\beta_8 FALL_{akij}+\beta_9 AMERICA_{akij}+\beta_{10} AFRICA_{akij} + +\beta_{11} ASIA_{akij}+\beta_{12} OCEANIA_{akij}+\beta_{13} CHRISTMAS_{akij} + \beta_{14}\log_{10} AUTHORS_{akij} + +\beta_{15}\log_{10} HDI_{akij}+\beta_{16}\log_{10} LTO_{akij} + \varepsilon_{akij} \quad (11)$$

The meaning of the indexes *a*, *k*, *i*, *j*, are the same as in (3), underlining that the AUTHORS, HDI and LTO will vary only by index *a*, CHRISTMAS by particular cases of index *i*.

When the consolidated dataset is considered, the models are the same, the only thing which differs from the models 6-11 is the computations of the dependent variable, y. In this case based on the relation 2.2 and the applied transformation the variable will be $y^{**}_{aki}$

The approach for the model design is from simple to complex adding, step by step, specific factors. Hence, at first step (6) the days of the week are combined with seasons while at the second step (7) continents are added. Furthermore, one by one the control factors are inserted: CHRISTMAS (8), AUTHORS (9), HDI (10) and LTO (11).

In order to check potential behavior changes over the time, the dataset is also analyzed in a longitudinal way. The covered timespan is split in five roll windows as follows: [2000-2004], [2005-2007], [2008-2010], [2011-2013], and [2014-2016]. The models (3-11) are run on every roll

window. The same procedure is followed when data set relies only on the individual journals (Nature, Cell, Physica A and PLOS ONE).

The regression models were validated via classical tests for: (i) model itself, (ii) estimated parameters (iii) residuals and (iv) multicollinearity. For all tests, the maximum significance level was set to be equal to 0.1.

*e) The Panel regression*

We define a particular structure of panel as follows: the time component is given by T=2 435 consecutive days from 01.01.2010 to 31.08.2016 and the cross-section is represented by top N=11 countries. Now the characteristics are referred a cell of type (*c*, *t*), the *c* being the index for countries and $t$ for the time(day). The $RUD_{kij}$ defined in (2.1. and 2.2.) becomes $RUD_{ct}$, the variation of day of the week *k* in the week *i* is now contained in *c* index.
With the experience of previous models, in line with (Orazbayev, 2017), a test and a control for other factors is performed. The following model is used:

$$log_{10}RUD_{ct} = \beta_{0c} + \beta_1 MON_{ct} + \beta_2 TUE_{ct} + \beta_3 WED_{ct} + \beta_4 THU_{ct} + \beta_5 WEEKEND_{ct} + \\ + \beta_6 \log_{10} AUTHORS_{ct} + \beta_7 CHRISTMAS_{ct} + \beta_8 FALL_{ct} + \\ + \beta_9 SUMMER_{ct} + \beta_{10} SPRING_{ct} + \beta_{11} log_{10} HDI_{cj} + \beta_{12} log_{10} LTO_{cj} + \beta_{13} log_{10} t + u_c + \\ \lambda_{ct} \qquad (12)$$

In order to keep a relevant daily granularity, we remove some parts of the dataset. First, from the available set of countries, only the top 11 countries were chosen for which total number of papers is more than 77% out of the total papers. The rest of 23% dataset was formed by countries which have a very small number of articles. Second, in the beginning of the period (2000-2009) in most of the days, the number of submitted papers was zero. Therefore a minimal daily granularity was not possible for this specific timespan. Hence, after timespan reduction as well as countries' removal, the dataset used for panel regression records 120 258 papers representing more than two thirds out of the initial 178 427 papers, accumulated in a *N*x*T* =26 785 panel cells as defined above in this sub-section.
In (12) the dependent variable is transformed:

$$log_{10}RUD_{ct} = \frac{1}{p}\sum_{a=1}^{p} log_{10}RUD_{act} \qquad (13)$$

where $p$ is the number of articles published in the day $t$ from country $c$.
The HDI and LTO are covariates specific to a country *j*. They were considered as fixed in time. The number of authors is also calculated as a mean, since in a specific day t as there are more articles with different number of authors.
Hence

$$log_{10}AUTHORS_{ct} = \frac{1}{p}\sum_{a=1}^{p} log_{10}AUTHORS_{act} \qquad (14)$$

In (12) the $u_c$ is the cross-sectional random component which measures the impact of other unknown factors associated with the results.

$$\varepsilon_{ct} = u_c + \lambda_{ct} \qquad (15)$$

denotes the total perturbation component

Moreover, in particular cases, we have multiple articles from country *j* in a specific day *t*. We denote this factor as $n_{ct}$ and is treated as a weighting component. When used this component contains a multiplication of each value of continuous variable $Y_{ct}$ or $X_{ct}$ with the $n_{jt}$ value. Before running the estimation, the dependent variable was tested for the presence of unit root using specific panel data tests using LLC, IPS-WStat, ADF-Statistic. The null hypothesis of unit root presence was rejected each time.

### 2.5. The use of GIS (Geographic Information System) tools

To obtain more insight into the geographical distribution of the manuscripts from the dataset GIS techniques and tools were used.

Spatial data was obtained from WWW5 (2017). This contains at the 0 level the country shape. Data from POP2000, which stores the country's population was recorded for the year 2000. In this way we built a spatial database with all data (spatial and nonspatial) and so could display all the data on a map. Graduated colors allow us to emphasize the intensity of publication activity at the country level. The ArcGIS software package then implemented the spatial database and enabled us to build customized geoprocessing tools for the spatial analysis. Thus, we have three scenarios for spatial analysis:

- Compute and display the total number of papers published by corresponding authors for each country;
- Compute and display the total number of papers published by corresponding authors per population for each country (as being shown in Figure SI.1. – Supplementary Information);
- Compute and display the Localization Quotient (LQ) in order to measure the intensity of publication activity from a country taking into consideration the moment of paper submission.

We adapted a model (Furtună *et al.*, 2013) for the Localization Quotient indicator to determine the intensity of publication activity in a country relative to the whole world.

The time period can be the day, week, month or year of submission or any combination of these combined with journal filtering. Our geoprocessing tool computed the Localization Quotient indicator and displayed it on the map. The model uses as input parameters the selection expression and the number of classes. A complex expression is based on logical operators that allow to establish nonstandard time period. For instance, to select only papers submitted in TUE, WEN, THU week days, the expression is:

*[received_week_day] = 2 OR [received_week_day] = 3 OR [received_week_day] = 4*

In order to select papers submitted SAT and SUN week days and only from *Cell* journal, the expression is:

*[received_week_day] = 2 OR [received_week_day] = 3 AND [idj] = 3*

where *idj* is the field for journal identification and the 3 value identify the *Cell* journal.

Number of classes is used to divide the countries in many groups, each country from a group will be filled with the same color. The statistical method for data classification is *Natural breaks* (WWW6, 2017) This method forms the groups so that the variance of the values within the class is minimal and the variation of the values among classes is maximum.

The geoprocessing model is illustrated in figure SI.2. (from the supplementary information). It uses predefined operation from *ArcToolbox* component and customized operations defined to compute the Localization Quotient indicator (LQ) and to apply the graduated colors symbology on maps (Simb_GC). In order to define the custom operators, it was used the Python language to write de source code, which is available within Supplementary Information section.

## 3. Data analysis

### 3.1. Preliminary descriptive inspection

Thanks to electronically available information we were able to study large data sets. However, the data sets only include information about papers that were accepted. This is in contrast to the smaller data set used by Hartley & Cabanac (2017) and Ausloos *et al.* (2016) which also included information about rejected and withdrawn papers. There is no information about rejected/ withdrawn papers in our set.

We find that of the accepted papers, more were submitted on week days than on weekends (Figure 2).

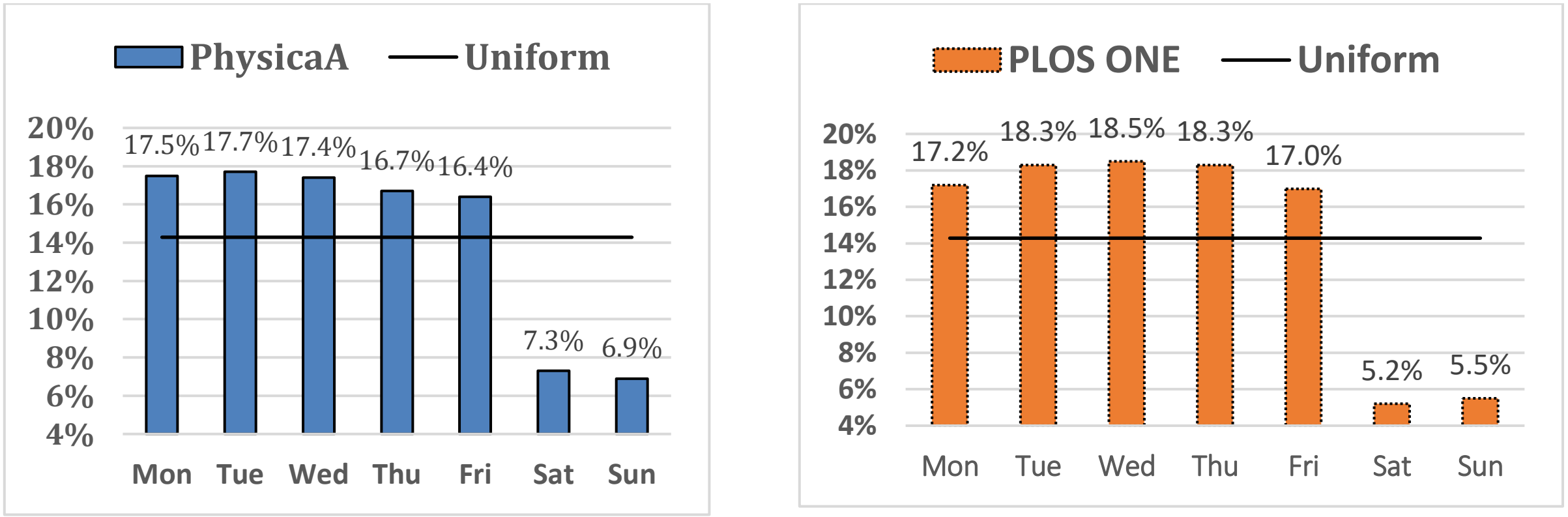

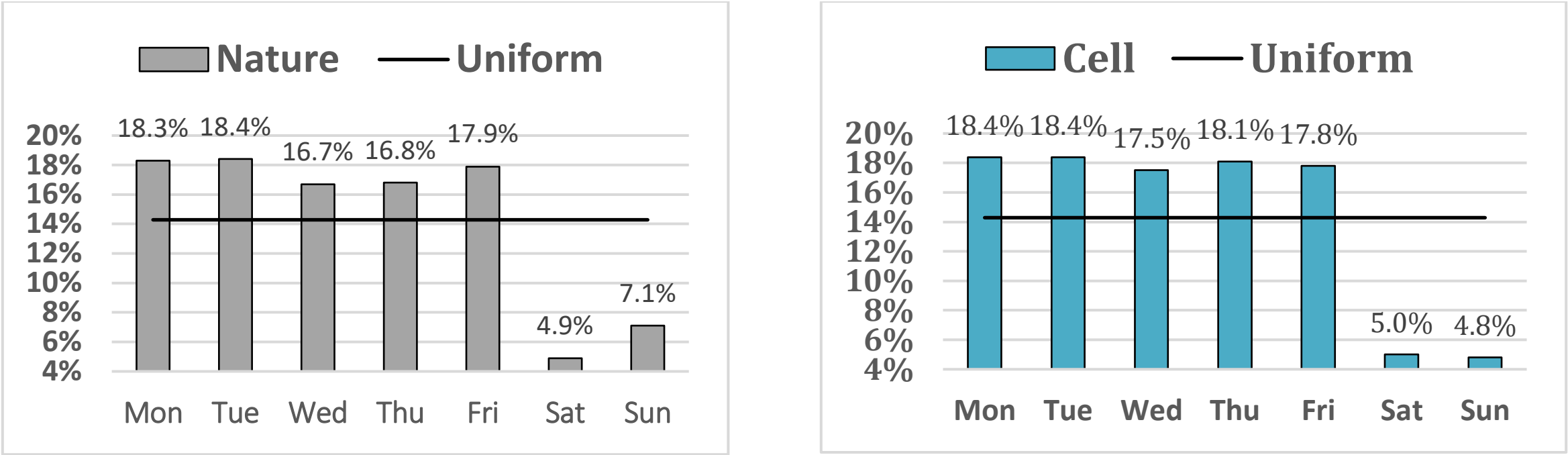

**Figure 2.** The share of papers submitted on each day of the week for the examined journals

**Note:** Percentages were computed for each journal. Chi Square values were computed based on absolute figures for each journal. It is found that the grouping factor (day of the week) is statistically significant (after performing a Chi Square test: p value is less than 0.01), - no matter which journal. The computed values for Chi Square test were: 994.3 (Physica A), 25 321.55 (PLOS ONE), 648.00 (Nature) and 657.82 (Cell). However, if weekends are excluded, only for PLOS ONE there are statistically significant differences among the days. The computed values for Chi Square test being: 7.47 (Physica A), 162.15 (PLOS ONE), 6.85 (Nature) and 1.34 (Cell) while the critical value is: $\chi^2_{0.05,4}$=9.49.

Most of the accepted manuscripts are submitted on Tuesdays for Physica A (17.7%) and Nature (18.4%), On Wednesdays for PLOS ONE (18.5%) while for Cell, Mondays and Tuesdays register 18.4% each. Weekends' submissions are more rarely being under 10% for Cell, around 11% for PLOS ONE, 12% for Nature and more than 14% in case of Physica A. No matter which journal, there are 2-3 times more published papers submitted in any week-day when compare to those which have been submitted Saturday or Sunday.

The geographic location of the papers' corresponding authors (PCA) is another important dimension to be analyzed. Still, this is far from the current research topic. To put our data into context and as an indirect validation that our consolidated sample is consistent, a brief description from the geographical point of view is available in the supplementary Information section.

### 3.2. Regression analysis

We performed multiple ***pooled*** or ***panel*** regression analysis to see clearly how variation in the data depended on the day of the week a paper was submitted. Description analysis (one to one) could also provide interesting information such as we show in figures 2 or SI.3 and SI.4. However, we believe that a regression model better emphasizes the reality since more than one factor can be inserted simultaneous within the model. We are aware that there are a lot of other cofounding variables (authors' gender, age, ethnic and religious affiliation, academic – or not – position, manuscript's number of references, time – hour/minute etc.) which are not included due to the data unavailability. Still, even with these limitations the analysis includes the regression tool in addition to the descriptive ones.

The initial estimations methods were classic OLS for article unit models, models 6-11.

Several tests such as Jarque Bera for testing the normality of residuals from regression, White or Bresuch-Pagan-Godfrey for heteroscedasticity effects of the residuals and Variance Inflation Factor (VIF) for testing the linear dependence of regressor**s** were performed. The classical assumptions of OLS were, in majority of the cases, rejected. Thus, the residuals are not coming from a normal distribution and usually they are heteroskedastic. Moreover, the non-commune behavior or even outliers are omnipresent, affecting the assumption and the quality of the OLS results.

Hence, in order to mitigate this problem, the dependent variable was transformed as we described above and the OLS method was replaced by Robust Least Square (RLS). For the RLS method we selected a Bisquare optimization function. For the scale estimates the median centered method (MAD) was used whereas for the method, the M-estimation. The tuning parameters was kept as proposed in the paper of Holland & Welsch (1977). Our option regarding the selection of objective function, scale estimated, M-estimation method and tuning constants takes into account the spread and number of the outliers (above and below the mean) and the fact that on large samples the objective function seems not to discriminate the model regarding power (Ozlem, 2011)

There is still an exception from such a rule: the robust regression method fails to estimate anything for the sample of manuscripts published by PLOS ONE. In this case, we provide only OLS estimation. When all results are analyzed we find solid evidence about consistency of the particular regression approaches, thus the existence of heteroskedasticities effects does not bias the main conclusions.

The Panel-EGLS (Wooldridge, 2002) with cross-section random effects was used for model 12.

All regressions (individual journals and consolidated data: Table 4 and 5 by time spans or by whole period) are valid (after applying ANOVA/ F test) with a very good significance (p value less than 0.01) while the levels of $R^2$ are varying on a wide range. This low proportion of variance for the dependent variable is expected since a limited number of factors were available for modelling. In addition, sometimes in case of such large datasets, information can be very noisy.

**Table 4.** Regression estimates for the dependent variables for consolidated data set and each journal for each model

| **A. Consolidated data set** | | | | | | | | | |
|---|---|---|---|---|---|---|---|---|---|
| **Models** / **Variables/ characteristics** | **M1 (equation 3)** | **M2 (equation 4)** | **M3 (equation 5)** | **M4 (equation 6)** | **M5 (equation 7)** | **M6 (equation 8)** | **M7 (equation 9)** | **M8 (equation 10)** | **M9 (equation 11)** |
| *0* | *1* | *2* | *3* | *4* | *5* | *6* | *7* | *8* | *9* |
| **Intercept** | 0.237*** | 0.2682*** | 0.2779*** | 0.239*** | 0.238*** | 0.2268*** | 0.235*** | 0.234*** | 0.23*** |
| **Monday** (1 yes, 0 no) | 0.01*** | | | 0.01*** | 0.019*** | 0.011*** | 0.011*** | 0.0144*** | 0.014*** |
| **Tuesday** (1 yes, 0 no) | 0.087*** | | | 0.087*** | 0.088*** | 0.087*** | 0.087*** | 0.09*** | 0.089*** |
| **Wednesday** (1 yes, 0 no) | 0.098*** | | | 0.098*** | 0.098*** | 0.098*** | 0.098*** | 0.1*** | 0.1*** |
| **Thursday** (1 yes, 0 no) | 0.088*** | | | 0.088*** | 0.088*** | 0.087*** | 0.087*** | 0.09*** | 0.089*** |
| **Weekend** (1 yes, 0 no) | -0.539*** | | | -0.539*** | -0.538*** | -0.541*** | -0.541*** | -0.538*** | -0.538*** |
| **Adjusted $R^2$ (weighted)** | **0.609*** ** | | | | | | | | |
| **Spring** (1 yes, 0 no) | | 0.0006 | | 0.0015*** | 0.0014*** | 0.014*** | 0.0139*** | 0.014*** | 0.014*** |
| **Summer** (1 yes, 0 no) | | -0.0004 | | -0.0033*** | -0.0035*** | 0.0089*** | 0.008*** | 0.0091*** | 0.009*** |
| **Fall** (1 yes, 0 no) | | -0.0034*** | | -0.0029*** | -0.0029*** | 0.0095*** | 0.009*** | 0.009*** | 0.009*** |
| **Adjusted $R^2$ (weighted)** | | **0.00004*** ** | | **0.607*** ** | | | | | |
| **America** (1 yes, 0 no) | | | 0.0091*** | | 0.006*** | 0.0063*** | 0.0057*** | 0.0053*** | 0.001*** |
| **Africa** (1 yes, 0 no) | | | -0.0341*** | | | | | -0.014*** | -0.018*** |
| **Asia** (1 yes, 0 no) | | | -0.0461*** | | -0.006*** | -0.005*** | -0.005*** | -0.005*** | -0.003*** |
| **Oceania** (1 yes, 0 no) | | | -0.0391*** | | -0.01*** | -0.009*** | -0.009*** | -0.0105*** | -0.016*** |
| **Adjusted $R^2$ (weighted)** | | | **0.001*** ** | | **0.607*** ** | | | | |
| **Christmas** (1 yes, 0 no) | | | | | | 0.125*** | 0.125*** | 0.126*** | 0.126*** |
| **Adjusted $R^2$ (weighted)** | | | | | | **0.612*** ** | | | |
| **$log_{10}$AUTHORS** | | | | | | | -0.0047*** | -0.005*** | -0.0046*** |
| **Adjusted $R^2$ (weighted)** | | | | | | | **0.6126*** ** | | |
| **$log_{10}$HDI** | | | | | | | | -0.001*** | 0.0013*** |
| **Adjusted $R^2$ (weighted)** | | | | | | | | **0.609*** ** | |
| **$log_{10}$LTO** | | | | | | | | | -0.013*** |
| **Adjusted $R^2$** | | | | | | | | | **0.602*** ** |

| B. Physica A: Statistical mechanics and its applications | | | | | | | | | |
|---|---|---|---|---|---|---|---|---|---|
| *0* | *1* | *2* | *3* | *4* | *5* | *6* | *7* | *8* | *9* |
| **Intercept** | 0.200*** | 0.183*** | 0.2033*** | 0.200*** | 0.213** | 0200*** | 0.212*** | 0.214*** | 0.227*** |
| **Monday** (1 yes, 0 no) | 0.020** | | | 0.020** | 0.020** | 0.018** | 0.018** | 0.018** | 0.016* |
| **Tuesday** (1 yes, 0 no) | 0.015* | | | 0.015* | 0.015* | 0.0147 | 0.014 | 0.014 | 0.012 |
| **Wednesday** (1 yes, 0 no) | 0.013 | | | 0.014 | 0.014 | 0.014 | 0.013 | 0.014 | 0.011 |
| **Thursday** (1 yes, 0 no) | 0.007 | | | 0.007 | 0.007 | 0.0069 | 0.006 | 0.006 | 0.003 |
| **Weekend** (1 yes, 0 no) | -0.215*** | | | -0.215*** | -0.212*** | -0.213*** | -0.213*** | -0.214*** | -0.219*** |
| **Adjusted $R^2$ (weighted)** | **0.104*** ** | | | | | | | | |
| **Spring** (1 yes, 0 no) | | -0.007 | | -0.006 | -0.005 | 0.008 | 0.008 | 0.009 | 0.008 |
| **Summer** (1 yes, 0 no) | | -0.003 | | 0.002 | 0.002 | 0.016** | 0.016** | 0.017** | 0.017** |
| **Fall** (1 yes, 0 no) | | 0.0001 | | 0.0006 | 0.0004 | 0.014* | 0.014* | 0.015* | 0.016** |
| **Adjusted $R^2$ (weighted)** | | **0.0001*** ** | | **0.104*** ** | | | | | |
| **America** (1 yes, 0 no) | | | -0.0092 | | -0.006 | -0.006 | -0.005 | -0.003 | 0.009 |
| **Africa** (1 yes, 0 no) | | | -0.0041 | | 0.02 | 0.019 | 0.02 | 0.029 | 0.06** |
| **Asia** (1 yes, 0 no) | | | -0.0437*** | | -0.025*** | -0.025*** | -0.022*** | -0.017** | -0.019** |
| **Oceania** (1 yes, 0 no) | | | -0.0594*** | | -0.068*** | -0.066*** | -0.066*** | -0.066*** | -0.069*** |
| **Adjusted $R^2$ (weighted)** | | | **0.007*** ** | | **0.107*** ** | | | | |
| **Christmas** (1 yes, 0 no) | | | | | | 0.081*** | 0.08*** | 0.081*** | 0.082*** |
| **Adjusted $R^2$ (weighted)** | | | | | | **0.111*** ** | | | |
| **$\log_{10}$AUTHORS** | | | | | | | -0.0146*** | -0.0146*** | -0.012** |
| **Adjusted $R^2$ (weighted)** | | | | | | | **0.111*** ** | | |
| **$\log_{10}$HDI** | | | | | | | | 0.03 | 0.049 |
| **Adjusted $R^2$ (weighted)** | | | | | | | | **0.113*** ** | |
| **$\log_{10}$LTO** | | | | | | | | | 0.041** |
| **Adjusted $R^2$** | | | | | | | | | **0.116*** ** |

| C. PLOS ONE | | | | | | | | | |
|---|---|---|---|---|---|---|---|---|---|
| *0* | *1* | *2* | *3* | *4* | *5* | *6* | *7* | *8* | *9* |
| **Intercept** | 0.308*** | 0.338*** | 0.343*** | 0.326*** | 0.328*** | 0.291*** | 0.292*** | 0.294*** | 0.29*** |
| **Monday** (1 yes, 0 no) | 0.021*** | | | 0.021*** | 0.021*** | 0.021*** | 0.021*** | 0.022*** | 0.021*** |
| **Tuesday** (1 yes, 0 no) | 0.11*** | | | 0.11*** | 0.111*** | 0.111*** | 0.111*** | 0.112*** | 0.111*** |
| **Wednesday** (1 yes, 0 no) | 0.126*** | | | 0.126*** | 0.126*** | 0.127*** | 0.128*** | 0.128*** | 0.127*** |
| **Thursday** (1 yes, 0 no) | 0.113*** | | | 0.113*** | 0.114*** | 0.114*** | 0.114*** | 0.114*** | 0.114*** |
| **Weekend** (1 yes, 0 no) | -0.481*** | | | -0.481*** | -0.478*** | -0.48*** | -0.48*** | -0.48*** | -0.481*** |
| **Adjusted $R^2$** | **0.352*** ** | | | | | | | | |
| **Spring** (1 yes, 0 no) | | -0.015*** | | -0.02*** | -0.02*** | 0.017*** | 0.017*** | 0.017*** | 0.017*** |
| **Summer** (1 yes, 0 no) | | -0.018*** | | -0.024*** | -0.025*** | 0.013*** | 0.013*** | 0.013*** | 0.013*** |
| **Fall** (1 yes, 0 no) | | -0.021*** | | -0.025*** | -0.025*** | 0.013*** | 0.013*** | 0.013*** | 0.013*** |
| **Adjusted $R^2$** | | **0.001*** ** | | **0.353*** ** | | | | | |
| **America** (1 yes, 0 no) | | | 0.001 | | 0.005*** | 0.004*** | 0.004*** | 0.004*** | -0.002 |
| **Africa** (1 yes, 0 no) | | | -0.04*** | | -0.014*** | -0.016*** | -0.016*** | -0.006 | -0.014* |
| **Asia** (1 yes, 0 no) | | | -0.058*** | | -0.013*** | -0.015*** | -0.015*** | -0.011*** | -0.008*** |
| **Oceania** (1 yes, 0 no) | | | -0.043*** | | -0.013*** | -0.012*** | -0.012*** | 0.013*** | -0.02*** |
| **Adjusted $R^2$** | | | **0.008*** ** | | **0.354*** ** | | | | |
| **Christmas** (1 yes, 0 no) | | | | | | 0.193*** | 0.193*** | 0.192*** | 0.192*** |
| **Adjusted $R^2$** | | | | | | **0.369*** ** | | | |
| **$\log_{10}$AUTHORS** | | | | | | | -0.0009 | -0.001 | -0.0003 |
| **Adjusted $R^2$** | | | | | | | **0.369*** ** | | |
| **$\log_{10}$HDI** | | | | | | | | 0.051*** | 0.063*** |
| **Adjusted $R^2$** | | | | | | | | **0.367*** ** | |
| **$\log_{10}$LTO** | | | | | | | | | -0.018*** |
| **Adjusted $R^2$** | | | | | | | | | **0.368*** ** |
| **D. Nature** | | | | | | | | | |
| *0* | *1* | *2* | *3* | *4* | *5* | *6* | *7* | *8* | *9* |
| **Intercept** | 0.313*** | 0.294*** | 0.2607*** | 0.32*** | 0.313*** | 0.303*** | 0.303*** | 0.305*** | 0.298*** |
| **Monday** (1 yes, 0 no) | 0.073 | | | 0.007 | 0.007 | 0.008 | 0.008 | 0.008 | 0.008 |
| **Tuesday** (1 yes, 0 no) | -0.001* | | | -0.0009 | -0.001 | -0.0008 | -0.0008 | -0.0009 | -0.0009 |
| **Wednesday** (1 yes, 0 no) | -0.016*** | | | -0.017** | -0.0169** | -0.016** | -0.016** | -0.016** | -0.0159** |
| **Thursday** (1 yes, 0 no) | -0.033*** | | | -0.032*** | -0.032*** | -0.030*** | -0.030*** | -0.030*** | -0.030*** |
| **Weekend** (1 yes, 0 no) | -0.188*** | | | -0.188*** | -0.188*** | -0.187*** | -0.187*** | -0.187*** | -0.187*** |
| **Adjusted $R^2$ (weighted)** | **0.19*** ** | | | | | | | | |
| **Spring** (1 yes, 0 no) | | -0.021*** | | -0.017*** | -0.018*** | -0.008 | -0.008 | -0.008 | -0.008 |
| **Summer** (1 yes, 0 no) | | -0.005 | | -0.004 | -0.004 | 0.004 | 0.004 | 0.004 | 0.004 |
| **Fall** (1 yes, 0 no) | | -0.009 | | -0.005 | -0.005 | 0.003 | 0.003 | 0.003 | 0.003 |
| **Adjusted $R^2$ (weighted)** | | **0.003*** ** | | **0.19*** ** | | | | | |
| **America** (1 yes, 0 no) | | | 0.0073 | | 0.011** | 0.011** | 0.011** | 0.011** | 0.0002 |
| **Africa** (1 yes, 0 no) | | | 0.0033 | | -0.001 | 0.002 | 0.002 | 0.002 | 0.0238 |
| **Asia** (1 yes, 0 no) | | | 0.0041 | | 0.009 | 0.01 | 0.01 | 0.01 | 0.015 |
| **Oceania** (1 yes, 0 no) | | | -0.0046 | | 0.01 | 0.01 | 0.01 | 0.01 | -0.003 |
| **Adjusted $R^2$ (weighted)** | | | **0.001*** ** | | **0.192*** ** | | | | |
| **Christmas** (1 yes, 0 no) | | | | | | 0.07*** | 0.07*** | 0.07*** | 0.07*** |
| **Adjusted $R^2$ (weighted)** | | | | | | **0.197*** ** | | | |
| **$\log_{10}$AUTHORS** | | | | | | | 0.0001 | 0.0001 | 0.000019 |
| **Adjusted $R^2$ (weighted)** | | | | | | | **0.197*** ** | | |
| **$\log_{10}$HDI** | | | | | | | | 0.021 | 0.018 |
| **Adjusted $R^2$ (weighted)** | | | | | | | | **0.197*** ** | |
| **$\log_{10}$LTO** | | | | | | | | | -0.029 |
| **Adjusted $R^2$** | | | | | | | | | **0.197*** ** |
| **E. Cell** | | | | | | | | | |
| *0* | *1* | *2* | *3* | *4* | *5* | *6* | *7* | *8* | *9* |
| **Intercept** | 0.3001*** | 0.3107*** | 0.2964*** | 0.3132*** | 0.3063*** | 0.2941*** | 0.2964*** | 0.288*** | 0.2756*** |
| **Monday** (1 yes, 0 no) | 0.0421*** | | | 0.0428*** | 0.0423*** | 0.0413*** | 0.0412*** | 0.0414*** | 0.0414*** |
| **Tuesday** (1 yes, 0 no) | 0.0149 | | | 0.0312 | 0.0135 | 0.0127 | 0.0126 | 0.0128 | 0.0129 |
| **Wednesday** (1 yes, 0 no) | 0.0055 | | | 0.0037 | 0.0036 | 0.002 | 0.0019 | 0.0019 | 0.002 |
| **Thursday** (1 yes, 0 no) | 0.003 | | | 0.0023 | 0.0028 | 0.0025 | 0.0025 | 0.002 | 0.0018 |
| **Weekend** (1 yes, 0 no) | -0.1416*** | | | -0.1428*** | -0.143*** | -0.1431*** | -0.1431*** | -0.143*** | -0.1431*** |
| **Adjusted $R^2$ (weighted)** | **0.0468*** ** | | | | | | | | |
| **Spring** (1 yes, 0 no) | | -0.0257** | | -0.0275** | -0.0269** | -0.0136 | -0.0136 | -0.0135 | -0.0135 |
| **Summer** (1 yes, 0 no) | | -0.0286** | | -0.0274** | -0.027** | -0.0136 | -0.0136 | -0.0136 | -0.0135 |
| **Fall** (1 yes, 0 no) | | 0.0116 | | 0.0119 | 0.0118 | 0.0251** | 0.0251** | 0.0252** | 0.0254** |
| **Adjusted $R^2$ (weighted)** | | **0.0059*** ** | | **0.0536*** ** | | | | | |
| **America** (1 yes, 0 no) | | | -0.0002 | | 0.0059 | 0.0054 | 0.0053 | 0.0059 | -0.0212 |
| **Africa** (1 yes, 0 no) | | | 0.0709 | | 0.088 | 0.0869 | 0.0864 | 0.0569 | 0.045 |
| **Asia** (1 yes, 0 no) | | | 0.0271 | | 0.0307* | 0.0303* | 0.0303* | 0.022 | 0.0292 |
| **Oceania** (1 yes, 0 no) | | | 0.0758 | | 0.0607 | 0.0609 | 0.0612 | 0.0637 | 0.0294 |
| **Adjusted $R^2$ (weighted)** | | | **0.0017*** ** | | **0.0548*** ** | | | | |
| **Christmas** (1 yes, 0 no) | | | | | | 0.0842*** | 0.0842*** | 0.0845*** | 0.0846*** |
| **Adjusted $R^2$ (weighted)** | | | | | | **0.0597*** ** | | | |
| **$log_{10}$AUTHORS** | | | | | | | -0.001 | -0.001 | -0.0007 |
| **Adjusted $R^2$ (weighted)** | | | | | | | **0.0597*** ** | | |
| **$log_{10}$HDI** | | | | | | | | -0.0934 | -0.0713 |
| **Adjusted $R^2$ (weighted)** | | | | | | | | **0.0597*** ** | |
| **$log_{10}$LTO** | | | | | | | | | -0.0697 |
| **Adjusted $R^2$** | | | | | | | | | **0.0599*** ** |

**Note:** * Statistically significant at level 10%
** Statistically significant at level 5%
*** Statistically significant at level 1%

The first point we notice is that once the variables for day of the week are introduced, the other three groups (seasons, continents and controls) add very little new information. After each step (variable or a group of variables entered in the model) further increase of the explained variance (measured by adjusted $R^2$) for the dependent variable is always very small.

Second, no matter if it is about consolidated data set or journal specific or roll window, Weekend found to be with relevant impact both statistical and practical. Thus, every time the regression parameters were negative for that factor while statistically significant at the level 1%. Other variables testing the day of the week effect (Monday, Tuesday, Wednesday or Thursday) were more volatile.

The negative influence of Weekends submissions seems to have a rational explanation: the quality of the papers might be thought to be increased after supplementary checks performed by research teams or their peers. The "negative" accepted paper submission effect during the week-end might be related to geography. There is an important share of population around the world (e. g. Muslim) which rest on Friday and works on Sunday. Of course, one might say that an important proportion of the people are not doing anything professional on weekends, neither writing, researching nor submitting.

Thirdly, the estimated regression parameters for Christmas are always positive (no matter if it is about the consolidated data set or specific journal samples). With regard to this outcome, we note that scholars tend to submit papers during Christmas time more often than in other periods.

Apart of the negative Weekends submissions effect, a closer look at the day of the week group of variables is necessarily since, as we mentioned above this is the most important one. Due to the size effect, PLOS ONE sample seems to drive the outcomes of the consolidated dataset. Indeed, for this two samples, the regression's coefficients for all factors (Monday, Tuesday, Wednesday and Thursday) are synchronized: all of them are positive while their effective values are 4-5 times bigger for Tuesday, Wednesday and Thursday when compare to Monday. Since Friday was the reference point for this group, this means that for PLOS ONE accepted submissions are most likely to happen Tuesday, Wednesday and Thursday and, in a smaller amount, Monday. This is an analytical confirmation of the visual information presented for PLOS ONE earlier (figure 2). For Nature sample, there are two factors (Wednesday and Thursday) which are, while statistically significant, negative. The effective values for the regression's coefficients in this case are 4-10 times smaller while compare to Weekends. This means that apart to weekend, accepted submissions to Nature occurs usually not Wednesday or Thursday. The Cell and Physica A sample shown a clear positive Monday effect while for some simpler (not with so many control factors inserted) models, for Physica A a positive Tuesday effect is visible.

As mentioned in the material and methods section, the seasonality is checked systematically. As we mention above, this variables group is adding only a tiny fraction of new information within all models. Let us split again our comments, first to look to the consolidated data set and its main driver (the PLOS ONE sample) together and after to have an opinion about the other three samples (Physica A, Nature and Cell). Due to the very large number of cases, for the consolidated and PLOS ONE data sets, every seasonal factor which is inserted in the model prove to be statistical significant. Still, there is a sign volatility (models M1-M5 versus M6-M9). This is a classical sign for multicollinearity. Indeed, the correlations matrix (Table_SI.2 from Supplementary Information) for this samples shows negative Pearson's correlation coefficients between seasons with effective values between 0.33 and 0.35 and also a negative relationship to Christmas (correlation coefficients around 0.13-0.14). In this context, the sign volatility between models M1-M5 and M6-M9 appears to be natural since the control factor (Christmas) is present within models M6-M9. The outcomes for the other three samples (Physica A, Nature and Cell) show the same signs volatility. In addition, sometimes there is also a statistical significance volatility (M1-M5 versus M6-M9). The main reason for this volatility is the same: a persistent correlation between seasonal and Christmas factors. Also, the statistical significance volatility could occur due to smaller samples (size effect) under weak factors' correlations within the dependent variables.

The geographical factors from the regression models (America, Africa, Asia, and Oceania while Europe is the reference) are many times statistically insignificant. This geographic volatility (in terms of how many times a factor is statistical significant and how many times there is a change in the sign of the estimated parameter) needs a deeper analysis. The most probable explanation for this volatility is similar to the seasonal factors. There is a sample size effect: as long as the samples are bigger (consolidated data set, PLOS ONE and, in a lesser amount, Physica A) the signs and statistical significance volatility remains lower. Due to the weaker correlation to the dependent variable and persistent negative correlations between geographical factors, when smaller samples are analyzed the outcomes tend to be volatile.

Due to the large number of models and samples when one is multiplying using the predefined roll windows, the outcomes (tens of pages of information) exceed the limited space available within a regular paper. Therefore, all detailed outcomes of the regression models for each roll window are available within Table SI.3. from the Supplementary Information while a brief visual presentation about regression coefficients' signs and their significance is available within figure 3.

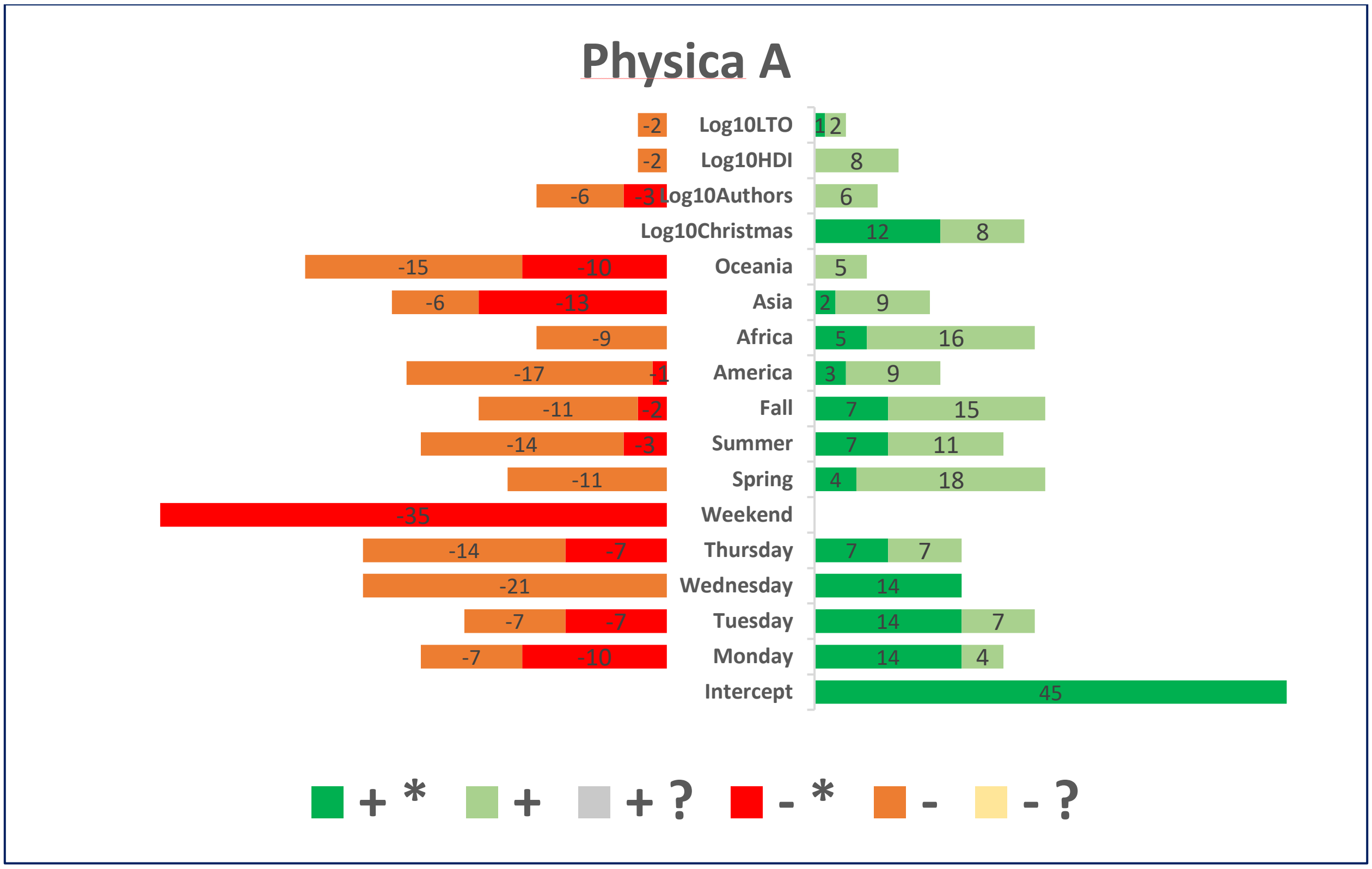
Physica A
-2 Log10LTO 1 2
-2 Log10HDI 8
-6 -3 Log10Authors 6
Log10Christmas 12 8
-15 -10 Oceania 5
-6 -13 Asia 2 9
-9 Africa 5 16
-17 -1 America 3 9
-11 -2 Fall 7 15
-14 -3 Summer 7 11
-11 Spring 4 18
-35 Weekend
-14 -7 Thursday 7 7
-21 Wednesday 14
-7 -7 Tuesday 14 7
-7 -10 Monday 14 4
Intercept 45
+ * + + ? - * - - ?

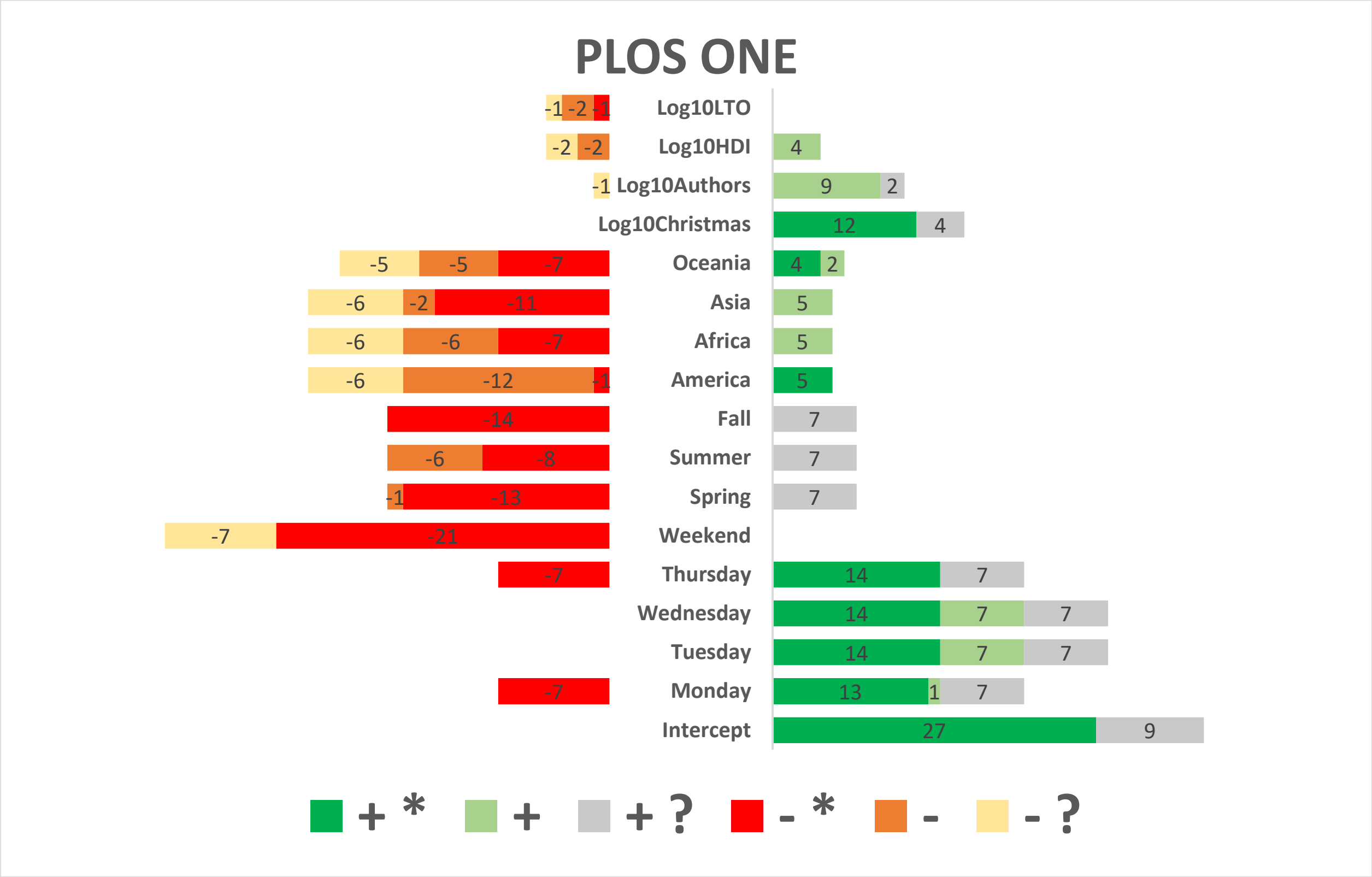
PLOS ONE
-1 -2 -1 Log10LTO
-2 -2 Log10HDI 4
-1 Log10Authors 9 2
Log10Christmas 12 4
-5 -5 -7 Oceania 4 2
-6 -2 -11 Asia 5
-6 -6 -7 Africa 5
-6 -12 -1 America 5
-14 Fall 7
-6 -8 Summer 7
-1 -13 Spring 7
-7 -21 Weekend
-7 Thursday 14 7
Wednesday 14 7 7
Tuesday 14 7 7
-7 Monday 13 1 7
Intercept 27 9
+ * + + ? - * - - ?

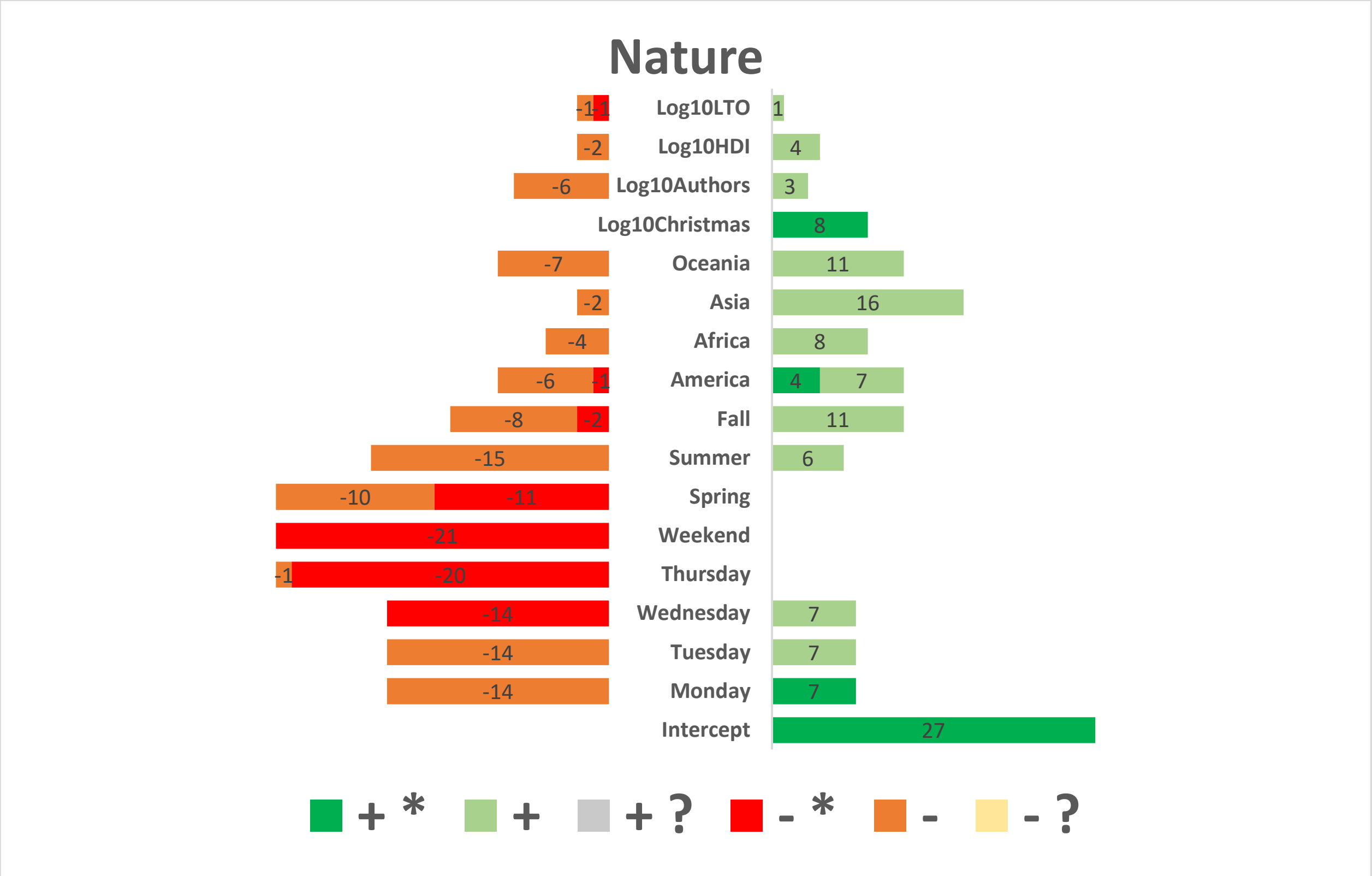

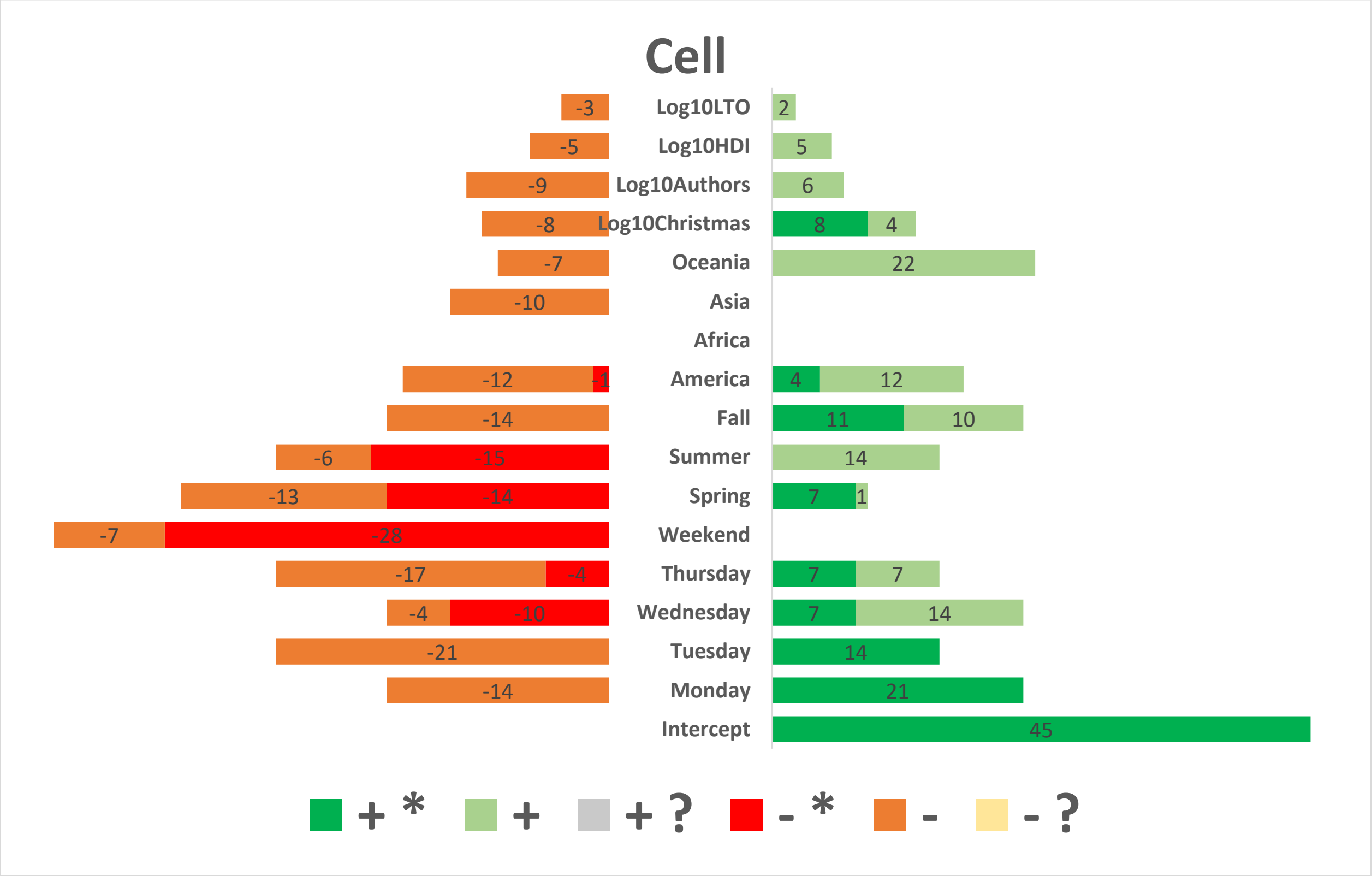

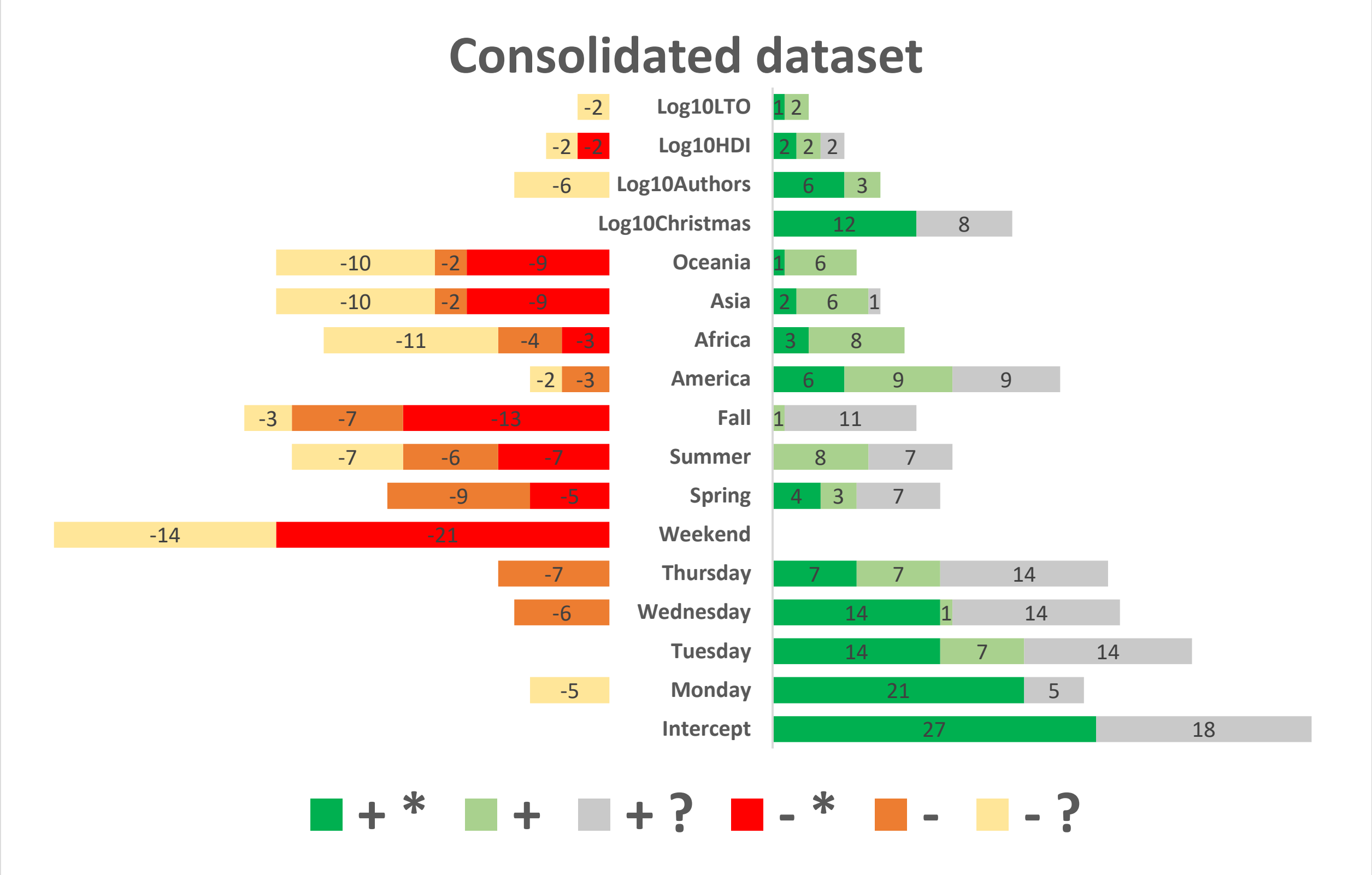

**Figure 3.** Distributions of regression coefficients' signs

**Note:** "+ *" denotes a statistically significant positive coefficient, "+" denotes a positive coefficient, "+ ?" denotes a coefficient for which the standard error could not be computed and, therefore, the t test for testing statistical significance is unavailable. The notations for negative coefficients are similar.

Since the samples' size for each roll window became smaller and smaller, as expected, sometimes the volatility of the regression coefficients or their statistical significance became greater. Even so, the main conclusions: most important factor (Weekend); most important group of factors (days of the week) and most important control factor (Christmas) still stand.

The results achieved after the panel analysis are convergent with those achieved under the unstructured sample. Here another proof for a significant impact of the week-day on the dependent variable emerges. Also it can be highlighted that other factors such as LTO or HDI are rather non-significant. The unobserved cross-sectional factors are found to be not relevant in the panel structure, as their share in total variation of the unobserved factors is almost zero (table 5.)

**Table 5.** Panel regression estimates for the dependent variables for consolidated data set and PLOS ONE

| Covariates | Consolidated dataset | PLOS ONE dataset |
|---|---|---|

| | No weighted | Weighted by $n_{ct}$ | No weighted | Weighted by $n_{ct}$ |
|---|---|---|---|---|
| **Intercept** | 0.267*** | 0.618*** | 0.306*** | 0.507*** |
| **Monday** (1 yes, 0 no) | - | - | - | - |
| **Tuesday** (1 yes, 0 no) | 0.069*** | 0.374*** | 0.094*** | 0.547*** |
| **Wednesday** (1 yes, 0 no) | 0.078*** | 0.432*** | 0.113*** | 0.657*** |
| **Weekend** (1 yes, 0 no) | -0.374*** | -0.860*** | -0.357*** | -0.792*** |
| **$log_{10}$AUTHORS** | -0.001 | 0.028*** | 0.002 | 0.049*** |
| **Christmas** (1 yes, 0 no) | - | - | - | - |
| **Spring** (1 yes, 0 no) | -0.011*** | -0.013 | -0.014*** | -0.017 |
| **Summer** (1 yes, 0 no) | -0.012*** | 0.006 | -0.020*** | -0.011 |
| **Fall** (1 yes, 0 no) | - | - | - | - |
| **Log10HDI** | 0.135 | -1.90*** | 0.108 | -2.255*** |
| **Log10LTO** | 0.020 | -0.454*** | 0.019 | -0.500*** |
| **Ln(t)=(TREND)** | -0.012*** | -0.099*** | -0.004*** | -0.083*** |
| **Adjusted $R^2$** | 0.499 | 0.620 | 0.429 | 0.589 |
| $\rho = \frac{\sigma_{uc}^2}{\sigma_{\varepsilon ct}^2}$ | 0.002 | 0.0009 | 0.0011 | 0.0004 |
| **Cases** | 26785 | 26785 | 26785 | 26785 |

**Note:** * Statistically significant at level 10%
** Statistically significant at level 5%
*** Statistically significant at level 1%
Log10 denotes, logarithm in base 10, and ln the natural logarithm
The results from Table 5 are based on the model 12 with the component mentioned in 13-15

### 3.3. Spatial analysis

The regression methods signaled (on the consolidated dataset) that there are some patterns within the propensity of submitting papers Tuesday and Wednesday (while Friday was the reference day). On the other hand, accepted manuscripts submissions tend to be lower (negative regression coefficients) especially for weekends but also slightly so for Monday. Therefore, we create two daily intervals: Tuesday-Thursday and Saturday-Monday. With a methodology presented in Materials and Methods section an indicator: Localization Quotient (LQ) is introduced using GIS tools for both above-mentioned daily intervals. The outcomes are interesting and shown in the Figure 4 and Figure 5. The figures are complementary. In figure 4, countries which tend to submit papers during Tuesday-Thursday (TUE-THU) with a greater intensity than the world average submission rate for the same interval (TUE-THU) are highlighted by dark brown. This group comprises countries like: Belarus, Belize, Benin, Bosnia and Herzegovina, Greenland, Haiti, Iraq, Mauritania, Namibia, Nicaragua, and Syria. Every continent except Oceania is represented in this list. As expected, most of the important countries (who, by size, have an important weight over the world average) register a level of LQ between 93.33% and 136.5% of the world average TUE-THU submission rate.

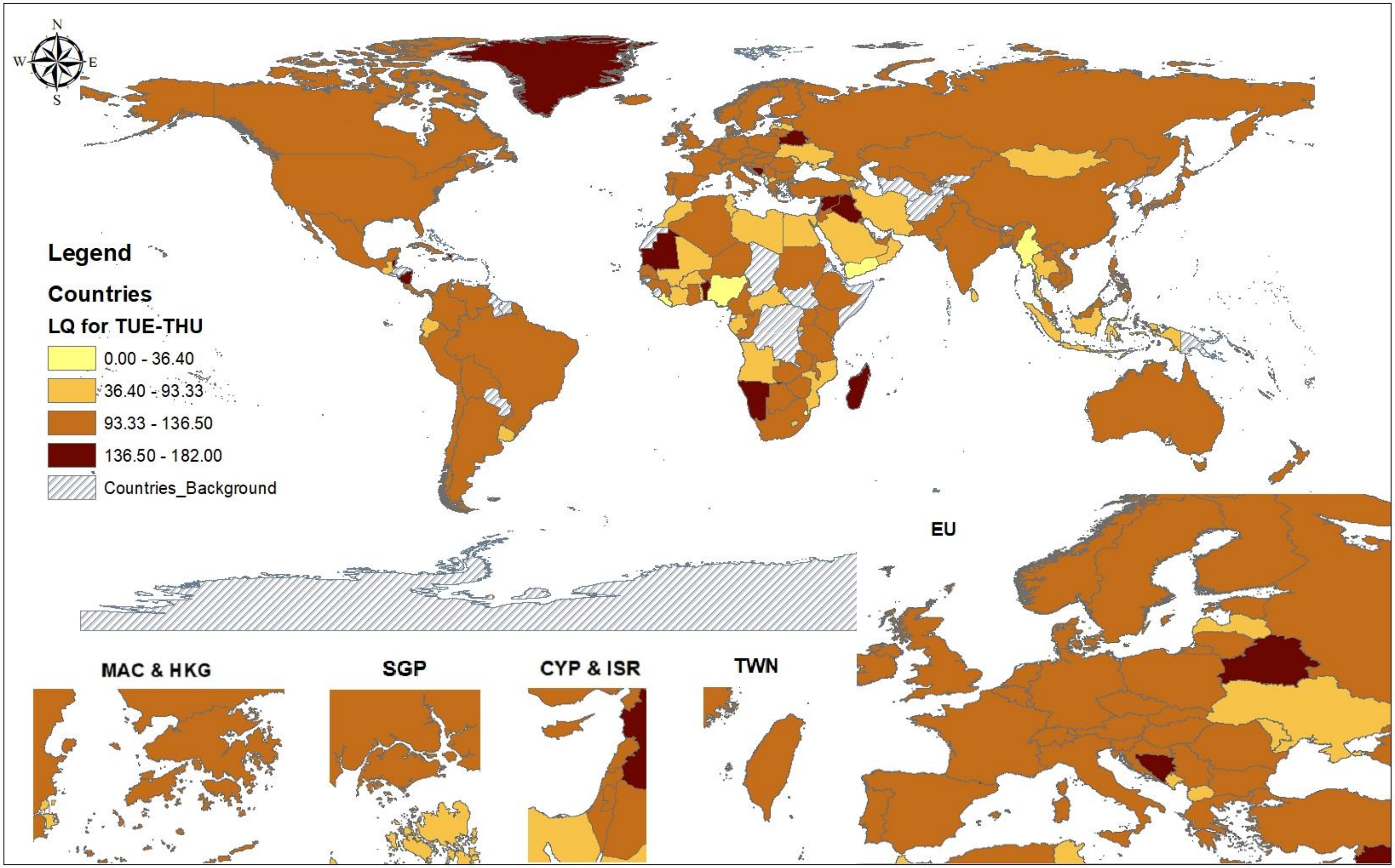

**Figure 4.** Distribution of the LQ for TUE-THU by PCA's country of origin within the consolidate dataset (2001-2016)

**Note:** The measurement unit for intervals is the percentage (%). The consolidated dataset rely on papers from: Physica A, PLOS ONE, Nature and Cell. LQ=Localization Quotient (the share of the papers submitted in day k in a country, divided by share of the papers submitted in same – k – day worldwide); TUE-THU= Tuesday to Thursday interval; PCA= Papers' Corresponding Authors; MAC=Macao (China); HKG=Hong Kong (China); SGP=Singapore; CYP=Cyprus; ISR=Israel; TWN=Taiwan and EU=Europe. Countries marked with shading lines record no information in our dataset.

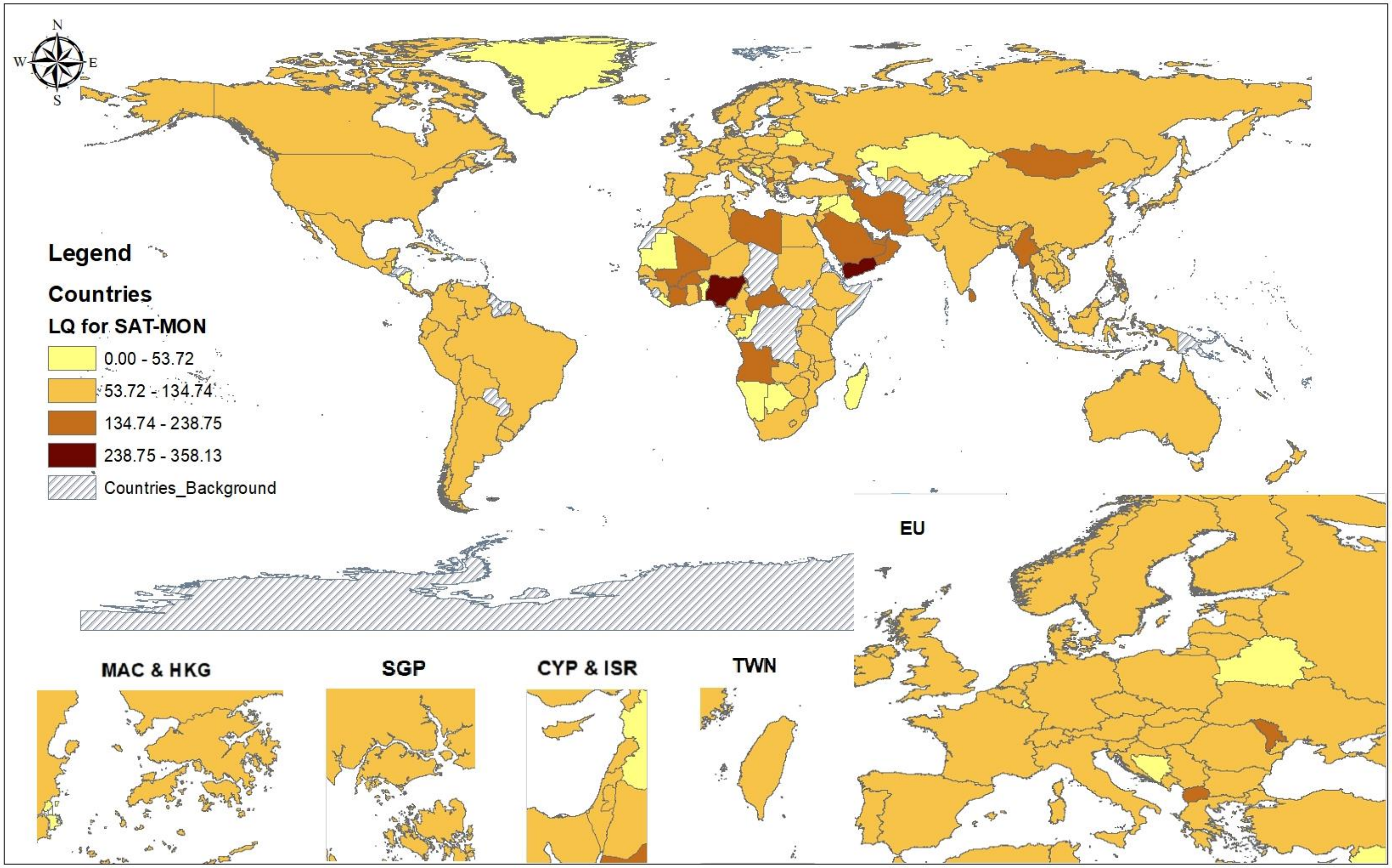

**Figure 5.** Distribution of the LQ for SAT-MON by PCA's country of origin within the consolidate dataset (2001-2016)

**Note:** The measurement unit for intervals is the percentage (%). The consolidated dataset relies on papers from: Physica A, PLOS ONE, Nature and Cell. LQ=Localization Quotient (the share of the papers submitted in day k in a country, divided by share of the papers submitted in same – k – day worldwide); SAT-MON= Saturday to Monday interval; PCA= Papers' Corresponding Authors; MAC=Macao (China); HKG=Hong Kong (China); SGP=Singapore; CYP=Cyprus; ISR=Israel; TWN=Taiwan and EU=Europe. Countries marked with shading lines record no information in our dataset.

A greater heterogeneity of the countries distribution by LQ designed for Saturday-Monday (SAT-MON) interval is easy to be seen in Figure 5. Countries that register low levels on the previous map are now concentrated in two leading groups. The first one is formed by two countries: Nigeria and Yemen. Their propensity of submitting papers during SAT-MON interval is more than twice as big as the world average. The second group register a propensity of submitting papers during SAT-MON interval greater than world average by a multiplier between 1.35 and 2.39. In this group countries like: Angola, Armenia, Burkina Faso, Central African Republic, Georgia, Iran, Ivory Coast, Libya, Mali, Macedonia, Moldova, Mongolia, Myanmar, Oman, Saudi Arabia, and Sri Lanka are included. Many of these countries belonging to the lead groups are known for their important Muslim population. Further research should be done on this topic.

## 4. Conclusions and discussion

In line with other reports (Cabanac & Hartley, 2013; Campos-Arceiz *et al.*, 2013; Hartley & Cabanac, 2017; Ausloos *et al.*, 2016; Ausloos *et al.*, 2017) our results confirm a DWE for published papers. Papers accepted for publication in the four journals included in our analysis are more likely to have been submitted on a week-day than on a day in the week-end. There are 2-3 times more published papers submitted in any week-day when compare to those which have been submitted Saturday or Sunday. The most likely week-day for historical accepted submissions differs from journal to journal. In the case of PLOS ONE there is a group of three days: Tuesday, Wednesday and Thursday; for Physica A there are two consecutive days: Monday and Tuesday; for Nature, there are two days: Friday and Monday while for Cell only Monday.

Most of the seasonal factors proved to be statistical insignificant. However, the Christmas period ($20^{th}$ of December to $10^{th}$ of January) has a statistically significant positive impact. More papers tend to be submitted in that time interval.

The geographical dimension (papers' corresponding authors' affiliation country) brings new information concerning how current research is done (descriptive: GIS and analytical – a mix of log-log and semi-log based on undated and panel structured data regression models). Further work on this aspect would be interesting even if most geographical driven factors from the regression models provide a weak explanatory power.

In addition to the day of the week, time is included here and examined from different perspective, namely the time trends in the data. The most important factors for the main topic of the manuscript (Week-days and Week-end) tend to be stable. Of course, some fluctuations occur for other factors. The yearly time span from our data set is not long enough to use other specific time series techniques. This is a topic for further study.

## Supplementary information

All information about each annual dataset on which the current paper relies is available from Excel file: Table_SI.1.xlsx. The customized script for all four journals analyzed in the current paper is available in the following Word file: SI_Scraper.docx. The Python's scripts are available in Word File: SI_Python_Scripts.docx. The correlations matrix for all samples are available within the Word file: Table_SI.2.docx. The regression analysis outcomes for roll windows are available within: Table_SI.3.docx. The detailed formula of Kurtosis is available within SI_Kurtosis.docx. The Geographical presentation of our consolidated dataset is available in: SI_Geographical.docx. Four supplementary figures (SI.1., SI.2., SI.3. and SI.4.) are available in Word Files: Figure_SI_1.docx, Figure_SI_2.docx, Figure_SI_3.docx and Figure_SI_4.docx .

## Acknowledgments

We acknowledge support from COST Action TD1210 “Analyzing the dynamics of information and knowledge landscapes (KNOWeSCAPE)”. We are grateful to Alexandru Agapie, David Berman, Alexandru Isaic-Maniu, Tudorel Andrei, Gurjeet Dhesi, Sebastian Buhai and Babar Syed for comments on an earlier draft. A preliminary version of the paper was presented to: (i) Cluj Economics and Business Seminar Series (CEBSS), Fall 2017 Session, Babes-Bolyai University and (ii) Annual Conference of the Romanian Academic Economists from Abroad (ERMAS), The 5th edition: 25-27 July 2018, "A. I. Cuza" University, Iasi. Thanks to the consistent feedback offered during the presentations from Cristian Litan, Alexandru Todea, Dorina Lazar, Marcel Voia and Cristian Dragos. David Berman, Peter Richmond and Roxana Herteliu-Iftode were our proof-readers. Finally, we thank to Teresa Dudley (Nature), Jared Graves (Cell), Kenneth A. Dawson (Physica A), and Nick Simon (PLOS ONE) for help with the metadata.

## Author contribution

CEB obtained data using a self-designed scraper. CEB, CH, MD, and BVI performed data analysis and manuscript design in its initial and revised versions. CH coordinated the team work.

## Competing financial interests

CEB, CH, MD, and BVI have nothing to declare regarding competing financial interests.

**Table SI.1.** Parsed papers’ sample by journal and year of publication

| Year | Physica A | | | | PLOS ONE | | | | Nature | | | | Cell | | | |
|---|---|---|---|---|---|---|---|---|---|---|---|---|---|---|---|---|
| | items within WoS | citable items within JCR | items in dataset | share of sample | items within WoS | citable items within JCR | items in dataset | share of sample | items within WoS | citable items within JCR | items in dataset | share of sample | items within WoS | citable items within JCR | items in dataset | share of sample |
| *0* | *1* | *2* | *3* | *4=100*3/2* | *5* | *6* | *7* | *8=100*7/6* | *9* | *10* | *11* | *12=100*11/10* | *13* | *14* | *15* | *16=100*15/14* |
| 2016 | 1071 | 1051 | 152 | 14.5% | 23039 | 22492 | 18035 | 80.2% | 2740 | 904 | 299 | 33.1% | 699 | 457 | 137 | 30.0% |
| 2015 | 798 | 793 | 885 | 111.6% | 29807 | 28114 | 28100 | 100.0% | 2737 | 897 | 729 | 81.3% | 656 | 437 | 351 | 80.3% |
| 2014 | 866 | 860 | 832 | 96.7% | 31482 | 30040 | 30051 | 100.0% | 2561 | 862 | 834 | 96.8% | 611 | 436 | 356 | 81.7% |
| 2013 | 659 | 653 | 752 | 115.2% | 31502 | 31496 | 31505 | 100.0% | 2605 | 857 | 830 | 96.8% | 593 | 432 | 358 | 82.9% |
| 2012 | 685 | 674 | 601 | 89.2% | 23456 | 23406 | 23440 | 100.1% | 2651 | 869 | 833 | 95.9% | 572 | 415 | 352 | 84.8% |
| 2011 | 487 | 484 | 670 | 138.4% | 13786 | 13781 | 13772 | 99.9% | 2591 | 841 | 774 | 92.0% | 553 | 338 | 260 | 76.9% |
| 2010 | 624 | 617 | 632 | 102.4% | 6729 | 6714 | 6749 | 100.5% | 2577 | 862 | 354 | 41.1% | 573 | 319 | 243 | 76.2% |
| 2009 | 539 | 536 | 580 | 108.2% | 4404 | 4263 | 4404 | 103.3% | | | | | 648 | 359 | 281 | 78.3% |
| 2008 | 703 | 692 | 613 | 88.6% | 2717 | | 2715 | | | | | | 648 | 348 | 267 | 76.7% |
| 2007 | 1030 | 987 | 1006 | 101.9% | 1230 | | 1248 | | | | | | 648 | 366 | 283 | 77.3% |
| 2006 | 850 | 835 | 640 | 76.6% | 137 | | 153 | | | | | | 646 | 352 | 281 | 79.8% |
| 2005 | 726 | 712 | 640 | 89.9% | | | | | | | | | 462 | 319 | 178 | 55.8% |
| 2004 | 863 | 854 | 657 | 76.9% | | | | | | | | | 378 | 288 | 229 | 79.5% |
| 2003 | 776 | 747 | 503 | 67.3% | | | | | | | | | 359 | 281 | 178 | 63.3% |
| 2002 | 709 | 683 | 527 | 77.2% | | | | | | | | | 362 | 350 | 23 | 6.6% |
| 2001 | 612 | 601 | 135 | 22.5% | | | | | | | | | | | | |
| **Total** | **11998** | **11779** | **9825** | **83.4%** | **168289** | **160306** | **160172** | **99.9%** | **18462** | **6092** | **4653** | **76.4%** | **8408** | **5497** | **3777** | **68.7%** |

**Note:** Green shaded cells are authors' estimates

### SI. Scraper description and source code

# Article.java

Java class that defines the article model. It is used to manage collected article data like title, URL, pages, journal, authors, received date, revised date, published date, start page and end page. The class acts like an entity model between the database and the processing routines.

```
package ro.ase.crowler.model;

import java.sql.PreparedStatement;
import java.sql.SQLException;
import java.sql.Statement;

import org.jsoup.nodes.Element;
import org.jsoup.select.Elements;

import ro.ase.crowler.DB;
import ro.ase.crowler.Parser;

public class Article {
	String title;
	String URL;
	String pages;
	int journalID;
	String authors;
	String history;
	boolean isParsed = false;
	String receivedDate;
	String revisedDate;
	String onlineDate;
	int startPage;
	int endPage;

	public String getReceivedDate() {
		return receivedDate;
	}

	public void setReceivedDate(String receivedDate) {
		this.receivedDate = receivedDate;
	}

	public String getRevisedDate() {
		return revisedDate;
	}

	public void setRevisedDate(String revisedDate) {
		this.revisedDate = revisedDate;
	}

	public String getOnlineDate() {
		return onlineDate;
	}

	public void setOnlineDate(String onlineDate) {
		this.onlineDate = onlineDate;
	}
```

```
	public int getStartPage() {
		return startPage;
	}

	public void setStartPage(int startPage) {
		this.startPage = startPage;
	}

	public int getEndPage() {
		return endPage;
	}

	public void setEndPage(int endPage) {
		this.endPage = endPage;
	}

	public String getHistory() {
		return history;
	}

	public void setHistory(String history) {
		this.history = history;

		// set dates for PLOS
		if (this.journalID == 1) {
			String[] dates = Parser.getPlosDates(history);
			this.setReceivedDate(dates[0]);
			this.setRevisedDate(dates[1]);
			this.setOnlineDate(dates[2]);
		}
	}

	public String getAuthors() {
		return authors;
	}

	public void setAuthors(String authors) {
		this.authors = authors;
	}

	public Article() {
	}

	public Article(String title, String uRL, String pages, int journalID) {
		super();
		this.title = title;
		URL = uRL;
		this.pages = pages;
		this.journalID = journalID;
	}

	@Override
	public String toString() {
		return String.format("Title: %s \n Authors: %s \n Pages: %s \n URL: %s \n History: %s \n Journal
ID: %d",this.title, this.authors, this.pages, this.URL, this.getHistory(), this.getJournalID());
	}

	public void getElementInfo(Element article) {
		for (Element info : article.children()) {
			// Debug.log("Article info:" + info.attr("class") + " - " +
			// info.text());
```

```
					if (info.hasClass("title")) {
						// Debug.log("Article URL:" +
						// info.child(0).child(0).attr("href"));
						this.setTitle(info.child(0).child(0).text());
						this.setURL(info.child(0).child(0).attr("href"));
					}

					if (info.hasClass("authors")) {
						// Debug.log("Article authors:" + info.text());
						this.setAuthors(info.text());
					}
					if (info.hasClass("source")) {
						// Debug.log("Article URL:" + info.text());
						this.setPages(info.text());
					}

				}
}
```

**public void getPLOSElementInfo(Element article) {**

```
				final String journalURL = "http://journals.plos.org";

				Elements info = article.getElementsByClass("list-title");
				if (info.size() > 0) {
					this.setTitle(info.get(0).attr("title"));
					this.setURL(journalURL + info.get(0).attr("href"));
				}

				info = article.getElementsByClass("authors");
				if (info.size() > 0) {
					String authors = "";
					for (Element author : info.get(0).select("span.author"))
						authors += author.text();
					this.setAuthors(authors);
				}

				/*
				 * info = article.getElementsByClass("date"); if(info.size()>0) {
				 * this.setHistory(info.get(0).text()); }
				 */

				/*
				 * for (Element info : article.children()) { // Debug.log(
				 * "Article info:" + info.attr("class") + " - " + // info.text());
				 *
				 * if (info.hasClass("list-title")) { //Debug.log(
				 * "Article URL:" + info.child(0).child(0).attr("href"));
				 * this.setTitle(info.attr("title")); this.setURL(info.attr("href")); }
				 *
				 * if (info.hasClass("authors")) { //Debug.log(
				 * "Article authors:" + info.text()); String authors = ""; for(Element
				 * author : info.select("span.author")) authors+=author.text();
				 * this.setAuthors(authors); } if (info.hasClass("date")) {
				 * //Debug.log("Article URL:" + info.text());
				 * this.setHistory(info.text()); //this.setPages(info.text()); }
				 *
				 * }
				 */
}
```

```
public void getNATUREElementInfo(Element article) {

        // final String journalURL = "http://journals.plos.org";

        Elements info = article.select("h1");
        if (info.size() > 0) {
                this.setTitle(info.select("a").text());

        }

        info = article.select("ul.links");
        if (info.size() > 0) {
                Element firstURL = info.select("li").get(0);
                this.setURL(firstURL.child(0).attr("href"));
        }

        info = article.getElementsByClass("authors");
        if (info.size() > 0) {
                String authors = "";
                for (Element author : info.get(0).select("li"))
                        authors += author.text() + ",";
                this.setAuthors(authors);
        }

}

public void save2DB(DB db, int journalID) throws SQLException {
        String sql = "INSERT INTO `crowler`.`article` (`idjournal`, `title`, `url`, `pages`, `parsed`, `authors`,
`history`, `received`, `revised`, `online`, `startPage`, `endPage`) VALUES (?, ?, ?, ?, ?, ?, ?, ?, ?, ?, ?, ?);";
    PreparedStatement stmt = db.conn.prepareStatement(sql, statement.RETURN_GENERATED_KEYS);
        stmt.setInt(1, journalID);
        if (this.getTitle() != null && this.getTitle().length() > 255)
                stmt.setString(2, this.getTitle().substring(0, 254));
        else
                stmt.setString(2, this.getTitle());
        stmt.setString(3, this.getURL());
        stmt.setString(4, this.getPages());
        stmt.setInt(5, 0);
        stmt.setString(6, this.getAuthors());
        stmt.setString(7, this.getHistory());

        stmt.setString(8, this.getReceivedDate());
        stmt.setString(9, this.getRevisedDate());
        stmt.setString(10, this.getOnlineDate());

        stmt.setInt(11, this.getStartPage());
        stmt.setInt(12, this.getEndPage());

        stmt.execute();
}

public String getTitle() {
        return title;
}

public void setTitle(String title) {
        this.title = title;
}

public String getURL() {
        return URL;
}
```

```
	public void setURL(String uRL) {
		URL = uRL;
	}

	public String getPages() {
		return pages;
	}

	public void setPages(String pages) {
		this.pages = pages;

		// set pages
		int noPages[] = Parser.getPages(pages);
		this.setStartPage(noPages[0]);
		this.setEndPage(noPages[1]);
	}

	public int getJournalID() {
		return journalID;
	}

	public void setJournalID(int journalID) {
		this.journalID = journalID;
	}

	public boolean isParsed() {
		return isParsed;
	}

	public void setParsed(boolean isParsed) {
		this.isParsed = isParsed;
	}


	public void getCellElementInfo(Element article) {

		final String journalURL = "http://www.cell.com";

		Elements info = article.getElementsByClass("article-details");
		if (info.size() > 0) {
			this.setTitle(info.get(0).getElementsByTag("a").text());
			this.setURL(journalURL + info.get(0).getElementsByTag("a").attr("href"));
		}

		info = article.getElementsByClass("authors");
		if (info.size() > 0) {
			String authors = article.text();
			this.setAuthors(authors);
		}
	}

}
```

# Affiliation.java

Another entity model class used to manage the authors affiliation

```
package ro.ase.crowler.model;

public class Affiliation {

	public final String id;
	private String address;
	private String country;
	private String order;


	public Affiliation(String id, String address, String country, String order) {
		super();
		this.id = id;
		this.address = address;
		this.country = country;
		this.order = order;
	}


	public String getAddress() {
		return address;
	}


	public void setAddress(String address) {
		this.address = address;
	}


	public String getCountry() {
		return country;
	}


	public void setCountry(String country) {
		this.country = country;
	}


	public String getOrder() {
		return order;
	}


	public void setOrder(String order) {
		this.order = order;
	}


	public String getId() {
		return id;
	}

}
```

# ArticleAffiliation.java

An entity model class used to manage the authors affiliation data

```
package ro.ase.crowler.model;

import java.sql.PreparedStatement;
import java.sql.ResultSet;
import java.sql.SQLException;
import java.sql.Statement;
import java.text.DateFormatSymbols;
import java.util.ArrayList;

import ro.ase.crowler.DB;

public class ArticleAffiliation {

	//public static final int JOURNAL_ID = 122016;
	public static final int JOURNAL_ID = 182016;

	public String title;
	ArrayList<Author> authors;


	String receivedDate;
	String revisedDate;
	String onlineDate;
	String URL;
	String pages;



	public String getReceivedDate() {
		return receivedDate;
	}

	public void setReceivedDate(String receivedDate) {
		this.receivedDate = receivedDate;
	}

	public String getRevisedDate() {
		return revisedDate;
	}

	public void setRevisedDate(String revisedDate) {
		this.revisedDate = revisedDate;
	}

	public String getOnlineDate() {
		return onlineDate;
	}

	public void setOnlineDate(String onlineDate) {
		this.onlineDate = onlineDate;
	}

	public String getURL() {
		return URL;
	}
```

```
public void setURL(String uRL) {
	URL = uRL;
}

public String getPages() {
	return pages;
}

public void setPages(String pages) {
	this.pages = pages;
}

public ArticleAffiliation(String title) {
	super();
	this.title = title;
	this.authors = new ArrayList<>();
}

public void addAuthor(Author author) {
	this.authors.add(author);
}

@Override
public String toString() {
	StringBuilder sb = new StringBuilder();
	sb.append("********************************");
	sb.append("\nTitle: " + title);
	sb.append("\nReceived:"+receivedDate);
	sb.append("\nAccepted:"+revisedDate);
	sb.append("\nPublished:"+onlineDate);
	sb.append("\nURL:"+URL);
	sb.append("\nNo pages:"+pages);
	sb.append("\nHistory:"+getHistory());
	sb.append("\nAuthors:"+getAuthors());

	for (Author author : authors)
		sb.append(author.toString());


	return sb.toString();
}

public String getHistory(){

	StringBuilder sb = new StringBuilder();

	String[] dates = this.receivedDate.split("/");
	if(dates.length>0){
		String monthString = new DateFormatSymbols().getMonths()[Integer.parseInt(dates[1])-1];
		sb.append("Received: "+monthString);
		sb.append(" "+dates[0]+", "+dates[2]+";");

	}

	dates = this.revisedDate.split("/");
	if(dates.length>0){
		String monthString = new DateFormatSymbols().getMonths()[Integer.parseInt(dates[1])-1];
		sb.append(" Accepted: "+monthString);
		sb.append(" "+dates[0]+","+dates[2]+";");

	}
```

```
		dates = this.onlineDate.split("/");
		if(dates.length>0){
			String monthString = new DateFormatSymbols().getMonths()[Integer.parseInt(dates[1])-1];
			sb.append(" Published: "+monthString);
			sb.append(" "+dates[0]+","+dates[2]);

		}

		return sb.toString();

	}
```

**public String getAuthors(){**

```
		StringBuilder sb = new StringBuilder();

		for(int i=0;i<authors.size();i++){
			sb.append(authors.get(i).getGivenNames()+" "+authors.get(i).getSurname());
			if(i!=authors.size()-1)
				sb.append(",");
		}

		return sb.toString();
	}
```

**public int getArticleId(DB db) throws SQLException {**

```
		int idArticle = 0;

		String shortTitle = "";
		if (this.title != null && this.title.length() > 255)
			shortTitle = this.title.substring(0, 254);
		else
			shortTitle = this.title;

		String sql = "SELECT idarticle,url FROM crowler.article where idjournal = 1 AND title like ?;";

		PreparedStatement stmt = db.conn.prepareStatement(sql);
		stmt.setString(1, shortTitle);


		ResultSet rs = stmt.executeQuery();

		if (rs.next()) {
			String URL = rs.getString("url");
			idArticle = rs.getInt("idarticle");

			System.out.println("Article " + title);
			System.out.println("URL " + URL);
			System.out.println("ID " + idArticle);
		}
		else
		{
			System.out.println("NOT found searching by title: "+title);

			sql = "SELECT idarticle,url FROM crowler.article where url like ?;";

			stmt = db.conn.prepareStatement(sql);

			String url = "%"+this.URL.split("/")[1]+"%";
			stmt.setString(1, url);
			rs = stmt.executeQuery();
			if (rs.next()) {
```

```
					String URL = rs.getString("url");
					idArticle = rs.getInt("idarticle");

					System.out.println("Article " + title);
					System.out.println("URL " + URL);
					System.out.println("ID " + idArticle);
				}
				else
				{
					System.out.println("NOT found by url: "+url);
				}
			}
			return idArticle;
		}
```

**public void save2DBAuthors(DB db, int paperID) throws SQLException {**

```
			for (Author author : this.authors) {
				// get the article ID

				int authorId = 0;

				String sql = "INSERT INTO `crowler`.`authors` (`idarticle`, `surname`,
					`givenNames`, `email`) VALUES (?, ?, ?, ?);";
				PreparedStatement stmt = db.conn.prepareStatement(sql,
					Statement.RETURN_GENERATED_KEYS);
				stmt.setInt(1, paperID);
				stmt.setString(2, author.getSurname());
				stmt.setString(3, author.getGivenNames());
				stmt.setString(4, author.getEmail());

				stmt.executeUpdate();

				ResultSet rs = stmt.getGeneratedKeys();
				if (rs.next()) {
					authorId = rs.getInt(1);
					for (Affiliation affiliation : author.affiliation) {

						sql = "INSERT INTO `crowler`.`affiliation` (`authorid`, `affId`,
							`address`, `country`,`order`) VALUES (?, ?, ?, ?, ?);";
						stmt = db.conn.prepareStatement(sql,
							Statement.RETURN_GENERATED_KEYS);
						stmt.setInt(1, authorId);
						stmt.setString(2, affiliation.getId());


						if (affiliation.getAddress() != null &&
							affiliation.getAddress().length() > 255)
							stmt.setString(3, affiliation.getAddress().substring(0, 254));
						else
							stmt.setString(3, affiliation.getAddress());

						stmt.setString(4, affiliation.getCountry());
						stmt.setString(5, affiliation.getOrder());
						stmt.execute();
					}
				}
				else
					throw new SQLException("Record NOT inserted");

			}
		}
```

```java
    public int save2DBArticle(DB db, int journalID) throws SQLException {

        int insertedArticleId = 0;

        String sql = "INSERT INTO `crowler`.`article` (`idjournal`, `title`, `url`, `pages`, `parsed`, `authors`,
`history`, `received`, `revised`, `online`, `startPage`, `endPage`) VALUES (?, ?, ?, ?, ?, ?, ?, ?, ?, ?, ?, ?);";
        PreparedStatement stmt = db.conn.prepareStatement(
            sql, Statement.RETURN_GENERATED_KEYS);
        stmt.setInt(1, journalID);
        if (this.title != null && this.title.length() > 255)
            stmt.setString(2, this.title.substring(0, 254));
        else
            stmt.setString(2, this.title);
        stmt.setString(3, this.getURL());
        stmt.setString(4, this.getPages());
        stmt.setInt(5, 0);
        stmt.setString(6, this.getAuthors());
        stmt.setString(7, this.getHistory());

        stmt.setString(8, this.getReceivedDate());
        stmt.setString(9, this.getRevisedDate());
        stmt.setString(10, this.getOnlineDate());

        stmt.setInt(11, 0);
        stmt.setInt(12, 0);

        stmt.executeUpdate();

        ResultSet rs = stmt.getGeneratedKeys();
        if (rs.next()) {
            insertedArticleId = rs.getInt(1);
        }

        return insertedArticleId;
    }
}
```

# Author.java

An entity model class for managing author data.

```
package ro.ase.crowler.model;

import java.awt.List;
import java.util.ArrayList;

public class Author {
	String surname;
	String givenNames;
	String email;
	ArrayList<Affiliation> affiliation;

	public Author(String surname, String givenNames) {
		super();
		this.surname = surname;
		this.givenNames = givenNames;
		this.email = "";
		this.affiliation = new ArrayList<>();
	}

	public String getSurname() {
		return surname;
	}

	public void setSurname(String surname) {
		this.surname = surname;
	}

	public String getGivenNames() {
		return givenNames;
	}

	public void setGivenNames(String givenNames) {
		this.givenNames = givenNames;
	}

	public String getEmail() {
		return email;
	}

	public void setEmail(String email) {
		this.email = email;
	}

	public ArrayList<Affiliation> getAffiliation() {
		return affiliation;
	}

	public void addAffiliation(Affiliation aff){
		this.affiliation.add(aff);
	}

	@Override
	public String toString(){

		StringBuilder sb = new StringBuilder();
		sb.append("\n\nName: "+surname+" "+givenNames);
```

```
		for(Affiliation aff : affiliation){
			sb.append("\n"+aff.getOrder()+" - "+aff.getId());
			sb.append("\nAddress:"+aff.getAddress());
			sb.append("\nCountry:"+aff.getCountry());
		}
		if(!email.isEmpty())
			sb.append("\nEmail:"+email);

		return sb.toString();
	}

}
```

# DB.java

Java class that manages the connection with the MySQL database used to store retrieved data

```
package ro.ase.crowler;

import java.sql.Connection;
import java.sql.DriverManager;
import java.sql.ResultSet;
import java.sql.SQLException;
import java.sql.Statement;

public class DB {

	public Connection conn = null;

	public DB() {
		try {
			Class.forName("com.mysql.jdbc.Driver");
			String url = "jdbc:mysql://localhost:3306/crowler";
			conn = DriverManager.getConnection(url, "root", "root");
			System.out.println("conn built");
		} catch (SQLException e) {
			e.printStackTrace();
		} catch (ClassNotFoundException e) {
			e.printStackTrace();
		}
	}

	public ResultSet runSql(String sql) throws SQLException {
		Statement sta = conn.createStatement();
		return sta.executeQuery(sql);
	}

	public boolean runSql2(String sql) throws SQLException {
		Statement sta = conn.createStatement();
		return sta.execute(sql);
	}

	@Override
	protected void finalize() throws Throwable {
		if (conn != null || !conn.isClosed()) {
			conn.close();
		}
	}
}
```

# Parser.java

Class that retrieves the history dates (received, revised and published dates) of the peer review process. The class processes the information gathered from different sources and extracts the needed dates in a standard format that can be used for further processing.

```
package ro.ase.crowler;
import java.text.ParseException;
import java.text.SimpleDateFormat;
import java.util.Locale;

import ro.ase.crowler.util.Debug;

public class Parser {

	public static String[] getPhysicaDates(String datesString){

		String receivedDate = null;
		String revisedDate = null;
		String onlineDate = null;


		if(datesString!=null && !datesString.equals("")){
			String[] dates = datesString.split(",");
			for(int i=0;i<dates.length;i++){
				if(dates[i].toLowerCase().contains("received")){
					String[] params = dates[i].trim().split(" ");
					if(params.length == 4){
						receivedDate = getDate(params[1],params[2],params[3]);
					}
				}
				if(dates[i].toLowerCase().contains("revised")){
					String[] params = dates[i].trim().split(" ");
					if(params.length == 4){
						revisedDate = getDate(params[1],params[2],params[3]);
					}
				}
				if(dates[i].toLowerCase().contains("online")){
					String[] params = dates[i].trim().split(" ");
					if(params.length == 5){
						onlineDate = getDate(params[2],params[3],params[4]);
					}
				}
			}
		}

		return new String[]{receivedDate,revisedDate,onlineDate};
	}

	public static String[] getPlosDates(String datesString){

		String receivedDate = null;
		String revisedDate = null;
		String onlineDate = null;

		if(datesString!=null && !datesString.equals("")){
			String[] dates = datesString.split(";");
			for(int i=0;i<dates.length;i++){
				if(dates[i].toLowerCase().contains("received")){
```

```
				String[] params = dates[i].trim().split(":");
				if(params.length == 2){
					receivedDate = getPLOSDate(params[1].trim());
				}
			}
			if(dates[i].toLowerCase().contains("accepted")){
				String[] params = dates[i].trim().split(":");
				if(params.length == 2){
					revisedDate = getPLOSDate(params[1].trim());
				}
			}
			if(dates[i].toLowerCase().contains("published")){
				String[] params = dates[i].trim().split(":");
				if(params.length == 2){
					onlineDate = getPLOSDate(params[1].trim());
				}
			}
		}
	}

	return new String[]{receivedDate,revisedDate,onlineDate};
}
```

**public static String getDate(String day, String month, String year){**

```
	String strOutput = null;


	SimpleDateFormat sdfmt1 = new SimpleDateFormat("MMMM dd, yyyy", Locale.US);
	SimpleDateFormat sdfmt2= new SimpleDateFormat("dd/MM/yyyy");
	java.util.Date dDate = null;
	try {
		dDate = sdfmt1.parse(month+" "+day+", "+year);
	} catch (ParseException e) {
		e.printStackTrace();
	}
	if(dDate!=null)
		{
			strOutput = sdfmt2.format(dDate);
			//Debug.log("Input date was "+strOutput);
		}
	return strOutput;
}
```

**public static String getPLOSDate(String dateString){**

```
	String strOutput = null;


	SimpleDateFormat sdfmt1 = new SimpleDateFormat("MMMM dd, yyyy", Locale.US);
	SimpleDateFormat sdfmt2= new SimpleDateFormat("dd/MM/yyyy");
	java.util.Date dDate = null;
	try {
		dDate = sdfmt1.parse(dateString);
	} catch (ParseException e) {
		e.printStackTrace();
	}
	if(dDate!=null)
		{
			strOutput = sdfmt2.format(dDate);
			//Debug.log("Input date was "+strOutput);
		}
```

```
		return strOutput;
	}

	public static String getNATUREDate(String dateString){

		String strOutput = null;


		SimpleDateFormat sdfmt1 = new SimpleDateFormat("yyyy-MM-dd");
		SimpleDateFormat sdfmt2= new SimpleDateFormat("dd/MM/yyyy");
		java.util.Date dDate = null;
		try {
			dDate = sdfmt1.parse(dateString);
		} catch (ParseException e) {
			e.printStackTrace();
		}
		if(dDate!=null)
			{
				strOutput = sdfmt2.format(dDate);
				//Debug.log("Input date was "+strOutput);
			}
		return strOutput;
	}

	public static int[] getPages(String pages){

		if(pages==null)
			return null;

		int startPage = 0;
		int endPage = 0;
		if(pages.toLowerCase().contains("pages")){
			pages = pages.toLowerCase().replace("pages", "").trim();
			String pagesNo[] = pages.split("-");
			if(pagesNo.length==2){
				try{
					startPage = Integer.parseInt(pagesNo[0]);
					endPage = Integer.parseInt(pagesNo[1]);
				}
				catch(Exception ex){
					//ex.printStackTrace();
					Debug.log("Unable to get pages for "+ex.getMessage());
				}
			}
		}

		return new int[]{startPage,endPage};
	}

}
```

# XMLParser.java

Java class that implements methods for parsing XML formatted files and extract needed data. The class has been used to extract data from the PLOS archive that contained published articles meta data structured in XML files.

```
package ro.ase.crowler;

import javax.xml.parsers.DocumentBuilderFactory;
import javax.xml.parsers.DocumentBuilder;
import org.w3c.dom.Document;
import org.w3c.dom.NodeList;
import org.xml.sax.SAXParseException;

import com.mysql.jdbc.Util;

import ro.ase.crowler.model.Affiliation;
import ro.ase.crowler.model.ArticleAffiliation;
import ro.ase.crowler.model.Author;
import ro.ase.crowler.util.Debug;
import ro.ase.crowler.util.Utility;

import org.w3c.dom.Node;
import org.w3c.dom.Element;

import java.io.BufferedReader;
import java.io.File;
import java.io.FileReader;
import java.io.FilenameFilter;
import java.util.HashMap;

public class XMLParser {

	private String folderPath;
	private static DB db = new DB();

	public XMLParser(String folderPath) {
		this.folderPath = folderPath;
	}

	public void listFiles(final String filterName) {

		int contor = 0;
		long startTime = System.currentTimeMillis();

		File root = new File(this.folderPath);

		// create new filename filter
		FilenameFilter fileNameFilter = null;
		if (filterName != null) {
			fileNameFilter = new FilenameFilter() {

				@Override
				public boolean accept(File dir, String name) {
					if (name.contains(filterName)) {
						return true;
					}
					return false;
```

```
				}
			};
		} else
			fileNameFilter = new FilenameFilter() {

				@Override
				public boolean accept(File dir, String name) {
					return true;
				}
			};

		for (File file : root.listFiles(fileNameFilter)) {
			Debug.log("File " + file.getName());
			contor++;
			System.out.println("Processing file " + contor);
			this.parseDoc(file);
		}

		long endTime = System.currentTimeMillis();
		long time = (endTime - startTime) / (1000 * 60);

		System.out.println("Proceesing time " + time + " minutes");
	}
```

**public void listErrorFiles(File errorFile, final String filterName) {**

```
		try {
			int contor = 0;
			long startTime = System.currentTimeMillis();

			try (BufferedReader br = new BufferedReader(new FileReader(
					errorFile))) {
				String line;
				while ((line = br.readLine()) != null) {
					if (line.startsWith("E")) {
						System.out.println("File " + line);
						if (line.contains(filterName)) {
							contor++;
							System.out.println("Processing file " + contor);
							this.parseDoc(new File(line));
						}
					}

				}
				br.close();
			}

			long endTime = System.currentTimeMillis();
			long time = (endTime - startTime) / (1000 * 60);

			System.out.println("Proceesing time " + time + " minutes");
		} catch (Exception e) {
			e.printStackTrace();
		}
	}
```

**public void parseDoc(File fXmlFile) {**

```
		try {

			Debug.log("----------------------------");
			String fileName = fXmlFile.getName();
			System.out.println("Parsing " + fileName);
```

```
DocumentBuilderFactory dbFactory = DocumentBuilderFactory.newInstance();
dbFactory.setFeature(
		"http://apache.org/xml/features/nonvalidating/load-external-dtd", false);
DocumentBuilder dBuilder = dbFactory.newDocumentBuilder();

Document doc = null;
try {
		doc = dBuilder.parse(fXmlFile);
} catch (SAXParseException e) {
		System.out.println("Trying to fix it ...");
		Utility.repairFile(fXmlFile);
		fXmlFile = new File(fXmlFile.getAbsolutePath());
		doc = dBuilder.parse(fXmlFile);
}

doc.getDocumentElement().normalize();

Debug.log("Root element :" + doc.getDocumentElement().getNodeName());

// check if research article
/*
 * NodeList articleSubject = doc.getElementsByTagName("subject"); if
 * (articleSubject != null && articleSubject.getLength() > 0) {
 * String subject = articleSubject.item(0).getTextContent();
 * Debug.log("Subject: "+subject); if(!subject.toLowerCase().equals(
 * "research article")) return; }
 */

if (!this.isResearchArticle(doc, "subject")) {
		return;
}

// get title
/*
 * String title = ""; NodeList articleTitle =
 * doc.getElementsByTagName("article-title"); if (articleTitle !=
 * null && articleTitle.getLength() > 0) { title =
 * articleTitle.item(0).getTextContent(); Debug.log("Title: " +
 * title); }
 */

String title = this.getTitle(doc, "article-title");
String receivedDate = this.getReceivedDate(doc, "date");
String acceptedDate = this.getAcceptedDate(doc, "date");
String publishedDate = this.getPublishedDate(doc, "pub-date");
String doi = this.getDOI(doc, "article-id");
String noPages = this.getNoPages(doc, "page-count");

ArticleAffiliation article = new ArticleAffiliation(title);

article.setOnlineDate(publishedDate);
article.setPages(noPages);
article.setURL(doi);
article.setRevisedDate(acceptedDate);
article.setReceivedDate(receivedDate);

// get affiliation
// build a map of unique affiliations
HashMap<String, Affiliation> affiliationMap = this.getAffiliations(doc, "aff");
```

```
// get corresponding
// build a map of corresponding addresses

HashMap<String, String> correspondingMap = this
                .getCorrespondingAddress(doc, "corresp");

// get authors
NodeList nAuthorsList = doc.getElementsByTagName("contrib");
if (nAuthorsList != null && nAuthorsList.getLength() > 0) {

        for (int temp = 0; temp < nAuthorsList.getLength(); temp++) {

                Node nNode = nAuthorsList.item(temp);

                Debug.log("\nCurrent Element :" + nNode.getNodeName());

                if (nNode.getNodeType() == Node.ELEMENT_NODE) {

                        Element eElement = (Element) nNode;

                        String authorType = eElement.getAttribute("contrib-type");

                        // Debug.log("\nCurrent element type:" +
                        // eElement.getAttribute("contrib-type"));

                        if (authorType.equals("author") &&
                                eElement.getElementsByTagName(
                                        "surname").getLength() > 0) {

                                String surname = "";
                                String givenName = "";

                                surname = eElement.getElementsByTagName(
                                        "surname").item(0).getTextContent();
                                givenName = eElement.getElementsByTagName(
                                        "given-names").item(0).getTextContent();

                                Debug.log("Author surname : " + surname);
                                Debug.log("Author given-names : " + givenName);

                                Author author = new Author(surname, givenName);

                                NodeList nAffList =
                                        eElement.getElementsByTagName("xref");
                                for (int i = 0; i < nAffList.getLength(); i++) {
                                        Node nAff = nAffList.item(i);
                                if (nAff.getNodeType() == Node.ELEMENT_NODE) {
                                                Element element = (Element) nAff;

                                                String affid = "";
                                                affid = element.getAttribute("rid");

                                                // get Affiliation
                                Affiliation affiliation = affiliationMap.get(affid);
                                if (affiliation != null)
                                        author.addAffiliation(affiliation);

                                        // get corresponding
                                        String email = correspondingMap.get(affid);
                                                if (email != null)
                                                        author.setEmail(email);
                                        }
```

```
                                        }

                                        // add author data
                                        article.addAuthor(author);

                                }
                        }
                }
            }

            // System.out.println(article.toString());

            int articleId = article.getArticleId(db);
            if (articleId == 0) {
                articleId = article.save2DBArticle(db,ArticleAffiliation.JOURNAL_ID);
            } else
                System.out.println("ARTICLE found " + title);
            if (articleId != 0) {
                System.out.println("New article added with ID = " + articleId);
                System.out.println("Saving authors data ....");
                article.save2DBAuthors(db, articleId);
            }

            Debug.log("----------------------------");

        } catch (Exception e) {

            Debug.logError(fXmlFile.getAbsolutePath());
            Debug.logError(e.getMessage());
            e.printStackTrace();
        }

        Debug.log("----------------------------");
}

/*
 *
 *
 * Internal method for checking if it is a research article
 */
boolean isResearchArticle(Document doc, String tagName) {
        // check if research article
        NodeList articleSubject = doc.getElementsByTagName(tagName);
        if (articleSubject != null && articleSubject.getLength() > 0) {
            String subject = articleSubject.item(0).getTextContent();
            Debug.log("Subject: " + subject);
            if (!subject.toLowerCase().equals("research article"))
                return false;
            else
                return true;
        } else
            return false;
}

/*
 *
 *
 * Internal method for getting the title
 */

String getTitle(Document doc, String tagName) {
```

```
		String title = "";
		NodeList articleTitle = doc.getElementsByTagName(tagName);
		if (articleTitle != null && articleTitle.getLength() > 0) {
			title = articleTitle.item(0).getTextContent();
			Debug.log("Title: " + title);
		}

		return title;
	}

	/*
	 *
	 * Internal method for getting the received date
	 */
```

**String getReceivedDate(Document doc, String tagName) {**

```
		// date
		String receivedDate = "";
		NodeList dateList = doc.getElementsByTagName(tagName);
		if (dateList != null && dateList.getLength() > 0) {

			for (int i = 0; i < dateList.getLength(); i++) {

				Element eElement = (Element) dateList.item(i);
				String type = eElement.getAttribute("date-type");
				if (type.equals("received")) {
					String day = eElement.getElementsByTagName("day").item(0)
							.getTextContent();
					String month = eElement.getElementsByTagName("month")
							.item(0).getTextContent();
					String year = eElement.getElementsByTagName("year").item(0)
							.getTextContent();

					receivedDate = day + "/" + month + "/" + year;
				}
			}
		}
		Debug.log("Received day: " + receivedDate);

		return receivedDate;
	}

	/*
	 *
	 * Internal method for getting the accepted date
	 */
```

**String getAcceptedDate(Document doc, String tagName) {**

```
		// date
		String acceptedDate = "";
		NodeList dateList = doc.getElementsByTagName(tagName);
		if (dateList != null && dateList.getLength() > 0) {

			for (int i = 0; i < dateList.getLength(); i++) {

				Element eElement = (Element) dateList.item(i);
				String type = eElement.getAttribute("date-type");
				if (type.equals("accepted")) {
					String day = eElement.getElementsByTagName("day").item(0)
							.getTextContent();
```

```
                String month = eElement.getElementsByTagName("month")
                        .item(0).getTextContent();
                String year = eElement.getElementsByTagName("year").item(0)
                        .getTextContent();

                acceptedDate = day + "/" + month + "/" + year;
            }
        }
    }
    Debug.log("Accepted day: " + acceptedDate);

    return acceptedDate;
}

/*
 *
 * Internal method for getting the accepted date
 */
```

**String getPublishedDate(Document doc, String tagName) {**

```
    // pub-date

    String onlineDate = "";
    NodeList dateList = doc.getElementsByTagName(tagName);
    if (dateList != null && dateList.getLength() > 0) {

        for (int i = 0; i < dateList.getLength(); i++) {

            Element eElement = (Element) dateList.item(i);
            String type = eElement.getAttribute("pub-type");
            if (type.equals("epub")) {
                String day = eElement.getElementsByTagName(
                    "day").item(0).getTextContent();
                String month = eElement.getElementsByTagName(
                    "month") .item(0).getTextContent();
                String year = eElement.getElementsByTagName(
                    "year").item(0).getTextContent();

                onlineDate = day + "/" + month + "/" + year;
            }
        }
    }
    Debug.log("Online day: " + onlineDate);

    return onlineDate;
}

/*
 *
 * Internal method for getting the article DOI
 */
```

**String getDOI(Document doc, String tagName) {**

```
    // article-id
    String doi = "";
    NodeList articleId = doc.getElementsByTagName(tagName);
    if (articleId != null && articleId.getLength() > 0) {

        for (int i = 0; i < articleId.getLength(); i++) {
```

```
				Element eElement = (Element) articleId.item(i);
				String attValue = eElement.getAttribute("pub-id-type");
				if (attValue.equals("doi"))
					doi = eElement.getTextContent();
			}
		}
		Debug.log("DOI: " + doi);

		return doi;
}

/*
 *
 * Internal method for getting the article no pages
 */
```

**String getNoPages(Document doc, String tagName) {**

```
		// page-count
		String noPages = "";
		NodeList pageCount = doc.getElementsByTagName(tagName);
		if (pageCount != null && pageCount.getLength() > 0) {

			Element eElement = (Element) pageCount.item(0);
			noPages = eElement.getAttribute("count");

		}
		Debug.log("Page count: " + noPages);

		return noPages;
}

/*
 *
 * Internal method for getting affiliation
 */
```

**HashMap<String, Affiliation> getAffiliations(Document doc, String tagName) {**

```
		HashMap<String, Affiliation> affiliationMap = new HashMap<>();

		NodeList nAffiliationList = doc.getElementsByTagName(tagName);
		if (nAffiliationList != null && nAffiliationList.getLength() > 0) {

			for (int temp = 0; temp < nAffiliationList.getLength(); temp++) {

				Node nNode = nAffiliationList.item(temp);

				Debug.log("\nCurrent Element :" + nNode.getNodeName());

				if (nNode.getNodeType() == Node.ELEMENT_NODE) {

					Element eElement = (Element) nNode;

					String affId = eElement.getAttribute("id");
					String affValue = "";
					String affAddress = "";
					String affCountry = "";

					// Debug.log("\nCurrent element type:" +
					// eElement.getAttribute("contrib-type"));
```

```
                        Debug.log("Aff id: " + affId);

                        if (eElement.getElementsByTagName("addr-line").getLength() > 0) {

                            if (eElement.getElementsByTagName("label").getLength() > 0)
                                affValue = eElement.getElementsByTagName("label")
                                        .item(0).getTextContent();
                            affAddress = eElement.getElementsByTagName("addr-line")
                                    .item(0).getTextContent();
                            affAddress = affAddress.trim();
                            Debug.log("Aff value: " + affValue);
                            Debug.log("Address: " + affAddress);

                            String value = eElement.getElementsByTagName(
                                "addr-line").item(0).getTextContent();
                            String[] values = value.split(",");

                            affCountry = values[values.length - 1].trim();

                            Debug.log("Country: " + affCountry);
                        }

                        // create Affiliation node
                        Affiliation affiliation = new Affiliation(affId,
                                affAddress, affCountry, affValue);
                        affiliationMap.put(affId, affiliation);
                    }
                }
            }

            return affiliationMap;
        }

        /*
         *
         *
         * Get corresponding addresses
         */
        HashMap<String, String> getCorrespondingAddress(Document doc, String tag) {
            HashMap<String, String> correspondingMap = new HashMap<>();

            NodeList nCorrList = doc.getElementsByTagName(tag);
            if (nCorrList != null && nCorrList.getLength() > 0) {
                for (int temp = 0; temp < nCorrList.getLength(); temp++) {

                    Node nNode = nCorrList.item(temp);

                    Debug.log("\nCurrent Element :" + nNode.getNodeName());

                    if (nNode.getNodeType() == Node.ELEMENT_NODE) {

                        Element eElement = (Element) nNode;

                        String corrId = eElement.getAttribute("id");
                        Debug.log("Corr id: " + corrId);

                        String corrEmail = "";
                        if (eElement.getElementsByTagName("email").getLength() > 0)
                            corrEmail = eElement.getElementsByTagName("email")
                                    .item(0).getTextContent();
                        Debug.log("Email: " + corrEmail);
```

```
                        correspondingMap.put(corrId, corrEmail);
                    }
                }
            }

            return correspondingMap;
        }

}
```

# Crawler.java

Java class that implements the crawler logic. The main objective of this class is to provide processing functions that are used to extract the data from different HTML or XML structures.

```
package ro.ase.crowler;

import java.io.IOException;
import java.net.SocketTimeoutException;
import java.sql.PreparedStatement;
import java.sql.ResultSet;
import java.sql.SQLException;
import java.sql.Statement;
import java.util.ArrayList;

import org.jsoup.HttpStatusException;
import org.jsoup.Jsoup;
import org.jsoup.nodes.Document;
import org.jsoup.nodes.Element;
import org.jsoup.select.Elements;

import ro.ase.crowler.model.Article;
import ro.ase.crowler.util.Debug;

public class Crowler {

        public static DB db = new DB();

                // process each article page for Physica A
                public static void processJournalArticle(String articleURL, Article articol)
                        throws SQLException, IOException
                {

                        // check if the given URL is already in database
                        String sql = "select * from article where URL = '" + articleURL + "'";
                        ResultSet rs = db.runSql(sql);
                        if (rs.next()) {
                                articol.setParsed(true);

                        } else {

                                // get useful information
                                boolean retry = true;
                                Document doc = null;
                                while (retry) {
                                        try {
                                                doc = Jsoup.connect(articleURL).timeout(10 * 1000).get();
                                                if (doc.text().contains("Physica A")) {
                                                        Debug.log(articleURL);
                                                        retry = false;
                                                }
                                        } catch (SocketTimeoutException ex) {
                                                Debug.log("SocketTimeoutException............retrying");
                                        }
                                }

                                Debug.log("Getting article content");

                                Elements articleHistory = doc.select("dl.articleDates");
```

```
			if (articleHistory != null) {
				String history = articleHistory.text();
				if (history.isEmpty() || history.equals(""))
					Debug.log("HISTORY: " + articleHistory.html());
				articol.setHistory(articleHistory.text());
			} else
				Debug.log("History not found");

			// get all links and recursively call the processPage method
			Elements articleContent = doc.select("div.article-content");

			for (Element item : articleContent) {
				// Debug.log("Processing "+link);
				for (Element info : item.children()) {
					// Debug.log("Article info:"+info.attr("class")+ "
					// - "+ info.text());
					if (info.hasClass("article-author-list")) {
						Elements authors = doc.select("span.author-name");
						for (Element author : authors)
							Debug.log("XXX Author:" + author.text());
					}
					if (info.hasClass("article-title"))
						Debug.log("XXX Article Title:" + info.text());
					if (info.hasClass("article-history") ||
						info.hasClass("articleDates")) {
					Debug.log("XXX Article History:" + info.text());
						articol.setHistory(info.text());
					}

				}
			}
		}
	}

	// process the article page for PLOS
	public static void processPLOSJournalArticle(String articleURL, Article articol)
		throws SQLException, IOException
	{

		// check if the given URL is already in database
		String sql = "select * from article where URL = '" + articleURL + "'";
		ResultSet rs = db.runSql(sql);
		if (rs.next()) {

		} else {

			// get useful information
			boolean retry = true;
			Document doc = null;
			while (retry) {
				try {
					doc = Jsoup.connect(articleURL).timeout(10 * 1000).get();
					if (doc.text().contains("PLOS")) {
						Debug.log(articleURL);
						retry = false;
					}
				} catch (SocketTimeoutException ex) {
					Debug.log("SocketTimeoutException...........retrying");
				}
			}
```

```
                Debug.log("Getting article content");

                Elements articleInfo = doc.select("div.articleinfo");

                if (articleInfo.size() > 0) {
                    for (Element para : articleInfo.get(0).select("p:contains(Received)")) {
                        if (para.text().contains("Received")) {

                            String text = para.text();
                            text = text.replace("<strong>", "");
                            text = text.replace("</strong>", "");

                            Debug.log("HISTORY: " + text);
                            articol.setHistory(text);
                            break;
                        }

                    }
                }
        }
}

// process the article page for Nature
public static void processNATUREJournalArticle(String articleURL, Article articol)
                throws SQLException, IOException
{

        // check if the given URL is already in database
        String sql = "select * from article where URL = '" + articleURL + "'";
        ResultSet rs = db.runSql(sql);
        if (rs.next()) {

        } else {

                // get useful information
                boolean retry = true;
                Document doc = null;
                while (retry) {
                    try {
                        doc = Jsoup.connect(articleURL).timeout(10 * 1000).get();
                        if (doc.text().contains("nature")) {
                            Debug.log(articleURL);
                            retry = false;
                        }
                    } catch (SocketTimeoutException ex) {
                        Debug.log("SocketTimeoutException............retrying");
                    }
                }

                Debug.log("Getting article content");

                Elements articleInfo = doc.select("dl.dates");
                if (articleInfo.size() > 0) {
                    Elements dates = articleInfo.select("time");
                    if (dates.size() >= 3) {
                        String received = dates.get(0).text();
                        String accepted = dates.get(1).text();
                        String online = dates.get(2).text();
                        Debug.log(String.format("HISTORY: Received %s, Accepted
                            %s, Published %s", received,accepted, online));
                        articol.setHistory(String.format("HISTORY: Received %s,
                        Accepted %s, Published %s", received, accepted, online));
```

```
                    articol.setReceivedDate(Parser.getNATUREDate(
                            dates.get(0).attr("datetime").trim()));
                    articol.setRevisedDate(Parser.getNATUREDate(
                            dates.get(1).attr("datetime").trim()));
                    articol.setOnlineDate(Parser.getNATUREDate(
                            dates.get(2).attr("datetime").trim()));
                }
            }
        }
}

// process the Nature issue page
public static void processNATUREJournalPage(String journalURL, int journalID)
        throws SQLException, IOException
{
        // check if the given URL is already in database
        String sql = "select * from Record where URL = '" + journalURL + "'";
        ResultSet rs = db.runSql(sql);
        if (rs.next()) {

        } else {
            // store the URL to database to avoid parsing again
            sql = "INSERT INTO `record` " + "(`URL`,`JournalID`) VALUES " + "(?,?);";
            PreparedStatement stmt = db.conn.prepareStatement(
                    sql, Statement.RETURN_GENERATED_KEYS);
            stmt.setString(1, journalURL);
            stmt.setInt(2, journalID);
            stmt.execute();

            // get useful information
            boolean retry = true;
            Document doc = null;
            while (retry) {

                try {
                    Thread.sleep(500);
                    doc = Jsoup.connect(journalURL).timeout(10 * 1000).get();

                    if (doc.text().contains("nature")) {
                        Debug.log(journalURL);
                    }

                    retry = false;
                } catch (Exception ex) {
                    retry = true;
                }
            }

            // get all links and recursively call the processPage method
            Elements articles = doc.select("ul.article-list>li");
            for (Element article : articles) {
                // Debug.log("Processing "+link);

                Article articol = new Article();
                articol.getNATUREElementInfo(article);

                if (articol.getURL() != null && articol.getURL().contains("http") &&
                        !articol.getURL().contains("full"))
                    processNATUREJournalArticle(articol.getURL(), articol);

                articol.setJournalID(journalID);
```

```
                Debug.log("Date articol \n " + articol.toString());
                if (articol.getHistory() != null) {
                    articol.save2DB(db, journalID);
                }

            }
        }
    }

    // process the PLOS issue page
    public static void processPLOSJournalPage(String journalURL, int journalID)
        throws SQLException, IOException
    {
        // check if the given URL is already in database
        String sql = "select * from Record where URL = '" + journalURL + "'";
        ResultSet rs = db.runSql(sql);
        if (rs.next()) {

        } else {
            // store the URL to database to avoid parsing again
            sql = "INSERT INTO `record` " + "(`URL`,`JournalID`) VALUES " + "(?,?);";
            PreparedStatement stmt = db.conn.prepareStatement(
                sql, Statement.RETURN_GENERATED_KEYS);
            stmt.setString(1, journalURL);
            stmt.setInt(2, journalID);
            stmt.execute();

            // get useful information
            boolean retry = true;
            Document doc = null;
            while (retry) {

                try {

                    Thread.sleep(500);
                    doc = Jsoup.connect(journalURL).timeout(10 * 1000).get();

                    if (doc.text().contains("PLOS")) {
                        Debug.log(journalURL);
                    }

                    retry = false;
                } catch (Exception ex) {
                    retry = true;
                }
            }

            // get all links and recursively call the processPage method
            Elements articles = doc.select("ul#search-results");
            for (Element article : articles) {
                // Debug.log("Processing "+link);

                Article articol = new Article();
                articol.getPLOSElementInfo(article);

                if (articol.getURL() != null)
                    processPLOSJournalArticle(articol.getURL(), articol);

                articol.setJournalID(journalID);

                Debug.log("Date articol \n " + articol.toString());
                if (articol.getHistory() != null) {
```

```
                articol.save2DB(db, journalID);
            }

        }
    }
}

// process the Nature issue page
public static void processJournalPage(String journalURL, int journalID)
    throws SQLException, IOException
{
    // check if the given URL is already in database
    String sql = "select * from Record where URL = '" + journalURL + "'";
    ResultSet rs = db.runSql(sql);
    if (rs.next()) {

    } else {
        // store the URL to database to avoid parsing again
        sql = "INSERT INTO `record` " + "(`URL`,`JournalID`) VALUES " + "(?,?);";
        PreparedStatement stmt = db.conn.prepareStatement(
            sql, Statement.RETURN_GENERATED_KEYS);
        stmt.setString(1, journalURL);
        stmt.setInt(2, journalID);
        stmt.execute();

        // get useful information
        boolean retry = true;
        Document doc = null;
        while (retry) {

            try {

                doc = Jsoup.connect(journalURL).timeout(10 * 1000).get();

                if (doc.text().contains("Physica A")) {
                    Debug.log(journalURL);
                }

                retry = false;
            } catch (Exception ex) {
                retry = true;
            }
        }

        // get all links and recursively call the processPage method
        Elements articles = doc.select("ul.article");
        for (Element article : articles) {
            // Debug.log("Processing "+link);

            Article articol = new Article();
            articol.getElementInfo(article);

            if (articol.getURL() != null)
                processJournalArticle(articol.getURL(), articol);

            articol.setJournalID(journalID);

            Debug.log("Date articol \n " + articol.toString());
            if (!articol.isParsed())
                articol.save2DB(db, journalID);
        }
    }
```

```
}

// process the Nature issue page
public static void reprocessJournalArticle() throws SQLException, IOException {
	// check if the given URL is already in database
	String sql = "select * from article where history = ''";
	ResultSet rs = db.runSql(sql);
	while (rs.next()) {
		String url = rs.getString("url");
		int id = rs.getInt("idarticle");

		// get useful information
		boolean retry = true;
		Document doc = null;
		while (retry) {
			try {
				doc = Jsoup.connect(url).timeout(10 * 1000).get();
				if (doc.text().contains("Physica A")) {
					Debug.log(url);
					retry = false;
				}
			} catch (SocketTimeoutException ex) {
				Debug.log("SocketTimeoutException............retrying");
			} catch (HttpStatusException ex) {
				Debug.log(ex.getMessage() + " - " + url);
				retry = false;
				continue;
			}
		}

		Debug.log("Getting article content");

		Elements articleHistory = null;
		if (doc != null)
			articleHistory = doc.select("dl.articleDates");

		if (articleHistory != null) {
			String history = articleHistory.text();
			if (history.isEmpty() || history.equals(""))
				Debug.log("HISTORY: " + articleHistory.html());
			else {
				sql = "update article SET history = ? where idarticle=?";
				PreparedStatement stmt = db.conn.prepareStatement(sql);
				stmt.setString(1, history);
				stmt.setInt(2, id);
				stmt.execute();
				Debug.log("Article with id = " + id + " updated with history " +
					history);
			}
		} else
			Debug.log("History not found");
	}
}

// reprocess pages for which the history data was missing
public static void reprocessJournalArticleHP() throws SQLException, IOException {
	// check if the given URL is already in database
	String sql = "select * from article where history <>''";
	ResultSet rs = db.runSql(sql);
	while (rs.next()) {
		String history = rs.getString("history");
		String pages = rs.getString("pages");
```

```
                int idJournal = rs.getInt("idjournal");
                int id = rs.getInt("idarticle");

                String[] dates = null;
                int[] page = null;

                if (idJournal == 1) {
                        dates = Parser.getPlosDates(history);
                        // page = Parser.getPages(pages);
                } else
                        if(idJournal >= 300){
                        dates = Parser.getPhysicaDates(history);
                        page = Parser.getPages(pages);
                }

                if (dates != null) {
                        sql = "update article SET received = ?, revised = ?, online = ?, startPage
                                = ?, endPage = ? where idarticle=?";
                        PreparedStatement stmt = db.conn.prepareStatement(sql);
                        stmt.setString(1, dates[0]);
                        stmt.setString(2, dates[1]);
                        stmt.setString(3, dates[2]);
                        if (page != null && page.length == 2) {
                                stmt.setInt(4, page[0]);
                                stmt.setInt(5, page[1]);
                        } else {
                                stmt.setInt(4, 0);
                                stmt.setInt(5, 0);
                        }
                        stmt.setInt(6, id);
                        stmt.execute();
                        Debug.log("Article with id = " + id + " updated with history " + history);
                }
        }
}


// process the Cell issue page
public static void processCellJournalPage(String journalIssueURL, int journalID)
        throws SQLException, IOException
{
        // check if the given URL is already in database
        String sql = "select * from Record where URL = '" + journalIssueURL + "'";
        ResultSet rs = db.runSql(sql);
        if (rs.next()) {

        } else {
                // store the URL to database to avoid parsing again
                sql = "INSERT INTO `record` " + "(`URL`,`JournalID`) VALUES " + "(?,?);";
                PreparedStatement stmt = db.conn.prepareStatement(
                        sql, Statement.RETURN_GENERATED_KEYS);
                stmt.setString(1, journalIssueURL);
                stmt.setInt(2, journalID);
                stmt.execute();

                // get useful information
                boolean retry = true;
                Document doc = null;
                while (retry) {

                        try {
```

```
					Thread.sleep(500);
					doc = Jsoup.connect(journalIssueURL).timeout(10 *
						1000).get();

					if (doc.text().contains("PLOS")) {
						Debug.log(journalIssueURL);
					}

					retry = false;
				} catch (Exception ex) {
					retry = true;
				}
			}

			// get all links and recursively call the processPage method
			Elements articles = doc.select("div.articleCitation");
			for (Element article : articles) {
				// Debug.log("Processing "+link);

				Article articol = new Article();
				articol.getCellElementInfo(article);

				if (articol.getURL() != null)
					processCellJournalArticle(articol.getURL(), articol);

				articol.setJournalID(journalID);

				Debug.log("Date articol \n " + articol.toString());
				if (articol.getHistory() != null) {
					articol.save2DB(db, journalID);
				}

			}
		}
	}

	// process the Cell article page
	public static void processCellJournalArticle(String articleURL, Article articol)
		throws SQLException, IOException
	{

		// check if the given URL is already in database
		String sql = "select * from article where URL = '" + articleURL + "'";
		ResultSet rs = db.runSql(sql);
		if (rs.next()) {

		} else {

			// get useful information
			boolean retry = true;
			Document doc = null;
			int contor = 0;
			while (retry && contor < 3) {
				try {
					doc = Jsoup.connect(articleURL).timeout(10 * 1000).get();
					if (doc.text().contains("cell")) {
						Debug.log(articleURL);
						retry = false;
						contor = 0;
					}
				} catch (SocketTimeoutException ex) {
					contor++;
```

```
                        Debug.log("SocketTimeoutException............retrying");
                    }
                }

                if(doc!=null){
                Debug.log("Getting article content");

                Elements articleInfo = doc.select("div.articleDates");

                String history = "";

                if (articleInfo.size() > 0) {
                    for (Element para : articleInfo.get(0).select("span.pubDatesRow")) {

                        if(para.hasText()){
                            if(!para.text().contains(";"))
                                history += para.text()+";";
                            else
                                history += para.text();
                        }

                    }
                    Debug.log("HISTORY: " + history);
                    articol.setHistory(history);
                }
                }
            }
        }


        // process each issue page for Physica A
        public static ArrayList<String> processVolumesPage(String URL, int id)
            throws SQLException, IOException
        {
            ArrayList<String> volumes = new ArrayList<>();

            // get useful information
            Document doc = Jsoup.connect(URL).timeout(10 * 1000).get();
            if (doc.text().contains("Physica A")) {
                Debug.log(URL);
            }

            // get all links and recursively call the processPage method
            Elements questions = doc.select("a[href]");
            for (Element link : questions) {
                // Debug.log("Processing "+link);
                if (link.attr("href").contains("/science/journal/03784371")) {
                    String volumeURL = link.attr("abs:href");
                    String[] paths = volumeURL.split("/");
                    if (paths.length > 7) {
                        volumes.add(volumeURL);
                        int volumeId = Integer.parseInt(paths[paths.length - 2]);
                        Debug.log("PROCESSING VOLUME " + volumeURL + " with
                            id " + volumeId);

                        try {
                            Thread.sleep(2000);
                        } catch (InterruptedException e) {
                            // TODO Auto-generated catch block
                            e.printStackTrace();
                        }
                        processJournalPage(volumeURL, volumeId);
```

```
                }
            }
            // processPage(link.attr("abs:href"));
        }

        return volumes;

    }

    // process the archive page for Physica A
    public static void processPage(String URL) throws SQLException, IOException {
        // check if the given URL is already in database
        String sql = "select * from Record where URL = '" + URL + "'";
        ResultSet rs = db.runSql(sql);
        if (rs.next()) {

        } else {
            // store the URL to database to avoid parsing again
            sql = "INSERT INTO `record` " + "(`URL`) VALUES " + "(?);";
            PreparedStatement stmt = db.conn.prepareStatement(sql,
                Statement.RETURN_GENERATED_KEYS);
            stmt.setString(1, URL);
            stmt.execute();

            // get useful information
            Document doc = Jsoup.connect(URL).get();

            if (doc.text().contains("Physica A")) {
                Debug.log(URL);
            }

            // get all links and recursively call the processPage method
            Elements questions = doc.select("a[href]");
            for (Element link : questions) {
                // Debug.log("Processing "+link);
                if (link.attr("href").contains("/journal/03784371"))
                    processPage(link.attr("abs:href"));
            }
        }
    }
}
```

# Utility.java

Utility class that contains different processing methods for extracting week days, country names and for correcting data formatting errors (some XML files where not well formatted).

```
package ro.ase.crowler.util;

import java.io.BufferedReader;
import java.io.BufferedWriter;
import java.io.File;
import java.io.FileNotFoundException;
import java.io.FileOutputStream;
import java.io.FileReader;
import java.io.IOException;
import java.io.OutputStreamWriter;
import java.io.PrintWriter;
import java.io.Writer;
import java.sql.PreparedStatement;
import java.sql.ResultSet;
import java.sql.SQLException;
import java.sql.Statement;
import java.text.DateFormat;
import java.text.ParseException;
import java.text.SimpleDateFormat;
import java.util.ArrayList;
import java.util.Calendar;
import java.util.Date;
import java.util.Locale;

import ro.ase.crowler.DB;

import com.mysql.fabric.xmlrpc.base.Array;

public class Utility {
	public static void repairFile(File file) throws FileNotFoundException, IOException
	{

		String oldName = file.getAbsolutePath();

		copyFile(file, oldName + ".old");

		File existingFile = new File(oldName + ".old");

		Writer out = new BufferedWriter(new OutputStreamWriter(
				new FileOutputStream(oldName), "UTF-8"));

		try (BufferedReader br = new BufferedReader(
				new FileReader(existingFile))) {
			String line;
			boolean isLineOk = false;
			while ((line = br.readLine()) != null) {

				if (!isLineOk && line.startsWith("<?xml"))
					isLineOk = true;

				if (isLineOk) {
					out.write(line);
				}

			}
```

```
			out.close();
			br.close();
		}

	}

	public static void copyFile(File file, String newFile) throws IOException
	{
		PrintWriter pw = new PrintWriter(new File(newFile));
		try (BufferedReader br = new BufferedReader(new FileReader(file))) {
			String line;
			while ((line = br.readLine()) != null) {
				pw.println(line);
			}
			pw.close();
			br.close();
		}
	}

	public static ArrayList<String> getCountryNames() {
		String[] locales = Locale.getISOCountries();
		ArrayList<String> countries = new ArrayList<>();

		for (String countryCode : locales) {

			Locale obj = new Locale("", countryCode);

			System.out.println("Country Code = " + obj.getCountry()
					+ ", Country Name = " + obj.getDisplayCountry());
			countries.add(obj.getDisplayCountry());
		}

		return countries;

	}

	public static void correctCountryNames(DB db, ArrayList<String> countries) {

		for (String country : countries) {

			System.out.println("Update for "+country);
			String sql = "UPDATE `crowler`.`affiliation` SET `country` = ?
				where country like '%"+country+"%'";
			PreparedStatement stmt;
			try {
				stmt = db.conn.prepareStatement(sql,
					Statement.RETURN_GENERATED_KEYS);
				stmt.setString(1, country);
				stmt.execute();

			} catch (SQLException e) {
				// TODO Auto-generated catch block
				e.printStackTrace();
			}

		}
	}

	public static void correctUSStateNames(DB db) {

		for (State country : State.values()) {
			System.out.println("Update for state "+country.getState());
```

```
			String sql = "UPDATE `crowler`.`affiliation` SET `country` = 'United States of America'
				where country like '%"+country.getState()+"%'";
			PreparedStatement stmt;
			try {
				stmt = db.conn.prepareStatement(sql,
					Statement.RETURN_GENERATED_KEYS);
				//stmt.setString(1, country.getState());
				stmt.execute();

			} catch (SQLException e) {
				// TODO Auto-generated catch block
				e.printStackTrace();
			}

		}
	}

	public static void insertCountryNames(DB db, ArrayList<String> countries) {

		for (String country : countries) {
			String sql = "INSERT INTO `crowler`.`countries` (name) VALUES(?)";
			PreparedStatement stmt;
			try {
				stmt = db.conn.prepareStatement(sql,
						Statement.RETURN_GENERATED_KEYS);
				stmt.setString(1, country);
				stmt.execute();

			} catch (SQLException e) {
				// TODO Auto-generated catch block
				e.printStackTrace();
			}

		}
	}

	public static void correctEmailAddressCell(DB db, String filter) {

		System.out.println("Update for filter" + filter);

		String sql = "SELECT authorId,idarticle,email, affiliation.id, country, address, affiliation.order as
affiliationOrder FROM crowler.authors, crowler.affiliation where authors.id=affiliation.authorId and  country like
'%"+filter+"%';";
		PreparedStatement stmt;

		try {
			stmt = db.conn.prepareStatement(sql);
			ResultSet rs = stmt.executeQuery();

			int contor = 0;

			while (rs.next()) {
				contor++;
				System.out.println("----------------- Processing article " + contor);
				int authorId = rs.getInt("authorId");
				int idArticle = rs.getInt("idarticle");
				int affId = rs.getInt("id");
				String country = rs.getString("country");
				int order = rs.getInt("affiliationOrder");

				if(order!=1){
					String sqlUpdate = "UPDATE `crowler`.`affiliation` SET
```

```
						`country` = '' where id = ?";
				PreparedStatement stmt2 = db.conn.prepareStatement(sqlUpdate);
				stmt2.setInt(1, affId);
				stmt2.execute();

				String sqlUpdateAuthor = "UPDATE `crowler`.`authors`
					SET `email` = ? where id = ?";
				PreparedStatement stmt3 =
					db.conn.prepareStatement(sqlUpdateAuthor);
				stmt3.setString(1, country);
				stmt3.setInt(2, authorId);
				stmt3.execute();
			}
		}

	} catch (SQLException e) {
		// TODO Auto-generated catch block
		e.printStackTrace();
	}
}


public static void extractDates(DB db) throws ParseException {

	System.out.println("Extract dates");

	String sql = "SELECT authorId,idarticle FROM crowler.aut_iso_detailed where online_year is null;";
	PreparedStatement stmt;

	try {
		stmt = db.conn.prepareStatement(sql);
		ResultSet rs = stmt.executeQuery();

		int contor = 0;

		while (rs.next()) {
			contor++;
			System.out.println("----------------- Processing article " + contor);



			int authorId = rs.getInt("authorId");
			int idArticle = rs.getInt("idarticle");

			System.out.println("Article ID "+idArticle);

			String sql_article = "SELECT idjournal,authors,received,revised,online,startPage,
endPage FROM crowler.article where idarticle = ?;";
			PreparedStatement stmt_article = db.conn.prepareStatement(sql_article);
			stmt_article.setInt(1, idArticle);

			ResultSet rs_article = stmt_article.executeQuery();

			if(rs_article.next()){

				String authors = rs_article.getString("authors");
				String receivedDate = rs_article.getString("received");
				String revisedDate = rs_article.getString("revised");
				String onlineDate = rs_article.getString("online");
				int startpage = rs_article.getInt("startPage");
				int endPage = rs_article.getInt("endPage");
				int journalid = rs_article.getInt("idjournal");
```

```
int receivedValues[] = Utility.getDataInfo(receivedDate);
int revisedValues[] = Utility.getDataInfo(revisedDate);
int online[] = Utility.getDataInfo(onlineDate);


        int authorsNo = 0;
        String[] authorsList = authors.split(",");
        if(authorsList.length != 0){
                String lastAuthor = authorsList[authorsList.length-1];
                if(lastAuthor.isEmpty() || lastAuthor.equals("") ||
                        lastAuthor.equals(" "))
                        authorsNo = authorsList.length-1;
                else
                        authorsNo = authorsList.length;
        }
        //System.out.println("Authors:" + authors);
        //System.out.println("Authors number: "+authorsNo);
        int pagesNo = endPage-startpage+1;
        //System.out.println("Pages number:"+pagesNo);


        String sqlUpdate = "UPDATE crowler.aut_iso_detailed SET
                received_week_day = ?, received_day = ?, "
                        + "received_month = ?, received_year = ?,"
                        + "revised_week_day = ?, revised_day =
                ?,revised_month = ?,revised_year = ?,"
                        + "online_week_day = ?, online_day =
                ?,online_month = ?,online_year = ?,"
                        + "authors_number = ? , pages_number = ? ,
                journalid = ? "
                        + "WHERE idarticle = ? AND authorid = ?;";
        PreparedStatement stmt_update =
                db.conn.prepareStatement(sqlUpdate);
        stmt_update.setInt(1, receivedValues[0]);
        stmt_update.setInt(2, receivedValues[1]);
        stmt_update.setInt(3, receivedValues[2]);
        stmt_update.setInt(4, receivedValues[3]);

        stmt_update.setInt(5, revisedValues[0]);
        stmt_update.setInt(6, revisedValues[1]);
        stmt_update.setInt(7, revisedValues[2]);
        stmt_update.setInt(8, revisedValues[3]);

        stmt_update.setInt(9, online[0]);
        stmt_update.setInt(10, online[1]);
        stmt_update.setInt(11, online[2]);
        stmt_update.setInt(12, online[3]);

        stmt_update.setInt(13, authorsNo);
        stmt_update.setInt(14, pagesNo);
        stmt_update.setInt(15, journalid);
        stmt_update.setInt(16, idArticle);
        stmt_update.setInt(17, authorId);

        stmt_update.execute();
    }

}

} catch (SQLException e) {
```

```
			// TODO Auto-generated catch block
			e.printStackTrace();
		}
	}

	public static int[] getDataInfo(String processedDate) throws ParseException {

		int[] values = new int[4];
		Date received = null;

		DateFormat df = new SimpleDateFormat("dd/MM/yyyy");
		try {
			received = df.parse(processedDate);
		} catch (Exception e) {
			values[0] = 0;
			values[1] = 0;
			values[2] = 0;
			values[3] = 0;
			return values;
		}

		// System.out.println(received.toLocaleString());

		String dayOfWeek = new SimpleDateFormat("EEEE", Locale.ENGLISH).format(received);
		// System.out.println(dayOfWeek); // Friday

		// System.out.println("Day number:"+Utility.getWeekDay(dayOfWeek));

		Calendar calendar = Calendar.getInstance();
		calendar.setTime(received);

		int receivedWeekDay = calendar.get(Calendar.DAY_OF_WEEK) - 1;
		if (receivedWeekDay == 0)
			receivedWeekDay = 7;
		int receivedDay = calendar.get(Calendar.DAY_OF_MONTH);
		int receivedMonth = calendar.get(Calendar.MONTH) + 1;
		int receivedYear = calendar.get(Calendar.YEAR);

		values[0] = receivedWeekDay;
		values[1] = receivedDay;
		values[2] = receivedMonth;
		values[3] = receivedYear;

		// System.out.println("Week Day:"+receivedWeekDay);
		// System.out.println("Day:"+receivedDay);
		// System.out.println("Month:"+receivedMonth);
		// System.out.println("Year:"+receivedYear);

		return values;

	}

	public static int getWeekDay(String weekday){
		switch(weekday){
		case "Monday":
			return 1;
		case "Tuesday":
			return 2;
		case "Wednesday":
			return 3;
		case "Thursday":
			return 4;
```

```
case "Friday":
    return 5;
case "Saturday":
    return 6;
case "Sunday":
    return 7;
default:
    throw new IllegalArgumentException();
}
}

}
```

## SI. Python Script for *LQ* tool that compute *Localization Quotient*:

```
import arcpy
import arcpy.da as da
tab_tari = arcpy.GetParameterAsText(0)
tab_date = arcpy.GetParameterAsText(1)
lts = [rnds for rnds in da.SearchCursor(tab_date, ("iso2", "FREQUENCY", "FREQUENCY_1") ) ]
st = 0
ssel = 0
for rd in lts:
	if rd[1] is not None: st += rd[1]
	if rd[2] is not None: ssel += rd[2]
rap_jos = float(ssel) / st * 100
for rd in lts:
	if rd[2] is None:
		rap_sus = 0.
	else:
		rap_sus = float(rd[2]) / rd[1] * 100
	coef_loc = rap_sus / rap_jos * 100
	with da.UpdateCursor(tab_tari, ("lq"), "ISO2 = '"+rd[0]+"'") as crs:
		for rdu in crs:
			rdu[0] = coef_loc
			crs.updateRow(rdu)
```

Python Script for *Simb_GC* tool that applies a graduated colors symbology on *Countries* layer:

```
import arcpy
tari = arcpy.GetParameterAsText(0)
nrcl = arcpy.GetParameter(1)
mxd = arcpy.mapping.MapDocument("CURRENT")
df = arcpy.mapping.ListDataFrames(mxd, "World")[0]
fsl = "d:/stat_art/simbol_GC.lyr"
sl = arcpy.mapping.Layer(fsl)
dl = arcpy.mapping.ListLayers(mxd, tari, df)[0]
arcpy.mapping.UpdateLayer(df, dl, sl, "TRUE")
if dl.symbologyType == "GRADUATED_COLORS":
	dl.symbology.numClasses = nrcl
	dl.symbology.valueField = "lq"
	arcpy.SetParameter(2,True)
else:
	arcpy.SetParameter(2,False)
arcpy.RefreshActiveView()
arcpy.RefreshTOC()
```

**Table SI.2.** Correlation matrix for the independent variables for consolidated data set and for each journal

**A. Consolidated data set**

| Variables | Mon | Tue | Wed | Thu | Week-end | Spring | Summer | Fall | America | Africa | Asia | Oceania | Christ-mas | $\log_{10}$AUTH | $\log_{10}$HDI | $\log_{10}$LTO |
|---|---|---|---|---|---|---|---|---|---|---|---|---|---|---|---|---|
| **Monday** | 1 | | | | | | | | | | | | | | | |
| **Tuesday** | -0.216** | 1 | | | | | | | | | | | | | | |
| **Wednesday** | -0.217** | -0.224** | 1 | | | | | | | | | | | | | |
| **Thursday** | -0.215** | -0.223** | -0.223** | 1 | | | | | | | | | | | | |
| **Weekend** | -0.160** | -0.165** | -0.166** | -0.162** | 1 | | | | | | | | | | | |
| **Spring** | -0.012** | 0.005* | 0.003 | 0.001 | -0.002 | 1 | | | | | | | | | | |
| **Summer** | 0.003 | -0.001 | 0.004 | 0 | -0.009** | -0.346** | 1 | | | | | | | | | |
| **Fall** | 0 | -0.001 | 0.002 | 0.002 | 0 | -0.332** | -0.332** | 1 | | | | | | | | |
| **America** | 0.011** | 0.007** | 0.007** | -0.002 | -0.046** | 0.005* | 0.008** | -0.001 | 1 | | | | | | | |
| **Africa** | -0.005* | 0.001 | -0.003 | 0.002 | 0.010** | 0 | -0.001 | 0 | -0.083** | 1 | | | | | | |
| **Asia** | -0.021** | -0.011** | -0.015** | -0.008** | 0.111** | -0.002 | -0.011** | 0.004 | -0.437** | -0.073** | 1 | | | | | |
| **Oceania** | 0.004 | 0.001 | 0.002 | 0.006* | 0.020** | -0.005* | -0.004 | 0.004 | -0.132** | -0.022** | -0.116** | 1 | | | | |
| **Christmas** | 0.005* | -0.006** | -0.014** | -0.002 | 0.023** | -.0132** | -0.132** | -0.126** | -0.005* | 0.004 | 0.019** | -0.004 | 1 | | | |
| **$\log_{10}$AUTH** | -0.002 | -0.001 | 0.004 | 0.003 | -0.004 | 0.001 | 0 | -0.005* | -0.050** | -0.007** | 0.041** | -0.035** | 0.003 | 1 | | |
| **$\log_{10}$HDI** | 0.019** | 0.009** | 0.013** | 0.005* | -0.084** | 0 | 0.013** | -0.004 | 0.277** | -0.435** | -0.573** | 0.136** | -0.019** | -0.006** | 1 | |
| **$\log_{10}$LTO** | -0.012** | -0.009** | -0.010** | -0.002 | 0.050** | -0.001 | -0.006* | -0.001 | -0.735** | -0.126** | 0.601** | -0.276** | 0.009** | 0.091** | -0.335** | 1 |

**B. Physica A: Statistical mechanics and its applications**

| Variables | Mon | Tue | Wed | Thu | Week-end | Spring | Summer | Fall | America | Africa | Asia | Oceania | Christ-mas | $\log_{10}$AUTH | $\log_{10}$HDI | $\log_{10}$LTO |
|---|---|---|---|---|---|---|---|---|---|---|---|---|---|---|---|---|
| **Monday** | 1 | | | | | | | | | | | | | | | |
| **Tuesday** | -0.214** | 1 | | | | | | | | | | | | | | |
| **Wednesday** | -0.211** | -0.213** | 1 | | | | | | | | | | | | | |
| **Thursday** | -0.207** | -0.208** | -0.205** | 1 | | | | | | | | | | | | |
| **Weekend** | -0.189** | -0.190** | -0.188** | -0.175** | 1 | | | | | | | | | | | |
| **Spring** | -0.01 | -0.007 | 0.037** | 0.01 | -0.011 | 1 | | | | | | | | | | |
| **Summer** | -0.011 | 0 | -0.027** | -0.006 | 0.013 | -0.329** | 1 | | | | | | | | | |
| **Fall** | 0.002 | 0 | 0.008 | 0.007 | 0.005 | -0.350** | -0.332** | 1 | | | | | | | | |
| **America** | 0.015 | 0.001 | -0.009 | 0.016 | -0.052** | -0.005 | 0.001 | 0.008 | 1 | | | | | | | |
| **Africa** | 0.006 | 0.001 | -0.007 | -0.005 | 0.030** | -0.019 | 0.006 | 0.013 | -0.078** | 1 | | | | | | |
| **Asia** | -0.046** | -0.002 | -0.008 | -0.012 | 0.101** | 0.008 | 0.003 | -0.007 | -0.487** | -0.122** | 1 | | | | | |
| **Oceania** | 0.013 | 0 | 0.009 | 0.005 | -0.024* | 0.006 | -0.013 | -0.012 | -0.070** | -0.018 | -0.110** | 1 | | | | |
| **Christmas** | 0.009 | 0.007 | -0.008 | -0.007 | 0.001 | -0.134** | -0.128** | -0.136** | 0.013 | 0.006 | 0.001 | -0.006 | 1 | | | |
| $log_{10}$AUTH | -0.015 | -0.014 | 0 | -0.007 | 0.016 | 0.004 | -0.012 | 0.008 | 0.019 | -0.005 | 0.112** | -0.027** | 0 | 1 | | |
| $log_{10}$HDI | 0.034** | 0.012 | 0.011 | 0.003 | -0.083** | 0.005 | 0.001 | -0.021* | 0.050** | -0.288** | -0.442** | 0.112** | -0.013 | -0.213** | 1 | |
| $log_{10}$LTO | -0.025* | -0.003 | 0.01 | -0.017 | 0.028** | 0.024* | -0.004 | -0.004 | -0.569** | -0.343** | 0.515** | -0.175** | -0.008 | 0.082** | -0.111** | 1 |

## C. PLOS ONE

| Variables | Mon | Tue | Wed | Thu | Week-end | Spring | Summer | Fall | America | Africa | Asia | Oceania | Christ-mas | $log_{10}$AUTH | $log_{10}$HDI | $log_{10}$LTO |
|---|---|---|---|---|---|---|---|---|---|---|---|---|---|---|---|---|
| Monday | 1 | | | | | | | | | | | | | | | |
| Tuesday | -0.216** | 1 | | | | | | | | | | | | | | |
| Wednesday | -0.217** | -0.225** | 1 | | | | | | | | | | | | | |
| Thursday | -0.216** | -0.224** | -0.225** | 1 | | | | | | | | | | | | |
| Weekend | -0.157** | -0.163** | -0.164** | -0.162** | 1 | | | | | | | | | | | |
| Spring | -0.013** | 0.006* | 0.002 | 0 | 0 | 1 | | | | | | | | | | |
| Summer | 0.004 | -0.001 | 0.006* | -0.001 | -0.010** | -0.347** | 1 | | | | | | | | | |
| Fall | 0.001 | -0.001 | 0.001 | 0.002 | 0 | -0.331** | -0.332** | 1 | | | | | | | | |
| America | 0.010** | 0.007** | 0.010** | -0.002 | -0.049** | 0.005* | 0.008** | -0.002 | 1 | | | | | | | |
| Africa | -0.006* | 0.001 | -0.003 | 0.003 | 0.008** | 0.001 | -0.002 | 0 | -0.082** | 1 | | | | | | |
| Asia | -0.020** | -0.012** | -0.016** | -0.007** | 0.114** | -0.002 | -0.011** | 0.004 | -0.428** | -0.074** | 1 | | | | | |
| Oceania | 0.003 | 0.002 | 0.001 | 0.006* | 0.023** | -0.006* | -0.003 | 0.005* | -0.133** | -0.023** | -0.121** | 1 | | | | |
| Christmas | 0.004 | -0.007** | -0.015** | 0 | 0.025** | -0.132** | -0.132** | -0.126** | -0.005* | 0.003 | 0.021** | -0.004 | 1 | | | |
| $log_{10}$AUTH | -0.001 | 0 | 0.004 | 0 | 0.004 | 0 | -0.005 | -0.002 | -0.108** | 0.002 | 0.096** | -0.046** | 0.005* | 1 | | |
| $log_{10}$HDI | 0.018** | 0.009** | 0.014** | 0.004 | -0.085** | 0 | 0.012** | -0.001 | 0.276** | -0.449** | -0.574** | 0.140** | -0.019** | -0.069** | 1 | |
| $log_{10}$LTO | -0.011** | -0.009** | -0.012** | -0.002 | 0.055** | -0.002 | -0.006* | 0 | -0.732** | -0.115** | 0.605** | -0.291** | 0.010** | 0.149** | -0.339** | 1 |

## D. Nature

| Variables | Mon | Tue | Wed | Thu | Week-end | Spring | Summer | Fall | America | Africa | Asia | Oceania | Christ-mas | $log_{10}$AUTH | $log_{10}$HDI | $log_{10}$LTO |
|---|---|---|---|---|---|---|---|---|---|---|---|---|---|---|---|---|
| Monday | 1 | | | | | | | | | | | | | | | |
| Tuesday | -0.224** | 1 | | | | | | | | | | | | | | |
| Wednesday | -0.212** | -0.212** | 1 | | | | | | | | | | | | | |
| Thursday | -0.212** | -0.213** | -0.201** | 1 | | | | | | | | | | | | |
| Weekend | -0.174** | -0.174** | -0.164** | -0.165** | 1 | | | | | | | | | | | |
| Spring | -0.011 | 0.015 | -0.013 | 0.013 | 0.006 | 1 | | | | | | | | | | |
| Summer | 0.006 | 0.008 | 0.008 | 0.004 | -0.019 | -0.363** | 1 | | | | | | | | | |
| Fall | -0.022 | -0.008 | 0.014 | 0.004 | 0.008 | -0.343** | -0.341** | 1 | | | | | | | | |
| America | 0.01 | 0.009 | -0.011 | -0.021 | 0.014 | -0.002 | -0.002 | 0.029* | 1 | | | | | | | |
| Africa | -0.016 | 0.001 | 0.003 | -0.015 | 0.008 | -0.005 | -0.005 | -0.003 | -0.038** | 1 | | | | | | |
| Asia | -0.009 | 0.007 | -0.009 | -0.016 | 0.022 | 0.002 | -0.004 | 0.009 | -0.346** | -0.01 | 1 | | | | | |
| Oceania | 0.017 | -0.027 | 0.015 | -0.009 | 0.021 | 0.014 | -0.023 | -0.006 | -0.168** | -0.005 | -0.044** | 1 | | | | |

| | | | | | | | | | | | | | | | |
|---|---|---|---|---|---|---|---|---|---|---|---|---|---|---|---|
| **Christmas** | 0.027 | -0.013 | -0.033* | -0.034* | 0.019 | -0.124** | -0.123** | -0.117** | 0.008 | -0.007 | -0.014 | 0.008 | 1 | | |
| **$log_{10}$AUTH** | -0.006 | -0.039** | 0.009 | 0.01 | 0.025 | -0.017 | -0.007 | -0.005 | -0.072** | 0.026 | 0.026 | 0.025 | 0.029 | 1 | |
| **$log_{10}$HDI** | 0.007 | 0.003 | -0.006 | 0.023 | -0.005 | -0.02 | 0.022 | -0.006 | 0.266** | -0.294** | -0.599** | 0.095** | 0.007 | -0.049** | 1 | |
| **$log_{10}$LTO** | -0.004 | 0.001 | -0.004 | 0.017 | -0.004 | -0.001 | 0.014 | -0.029 | -0.863** | -0.006 | 0.453** | -0.166** | -0.01 | 0.058** | -0.336** | 1 |

**E. Cell**

| Variables | Mon | Tue | Wed | Thu | Week-end | Spring | Summer | Fall | America | Africa | Asia | Oceania | Christ-mas | $log_{10}$AUTH | $log_{10}$HDI | $log_{10}$LTO |
|---|---|---|---|---|---|---|---|---|---|---|---|---|---|---|---|---|
| **Monday** | 1 | | | | | | | | | | | | | | | |
| **Tuesday** | -0.226** | 1 | | | | | | | | | | | | | | |
| **Wednesday** | -0.219** | -0.219** | 1 | | | | | | | | | | | | | |
| **Thursday** | -0.224** | -0.224** | -0.217** | 1 | | | | | | | | | | | | |
| **Weekend** | -0.157** | -0.157** | -0.152** | -0.155** | 1 | | | | | | | | | | | |
| **Spring** | 0.042* | -0.018 | -0.021 | -0.013 | -0.01 | 1 | | | | | | | | | | |
| **Summer** | -0.006 | -0.009 | -0.012 | 0.018 | 0.023 | -0.368** | 1 | | | | | | | | | |
| **Fall** | -0.041* | 0.007 | 0.031 | 0.015 | -0.02 | -0.338** | -0.337** | 1 | | | | | | | | |
| **America** | -0.004 | 0.011 | -0.003 | -0.014 | 0.056** | -0.002 | -0.012 | 0.019 | 1 | | | | | | | |
| **Africa** | -0.008 | -0.008 | -0.008 | -0.008 | -0.005 | 0.027 | -0.01 | -0.009 | -0.027 | 1 | | | | | | |
| **Asia** | 0.025 | -0.009 | -0.003 | -0.022 | 0.007 | -0.006 | -0.003 | 0.001 | -0.420** | -0.004 | 1 | | | | | |
| **Oceania** | 0.023 | -0.017 | 0.025 | -0.009 | -0.018 | -0.032 | -0.004 | 0.032 | -0.141** | -0.001 | -0.022 | 1 | | | | |
| **Christmas** | 0.01 | 0.01 | 0.023 | -0.019 | -0.009 | -0.124** | -0.124** | -0.114** | 0.005 | -0.003 | 0.003 | -0.002 | 1 | | | |
| **$log_{10}$AUTH** | -0.015 | -0.029 | -0.016 | 0.022 | 0.013 | 0.038* | -0.03 | -0.001 | -0.084** | -0.013 | .059** | .067** | 0.016 | 1 | | |
| **$log_{10}$HDI** | 0.013 | 0.027 | -0.015 | 0.003 | -0.008 | -0.013 | 0 | -0.003 | 0.285** | -0.154** | -0.542** | 0.069** | 0.003 | -0.057** | 1 | |
| **$log_{10}$LTO** | 0.001 | 0 | 0 | 0.008 | -.050** | 0.006 | 0.012 | -0.02 | -0.920** | 0.001 | 0.486** | -.093** | -0.004 | 0.079** | -0.289** | 1 |

**Note:** * Statistically significant at level 10%
** Statistically significant at level 5%
*** Statistically significant at level 1%

## 2000-2004 timespan

**Table SI.3.** Regression estimates for the dependent variables for consolidated data set and each journal for each model by time

| **A. Consolidated data set** | | | | | | | | | |
|---|---|---|---|---|---|---|---|---|---|
| **Models** / **Variables/ characteristics** | **M1 (equation 3)** | **M2 (equation 4)** | **M3 (equation 5)** | **M4 (equation 6)** | **M5 (equation 7)** | **M6 (equation 8)** | **M7 (equation 9)** | **M8 (equation 10)** | **M9 (equation 11)** |
| *0* | *1* | *2* | *3* | *4* | *5* | *6* | *7* | *8* | *9* |
| **Intercept** | 0.654*** | 0.653*** | 0.653*** | 0.668*** | 0.680*** | 0.657*** | 0.583*** | 0.622*** | 0.667*** |
| **Monday** (1 yes, 0 no) | 0.084*** | | | 0.085*** | 0.088*** | 0.088*** | 0.099*** | 0.099*** | 0.099*** |
| **Tuesday** (1 yes, 0 no) | 0.005 | | | 0.004 | 0.009 | 0.007 | 0.017 | 0.017 | 0.016 |
| **Wednesday** (1 yes, 0 no) | -0.0117 | | | -0.01 | -0.005 | -0.007 | 0.0001 | -0.0009 | -0.0018 |
| **Thursday** (1 yes, 0 no) | -0.0196 | | | -0.0189 | -0.019 | -0.021 | -0.012 | -0.011 | -0.015 |
| **Weekend** (1 yes, 0 no) | -0.467*** | | | -0.472*** | -0.469*** | -0.466*** | -0.441*** | -0.44*** | -0.44*** |
| **Adjusted $R^2$ (weighted)** | **0.116*** ** | | | | | | | | |
| **Spring** (1 yes, 0 no) | | -0.026 | | -0.02 | -0.02 | 0.002 | -0.0003 | -0.0004 | -0.006 |
| **Summer** (1 yes, 0 no) | | -0.019 | | -0.005 | -0.007 | 0.015 | 0.015 | 0.015 | 0.014 |
| **Fall** (1 yes, 0 no) | | -0.026 | | -0.021 | -0.024 | -0.001 | -0.0031 | 0.0008 | -0.005 |
| **Adjusted $R^2$ (weighted)** | | **0.001*** ** | | **0.12*** ** | | | | | |
| **America** (1 yes, 0 no) | | | 0.022 | | 0.012 | 0.01 | -0.018 | -0.005 | 0.044 |
| **Africa** (1 yes, 0 no) | | | 0.076 | | 0.016 | 0.104 | 0.111 | 0.219*** | 0.315*** |
| **Asia** (1 yes, 0 no) | | | -0.089*** | | -0.081*** | -0.080*** | -0.073*** | -0.045* | -0.058** |
| **Oceania** (1 yes, 0 no) | | | -0.159* | | -0.168** | -0.1674* | -0.143* | 0.148* | -0.067 |
| **Adjusted $R^2$ (weighted)** | | | **0.013*** ** | | **0.134*** ** | | | | |
| **Christmas** (1 yes, 0 no) | | | | | | 0.133*** | 0.131*** | 0.137*** | 0.134*** |
| **Adjusted $R^2$ (weighted)** | | | | | | **0.136*** ** | | | |
| **$log_{10}$AUTHORS** | | | | | | | 0.066*** | 0.058*** | 0.061*** |
| **Adjusted $R^2$ (weighted)** | | | | | | | **0.145*** ** | | |
| **$log_{10}$HDI** | | | | | | | | 0.292*** | 0.374*** |
| **Adjusted $R^2$ (weighted)** | | | | | | | | **0.146*** ** | |
| **$log_{10}$LTO** | | | | | | | | | 0.154** |
| **Adjusted $R^2$** | | | | | | | | | **0.151*** ** |
| **B. Physica A: Statistical mechanics and its applications** | | | | | | | | | |
| *0* | *1* | *2* | *3* | *4* | *5* | *6* | *7* | *8* | *9* |
| **Intercept** | 0.248*** | 0.251*** | 0.268*** | 0.247*** | 0.263*** | 0.248*** | 0.258*** | 0.260*** | 0.289*** |
| **Monday** (1 yes, 0 no) | 0.081*** | | | 0.081*** | 0.082*** | 0.082*** | 0.081*** | 0.081*** | 0.081*** |

## 2000-2004 timespan

| | | | | | | | | | |
|---|---|---|---|---|---|---|---|---|---|
| **Tuesday** (1 yes, 0 no) | 0.022 | | | 0.022 | 0.025 | 0.023 | 0.022 | 0.022 | 0.021 |
| **Wednesday** (1 yes, 0 no) | -0.00039 | | | -0.0011 | -0.0018 | -0.002 | -0.002 | -0.003 | -0.003 |
| **Thursday** (1 yes, 0 no) | -0.00029 | | | -0.0001 | -0.00069 | -0.002 | -0.003 | -0.003 | -0.006 |
| **Weekend** (1 yes, 0 no) | -0.227*** | | | -0.226*** | -0.224*** | -0.223*** | -0.223*** | -0.224*** | 0.0225*** |
| **Adjusted $R^2$ (weighted)** | **0.08*** ** | | | | | | | | |
| **Spring** (1 yes, 0 no) | | 0.0074 | | 0.0085 | 0.008 | 0.024 | 0.023 | 0.024 | 0.020 |
| **Summer** (1 yes, 0 no) | | 0.0073 | | 0.011 | 0.010 | 0.027 | 0.027 | 0.027 | 0.026 |
| **Fall** (1 yes, 0 no) | | -0.005 | | -0.009 | -0.0126 | 0.006 | 0.004 | 0.005 | 0.0003 |
| **Adjusted $R^2$ (weighted)** | | **0.0003*** ** | | **0.08*** ** | | | | | |
| **America** (1 yes, 0 no) | | | -0.011 | | -0.0159 | -0.017 | -0.016 | -0.0145 | 0.014 |
| **Africa** (1 yes, 0 no) | | | 0.072 | | 0.098** | 0.096** | 0.097** | 0.105* | 0.165** |
| **Asia** (1 yes, 0 no) | | | -0.052*** | | -0.047*** | -0.047*** | -0.047*** | -0.045** | -0.054*** |
| **Oceania** (1 yes, 0 no) | | | -0.097 | | -0.097* | -0.096* | -0.098* | -0.098* | -0.045 |
| **Adjusted $R^2$ (weighted)** | | | **0.008*** ** | | **0.088*** ** | | | | |
| **Christmas** (1 yes, 0 no) | | | | | | 0.113*** | 0.113*** | 0.114*** | 0.109*** |
| **Adjusted $R^2$ (weighted)** | | | | | | **0.092*** ** | | | |
| **log$_{10}$AUTHORS** | | | | | | | -0.029 | -0.027 | -0.019 |
| **Adjusted $R^2$ (weighted)** | | | | | | | **0.092*** ** | | |
| **log$_{10}$HDI** | | | | | | | | 0.053 | 0.150 |
| **Adjusted $R^2$ (weighted)** | | | | | | | | **0.092*** ** | |
| **log$_{10}$LTO** | | | | | | | | | 0.097*** |
| **Adjusted $R^2$** | | | | | | | | | **0.094*** ** |
| **E. Cell** | | | | | | | | | |
| *0* | *1* | *2* | *3* | *4* | *5* | *6* | *7* | *8* | *9* |
| **Intercept** | 0.3549*** | 0.4264*** | 0.3988*** | 0.4176*** | 0.4468*** | 0.4333*** | 0.4632*** | 0.4889*** | 0.5066*** |
| **Monday** (1 yes, 0 no) | 0.0658* | | | 0.0664** | 0.0626* | 0.0622* | 0.0604* | 0.061* | 0.06* |
| **Tuesday** (1 yes, 0 no) | -0.003 | | | -0.0118 | -0.0132 | -0.0152 | -0.0163 | -0.0163 | -0.017 |
| **Wednesday** (1 yes, 0 no) | 0.0419 | | | 0.0483 | 0.0482 | 0.0469 | 0.0453 | 0.0471 | 0.0466 |
| **Thursday** (1 yes, 0 no) | -0.049 | | | -0.0482 | -0.049 | -0.0512 | -0.0525 | -0.0522 | -0.0531 |
| **Weekend** (1 yes, 0 no) | -0.088 | | | -0.0542 | -0.0512 | -0.0572 | -0.0572 | -0.0556 | -0.0557 |
| **Adjusted $R^2$ (weighted)** | **0.04*** ** | | | | | | | | |
| **Spring** (1 yes, 0 no) | | -0.1281*** | | -0.1296*** | -0.13*** | -0.1155*** | -0.1148*** | -0.1142*** | -0.1149*** |
| **Summer** (1 yes, 0 no) | | -0.0791** | | -0.08** | -0.0789** | -0.0637* | -0.0637* | -0.0641* | -0.065* |
| **Fall** (1 yes, 0 no) | | -0.0312 | | -0.0351 | -0.0347 | -0.0206 | -0.0201 | -0.0213 | -0.0205 |
| **Adjusted $R^2$ (weighted)** | | **0.049*** ** | | **0.085*** ** | | | | | |

## 2000-2004 timespan

| | | | | | | | | | |
|---|---|---|---|---|---|---|---|---|---|
| **America** (1 yes, 0 no) | | | -0.0466* | | -0.0396 | -0.0394 | -0.0404 | -0.0416 | 0.001 |
| **Africa** (1 yes, 0 no) | | | N.A. | | N.A. | N.A. | N.A. | N.A. | N.A. |
| **Asia** (1 yes, 0 no) | | | -0.0073 | | -0.0032 | -0.0047 | -0.0048 | 0.0074 | -0.0044 |
| **Oceania** (1 yes, 0 no) | | | 0.1028 | | 0.1221 | 0.1215 | 0.121 | 0.1126 | 0.1683 |
| **Adjusted $R^2$ (weighted)** | | | **0.009*** ** | | **0.091*** ** | | | | |
| **Christmas** (1 yes, 0 no) | | | | | | 0.0628 | 0.0629 | 0.0615 | 0.0634 |
| **Adjusted $R^2$ (weighted)** | | | | | | **0.093*** ** | | | |
| **$\log_{10}$AUTHORS** | | | | | | | -0.0282 | -0.0273 | -0.0293 |
| **Adjusted $R^2$ (weighted)** | | | | | | | **0.934*** ** | | |
| **$\log_{10}$HDI** | | | | | | | | 0.7053 | 0.5234 |
| **Adjusted $R^2$ (weighted)** | | | | | | | | **0.094*** ** | |
| **$\log_{10}$LTO** | | | | | | | | | 0.1101 |
| **Adjusted $R^2$** | | | | | | | | | **0.096*** ** |

**Note:** * Statistically significant at level 10%

** Statistically significant at level 5%

*** Statistically significant at level 1%

N.A. – Due to the lack of variability (no papers having PCA – Papers' Corresponding Authors – from Africa) this factor was automatically removed from the model.

## 2005-2007 timespan

| A. Consolidated data set | | | | | | | | | |
|---|---|---|---|---|---|---|---|---|---|
| Variables/ characteristics \ Models | **M1 (equation 3)** | **M2 (equation 4)** | **M3 (equation 5)** | **M4 (equation 6)** | **M5 (equation 7)** | **M6 (equation 8)** | **M7 (equation 9)** | **M8 (equation 10)** | **M9 (equation 11)** |
| *0* | *1* | *2* | *3* | *4* | *5* | *6* | *7* | *8* | *9* |
| **Intercept** | 0.501*** | 0.516*** | 0.492*** | 0.524*** | 0.507*** | 0.494*** | 0.464*** | 0.476*** | 0.494*** |
| **Monday** (1 yes, 0 no) | 0.0528** | | | 0.053*** | 0.053*** | 0.055*** | 0.056*** | 0.053*** | 0.052** |
| **Tuesday** (1 yes, 0 no) | 0.074*** | | | 0.075*** | 0.076*** | 0.077*** | 0.078*** | 0.078*** | 0.075*** |
| **Wednesday** (1 yes, 0 no) | 0.043** | | | 0.48** | 0.049** | 0.05** | 0.05** | 0.047** | 0.046** |
| **Thursday** (1 yes, 0 no) | 0.026 | | | 0.03 | 0.03 | 0.033 | 0.034 | 0.033 | 0.029 |
| **Weekend** (1 yes, 0 no) | -0.460*** | | | -0.459*** | -0.459*** | -0.457*** | -0.455*** | -0.456*** | -0.463*** |
| **Adjusted $R^2$ (weighted)** | **0.121*** ** | | | | | | | | |
| **Spring** (1 yes, 0 no) | | -0.037* | | -0.046** | -0.046** | -0.035* | -0.035* | -0.033 | -0.028 |
| **Summer** (1 yes, 0 no) | | -0.002 | | -0.003 | -0.003 | 0.007 | 0.006 | 0.008 | 0.01 |
| **Fall** (1 yes, 0 no) | | -0.039** | | -0.045*** | -0.04** | -0.032* | -0.034* | -0.031* | -0.027 |
| **Adjusted $R^2$ (weighted)** | | **0.002*** ** | | **0.124*** ** | | | | | |
| **America** (1 yes, 0 no) | | | 0.026* | | 0.032** | 0.032** | 0.027* | 0.028* | 0.053** |
| **Africa** (1 yes, 0 no) | | | -0.033 | | 0.025 | 0.025 | 0.031 | 0.072 | 0.129* |
| **Asia** (1 yes, 0 no) | | | -0.029* | | 0.012 | 0.012 | 0.019 | 0.035* | 0.036* |
| **Oceania** (1 yes, 0 no) | | | 0.0019 | | 0.013 | 0.016 | 0.02 | 0.019 | 0.023 |
| **Adjusted $R^2$ (weighted)** | | | **0.003*** ** | | **0.124*** ** | | | | |
| **Christmas** (1 yes, 0 no) | | | | | | 0.055* | 0.056* | 0.06* | 0.061* |
| **Adjusted $R^2$ (weighted)** | | | | | | **0.125*** ** | | | |
| **$\log_{10}$AUTHORS** | | | | | | | 0.021*** | 0.02** | 0.021*** |
| **Adjusted $R^2$ (weighted)** | | | | | | | **0.127*** ** | | |
| **$\log_{10}$HDI** | | | | | | | | 0.103 | 0.145 |
| **Adjusted $R^2$ (weighted)** | | | | | | | | **0.128*** ** | |
| **$\log_{10}$LTO** | | | | | | | | | 0.075 |
| **Adjusted $R^2$** | | | | | | | | | **0.13*** ** |
| **B. Physica A: Statistical mechanics and its applications** | | | | | | | | | |
| *0* | *1* | *2* | *3* | *4* | *5* | *6* | *7* | *8* | *9* |
| **Intercept** | 0.176*** | 0.163*** | 0.204*** | 0.161*** | 0.174*** | 0.167*** | 0.196*** | 0.194*** | 0.221*** |
| **Monday** (1 yes, 0 no) | -0.0004 | | | 0.0009 | 0.00073 | 0.0002 | 0.0005 | -0.003 | -0.002 |
| **Tuesday** (1 yes, 0 no) | 0.056*** | | | 0.056*** | 0.0558*** | 0.054*** | 0.053*** | 0.054*** | 0.052*** |

## 2005-2007 timespan

| | | | | | | | | | |
|---|---|---|---|---|---|---|---|---|---|
| **Wednesday** (1 yes, 0 no) | 0.052*** | | | 0.052*** | 0.052*** | 0.052*** | 0.051*** | 0.048*** | 0.047** |
| **Thursday** (1 yes, 0 no) | 0.059*** | | | 0.058*** | 0.0589*** | 0.058*** | 0.055*** | 0.055*** | 0.051*** |
| **Weekend** (1 yes, 0 no) | -0.190*** | | | -0.192*** | -0.191*** | -0.191*** | -0.191*** | -0.193*** | -0.200*** |
| **Adjusted $R^2$ (weighted)** | **0.109*** ** | | | | | | | | |
| **Spring** (1 yes, 0 no) | | 0.008 | | -0.0011 | -0.0019 | 0.005 | 0.005 | 0.007 | 0.011 |
| **Summer** (1 yes, 0 no) | | 0.032** | | 0.038** | 0.038** | 0.045*** | 0.043*** | 0.045*** | 0.047*** |
| **Fall** (1 yes, 0 no) | | 0.033** | | 0.026* | 0.0261* | 0.033** | 0.032** | 0.035** | 0.038** |
| **Adjusted $R^2$ (weighted)** | | **0.003*** ** | | **0.114*** ** | | | | | |
| **America** (1 yes, 0 no) | | | -0.032** | | -0.0233 | -0.023 | -0.019 | -0.020 | 0.009 |
| **Africa** (1 yes, 0 no) | | | -0.015 | | 0.016 | 0.017 | 0.0068 | -0.002 | 0.077 |
| **Asia** (1 yes, 0 no) | | | -0.031** | | -0.0179 | -0.0169 | -0.011 | -0.013 | -0.013 |
| **Oceania** (1 yes, 0 no) | | | -0.037 | | -0.023 | -0.021 | -0.027 | -0.026 | -0.0398 |
| **Adjusted $R^2$ (weighted)** | | | **0.003*** ** | | **0.117*** ** | | | | |
| **Christmas** (1 yes, 0 no) | | | | | | 0.037 | 0.037 | 0.041 | 0.044 |
| **Adjusted $R^2$ (weighted)** | | | | | | **0.117*** ** | | | |
| **$\log_{10}$AUTHORS** | | | | | | | -0.086*** | -0.091*** | -0.088*** |
| **Adjusted $R^2$ (weighted)** | | | | | | | **0.123*** ** | | |
| **$\log_{10}$HDI** | | | | | | | | -0.058 | 0.061 |
| **Adjusted $R^2$ (weighted)** | | | | | | | | **0.125*** ** | |
| **$\log_{10}$LTO** | | | | | | | | | 0.095*** |
| **Adjusted $R^2$** | | | | | | | | | **0.132*** ** |
| **C. PLOS ONE** | | | | | | | | | |
| *0* | *1* | *2* | *3* | *4* | *5* | *6* | *7* | *8* | *9* |
| **Intercept** | 0.6366*** | 0.683*** | 0.599*** | 0.7180*** | 0.7192*** | 0.6781*** | 0.6756*** | 0.686*** | 0.681*** |
| **Monday** (1 yes, 0 no) | 0.0836** | | | 0.0897** | 0.0909** | 0.1026*** | 0.1026*** | 0.1046 | 0.103*** |
| **Tuesday** (1 yes, 0 no) | 0.0299 | | | 0.0347 | 0.0355 | 0.0474 | 0.0474 | 0.0470 | 0.0452 |
| **Wednesday** (1 yes, 0 no) | 0.017 | | | 0.0303 | 0.0290 | 0.0401 | 0.0400 | 0.0365 | 0.0388 |
| **Thursday** (1 yes, 0 no) | -0.107*** | | | -0.0936** | -0.0972** | -0.0865** | -0.0865** | -0.0897** | -0.0878** |
| **Weekend** (1 yes, 0 no) | -0.756*** | | | -0.7513*** | -0.7583*** | -0.750*** | -0.750*** | -0.751*** | -0.7498*** |
| **Adjusted $R^2$** | **0.157*** ** | | | | | | | | |
| **Spring** (1 yes, 0 no) | | -0.149*** | | -0.1437*** | -0.1444*** | -0.1128** | -0.1127*** | -0.1114*** | -0.1084*** |
| **Summer** (1 yes, 0 no) | | -0.042 | | -0.0526 | -0.0550* | -0.0235 | -0.0235 | -0.0232 | -0.0239 |
| **Fall** (1 yes, 0 no) | | -0.140*** | | -0.1257*** | -0.1289*** | -0.0975*** | -0.0975*** | -0.0946*** | -0.0914*** |
| **Adjusted $R^2$** | | **0.0145*** ** | | **0.172*** ** | | | | | |
| **America** (1 yes, 0 no) | | | -0.0126 | | -0.0111 | -0.0105 | -0.0103 | -0.0110 | -0.024 |

## 2005-2007 timespan

| | | | | | | | | | |
|---|---|---|---|---|---|---|---|---|---|
| **Africa** (1 yes, 0 no) | | | -0.0246 | | 0.0228 | 0.0239 | 0.0235 | 0.0939 | 0.0336 |
| **Asia** (1 yes, 0 no) | | | -0.0219 | | 0.0250 | 0.0257 | 0.0256 | 0.0423 | 0.0448 |
| **Oceania** (1 yes, 0 no) | | | 0.0757 | | 0.1416* | 0.1492* | 0.1496* | 0.1465** | 0.1309 |
| **Adjusted $R^2$** | | | **0.0007*** ** | | **0.176*** ** | | | | |
| **Christmas** (1 yes, 0 no) | | | | | | 0.1485** | 0.1483** | 0.1499*** | 0.1473** |
| **Adjusted $R^2$** | | | | | | **0.178*** ** | | | |
| **$log_{10}$AUTHORS** | | | | | | | 0.0033 | 0.0058 | 0.0014 |
| **Adjusted $R^2$** | | | | | | | **0.1779*** ** | | |
| **$log_{10}$HDI** | | | | | | | | 0.3234 | 0.3340 |
| **Adjusted $R^2$** | | | | | | | | **0.1787*** ** | |
| **$log_{10}$LTO** | | | | | | | | | -0.0345 |
| **Adjusted $R^2$** | | | | | | | | | **0.1774*** ** |
| **E. Cell** | | | | | | | | | |
| *0* | *1* | *2* | *3* | *4* | *5* | *6* | *7* | *8* | *9* |
| **Intercept** | 0.2898*** | 0.3369*** | 0.2826*** | 0.317*** | 0.2872*** | 0.2715*** | 0.3053*** | 0.3166*** | 0.2914*** |
| **Monday** (1 yes, 0 no) | 0.0569** | | | 0.054* | 0.055* | 0.0552* | 0.0531* | 0.0532* | 0.0523* |
| **Tuesday** (1 yes, 0 no) | 0.0545* | | | 0.051* | 0.0517* | 0.0523* | 0.0504* | 0.0504* | 0.0507* |
| **Wednesday** (1 yes, 0 no) | 0.0292 | | | 0.0287 | 0.0285 | 0.024 | 0.0224 | 0.0228 | 0.0236 |
| **Thursday** (1 yes, 0 no) | 0.0172 | | | 0.016 | 0.018 | 0.0204 | 0.0195 | 0.0198 | 0.0202 |
| **Weekend** (1 yes, 0 no) | -0.1329*** | | | -0.1368*** | -0.1396*** | -0.1376*** | -0.1385*** | -0.1386*** | -0.1393*** |
| **Adjusted $R^2$ (weighted)** | **0.045*** ** | | | | | | | | |
| **Spring** (1 yes, 0 no) | | -0.0228 | | -0.025 | -0.0236 | -0.0088 | -0.0074 | -0.0073 | -0.0087 |
| **Summer** (1 yes, 0 no) | | -0.0475* | | -0.044* | -0.0435* | -0.0284 | -0.029 | -0.0292 | -0.0296 |
| **Fall** (1 yes, 0 no) | | -0.0295 | | -0.030 | -0.0299 | -0.0151 | -0.015 | -0.0151 | -0.0172 |
| **Adjusted $R^2$ (weighted)** | | **0.005*** ** | | **0.051*** ** | | | | | |
| **America** (1 yes, 0 no) | | | 0.0329 | | 0.0358* | 0.0369* | 0.0365* | 0.0357* | -0.0342 |
| **Africa** (1 yes, 0 no) | | | N.A. | | N.A. | N.A. | N.A. | N.A. | N.A. |
| **Asia** (1 yes, 0 no) | | | 0.0674* | | 0.0699* | 0.0697* | 0.0711* | 0.0773* | 0.1037** |
| **Oceania** (1 yes, 0 no) | | | 0.022 | | 0.0061 | 0.0136 | 0.0192 | 0.0156 | -0.0743 |
| **Adjusted $R^2$ (weighted)** | | | **0.005*** ** | | **0.058*** ** | | | | |
| **Christmas** (1 yes, 0 no) | | | | | | 0.1171** | 0.1161** | 0.1162** | 0.1125** |
| **Adjusted $R^2$ (weighted)** | | | | | | **0.065*** ** | | | |
| **$log_{10}$AUTHORS** | | | | | | | -0.0312 | -0.0306 | -0.0287 |
| **Adjusted $R^2$ (weighted)** | | | | | | | **0.066*** ** | | |
| **$log_{10}$HDI** | | | | | | | | 0.3066 | 0.537 |

## 2005-2007 timespan

| | | | | | | | | | |
|---|---|---|---|---|---|---|---|---|---|
| **Adjusted $R^2$ (weighted)** | | | | | | | | **0.066*** ** | |
| **$\log_{10}$LTO** | | | | | | | | | -0.1766 |
| **Adjusted $R^2$** | | | | | | | | | **0.068*** ** |

**Note:** * Statistically significant at level 10%

** Statistically significant at level 5%

*** Statistically significant at level 1%

N.A. – Due to the lack of variability (no papers having PCA – Papers' Corresponding Authors – from Africa) this factor was automatically removed from the model.

## 2008-2010 timespan

| A. Consolidated data set | | | | | | | | | |
|---|---|---|---|---|---|---|---|---|---|
| Variables/ characteristics \ Models | **M1 (equation 3)** | **M2 (equation 4)** | **M3 (equation 5)** | **M4 (equation 6)** | **M5 (equation 7)** | **M6 (equation 8)** | **M7 (equation 9)** | **M8 (equation 10)** | **M9 (equation 11)** |
| *0* | *1* | *2* | *3* | *4* | *5* | *6* | *7* | *8* | *9* |
| **Intercept** | 0.327*** | 0.362*** | 0.335*** | 0.357*** | 0.355*** | 0.332*** | 0.327*** | 0.315*** | 0.316*** |
| **Monday** (1 yes, 0 no) | 0.026*** | | | 0.028*** | 0.028*** | 0.023*** | 0.023*** | 0.024*** | 0.023*** |
| **Tuesday** (1 yes, 0 no) | 0.062*** | | | 0.064*** | 0.065*** | 0.061*** | 0.061*** | 0.061*** | 0.060*** |
| **Wednesday** (1 yes, 0 no) | 0.078*** | | | 0.079*** | 0.080*** | 0.076*** | 0.076*** | 0.077*** | 0.077*** |
| **Thursday** (1 yes, 0 no) | 0.062*** | | | 0.062*** | 0.062*** | 0.059*** | 0.059*** | 0.059*** | 0.060*** |
| **Weekend** (1 yes, 0 no) | -0.584*** | | | -0.583*** | -0.583*** | -0.586*** | -0.586*** | -0.587*** | -0.586*** |
| **Adjusted $R^2$ (weighted)** | **0.316*** ** | | | | | | | | |
| **Spring** (1 yes, 0 no) | | -0.006 | | 0.0004 | 0.000003 | 0.026*** | 0.026*** | 0.027*** | 0.028*** |
| **Summer** (1 yes, 0 no) | | -0.058*** | | -0.053*** | -0.053*** | -0.027*** | -0.027*** | -0.026*** | -0.026*** |
| **Fall** (1 yes, 0 no) | | -0.067*** | | -0.058*** | -0.058*** | -0.032*** | -0.032*** | -0.031*** | -0.031*** |
| **Adjusted $R^2$ (weighted)** | | **0.008*** ** | | **0.325*** ** | | | | | |
| **America** (1 yes, 0 no) | | | -0.003 | | 0.006 | 0.005 | 0.006 | 0.0061 | 0.0093 |
| **Africa** (1 yes, 0 no) | | | -0.067*** | | -0.019 | -0.019 | -0.019 | -0.079*** | -0.068** |
| **Asia** (1 yes, 0 no) | | | -0.042*** | | 0.004 | 0.003 | 0.004 | -0.01 | -0.011 |
| **Oceania** (1 yes, 0 no) | | | -0.053*** | | -0.029** | -0.03** | -0.03** | -0.025* | -0.022 |
| **Adjusted $R^2$ (weighted)** | | | **0.03*** ** | | **0.325*** ** | | | | |
| **Christmas** (1 yes, 0 no) | | | | | | 0.14*** | 0.14*** | 0.142*** | 0.143*** |
| **Adjusted $R^2$ (weighted)** | | | | | | **0.333*** ** | | | |
| **$\log_{10}$AUTHORS** | | | | | | | 0.002 | 0.0024 | 0.0228 |
| **Adjusted $R^2$ (weighted)** | | | | | | | **0.333*** ** | | |
| **$\log_{10}$HDI** | | | | | | | | -0.122*** | -0.124*** |
| **Adjusted $R^2$ (weighted)** | | | | | | | | **0.333*** ** | |
| **$\log_{10}$LTO** | | | | | | | | | 0.0095 |
| **Adjusted $R^2$** | | | | | | | | | **0.331*** ** |
| **B. Physica A: Statistical mechanics and its applications** | | | | | | | | | |
| *0* | *1* | *2* | *3* | *4* | *5* | *6* | *7* | *8* | *9* |
| **Intercept** | 0.213*** | 0.184*** | 0.206*** | 0.219*** | 0.241*** | 0.219*** | 0.208*** | 0.208*** | 0.216*** |
| **Monday** (1 yes, 0 no) | -0.019 | | | -0.211 | -0.020 | -0.025 | -0.024 | -0.022 | -0.024 |
| **Tuesday** (1 yes, 0 no) | -0.0104 | | | -0.008 | -0.007 | -0.004 | -0.004 | -0.005 | -0.004 |

## 2008-2010 timespan

| | | | | | | | | | |
|---|---|---|---|---|---|---|---|---|---|
| **Wednesday** (1 yes, 0 no) | -0.015 | | | -0.018 | -0.016 | -0.019 | -0.018 | -0.013 | -0.013 |
| **Thursday** (1 yes, 0 no) | 0.031 | | | 0.027 | 0.027 | 0.025 | 0.025 | 0.026 | 0.026 |
| **Weekend** (1 yes, 0 no) | -0.237*** | | | -0.236*** | -0.231*** | -0.234*** | -0.234*** | -0.237*** | -0.238*** |
| **Adjusted $R^2$ (weighted)** | **0.131*** ** | | | | | | | | |
| **Spring** (1 yes, 0 no) | | 0.020 | | 0.021 | 0.021 | 0.045*** | 0.044*** | 0.046*** | 0.049*** |
| **Summer** (1 yes, 0 no) | | -0.041** | | -0.029* | -0.029* | -0.006 | -0.006 | -0.003 | -0.002 |
| **Fall** (1 yes, 0 no) | | -0.022 | | -0.018 | -0.018 | 0.005 | 0.003 | 0.004 | 0.008 |
| **Adjusted $R^2$ (weighted)** | | **0.009*** ** | | **0.136*** ** | | | | | |
| **America** (1 yes, 0 no) | | | -0.015 | | -0.014 | -0.015 | -0.018 | -0.017 | -0.007 |
| **Africa** (1 yes, 0 no) | | | -0.032 | | 0.008 | 0.007 | 0.0006 | 0.0049 | 0.006 |
| **Asia** (1 yes, 0 no) | | | -0.057*** | | -0.039*** | -0.038*** | -0.042*** | -0.039** | -0.042** |
| **Oceania** (1 yes, 0 no) | | | -0.116** | | -0.118*** | -0.124*** | -0.128*** | -0.128*** | -0.130** |
| **Adjusted $R^2$ (weighted)** | | | **0.013*** ** | | **0.144*** ** | | | | |
| **Christmas** (1 yes, 0 no) | | | | | | 0.132*** | 0.133*** | 0.128*** | 0.132*** |
| **Adjusted $R^2$ (weighted)** | | | | | | **0.155*** ** | | | |
| **$log_{10}$AUTHORS** | | | | | | | 0.034 | 0.032 | 0.030 |
| **Adjusted $R^2$ (weighted)** | | | | | | | **0.157*** ** | | |
| **$log_{10}$HDI** | | | | | | | | 0.019 | 0.027 |
| **Adjusted $R^2$ (weighted)** | | | | | | | | **0.159*** ** | |
| **$log_{10}$LTO** | | | | | | | | | 0.035 |
| **Adjusted $R^2$** | | | | | | | | | **0.164*** ** |
| **C. PLOS ONE** | | | | | | | | | |
| *0* | *1* | *2* | *3* | *4* | *5* | *6* | *7* | *8* | *9* |
| **Intercept** | 0.382*** | 0.4454*** | 0.4141*** | 0.4252*** | 0.430*** | 0.3996*** | 0.394*** | 0.3886*** | 0.387*** |
| **Monday** (1 yes, 0 no) | 0.0366*** | | | 0.0394*** | 0.0401*** | 0.0337*** | 0.0336*** | 0.0338*** | 0.0334*** |
| **Tuesday** (1 yes, 0 no) | 0.085*** | | | 0.0898*** | 0.090*** | 0.0847*** | 0.0847*** | 0.0850*** | 0.0844*** |
| **Wednesday** (1 yes, 0 no) | 0.122*** | | | 0.1238*** | 0.1243*** | 0.1206*** | 0.1206*** | 0.1204*** | 0.121*** |
| **Thursday** (1 yes, 0 no) | 0.089*** | | | 0.0896*** | 0.0901*** | 0.0860*** | 0.0860*** | 0.0867*** | 0.0870*** |
| **Weekend** (1 yes, 0 no) | -0.568*** | | | -0.5654*** | -0.563*** | -0.5663*** | -0.5662*** | -0.5666*** | -0.5661*** |
| **Adjusted $R^2$** | **0.264*** ** | | | | | | | | |
| **Spring** (1 yes, 0 no) | | -0.0189** | | -0.0102 | -0.0105 | 0.0244*** | 0.0244*** | 0.0254*** | 0.0257*** |
| **Summer** (1 yes, 0 no) | | -0.078*** | | -0.0718*** | -0.0718*** | -0.03693*** | -0.0369*** | -0.0363*** | -0.0361*** |
| **Fall** (1 yes, 0 no) | | -0.084*** | | -0.0763*** | -0.0761*** | -0.0411*** | -0.0411*** | -0.0413*** | -0.0409*** |
| **Adjusted $R^2$** | | **0.0096*** ** | | **0.273*** ** | | | | | |
| **America** (1 yes, 0 no) | | | -0.0182*** | | -0.0049 | -0.0053 | -0.0049 | -0.0046 | -0.0076 |

## 2008-2010 timespan

| | | | | | | | | | |
|---|---|---|---|---|---|---|---|---|---|
| **Africa** (1 yes, 0 no) | | | -0.0628*** | | -0.0212 | -0.0117 | -0.0124 | -0.0439 | -0.0549 |
| **Asia** (1 yes, 0 no) | | | -0.0677*** | | -0.0197 | -0.0212*** | -0.0213*** | -0.0301*** | -0.0269** |
| **Oceania** (1 yes, 0 no) | | | -0.0427** | | -0.0154 | -0.0166 | -0.0162 | -0.0134 | -0.0180 |
| **Adjusted $R^2$** | | | **0.0038*** ** | | **0.273*** ** | | | | |
| **Christmas** (1 yes, 0 no) | | | | | | 0.18156*** | 0.1815*** | 0.1852*** | 0.1859*** |
| **Adjusted $R^2$** | | | | | | **0.282*** ** | | | |
| **$log_{10}$AUTHORS** | | | | | | | 0.5256 | 0.0052 | 0.0063 |
| **Adjusted $R^2$** | | | | | | | **0.282*** ** | | |
| **$log_{10}$HDI** | | | | | | | | -0.1479 | -0.1104 |
| **Adjusted $R^2$** | | | | | | | | **0.2812*** ** | |
| **$log_{10}$LTO** | | | | | | | | | -0.0079 |
| **Adjusted $R^2$** | | | | | | | | | **0.2798*** ** |
| **D. Nature** | | | | | | | | | |
| *0* | *1* | *2* | *3* | *4* | *5* | *6* | *7* | *8* | *9* |
| **Intercept** | 0.345*** | 0.325*** | 0.315*** | 0.368*** | 0.377*** | 0.362*** | 0.348*** | 0.367*** | 0.342*** |
| **Monday** (1 yes, 0 no) | -0.008 | | | -0.0068 | -0.0072 | -0.003 | -0.0012 | -0.0015 | -0.0017 |
| **Tuesday** (1 yes, 0 no) | -0.0168 | | | -0.0158 | -0.016 | -0.0152 | -0.013 | -0.012 | -0.0137 |
| **Wednesday** (1 yes, 0 no) | -0.039** | | | -0.042*** | -0.042*** | -0.039** | -0.037** | -0.036** | -0.036** |
| **Thursday** (1 yes, 0 no) | -0.078*** | | | -0.0766*** | -0.0763*** | -0.073*** | -0.071*** | -0.0708*** | -0.0721*** |
| **Weekend** (1 yes, 0 no) | -0.166*** | | | -0.167*** | -0.167*** | -0.165*** | -0.164*** | -0.164*** | -0.163*** |
| **Adjusted $R^2$ (weighted)** | **0.147*** ** | | | | | | | | |
| **Spring** (1 yes, 0 no) | | -0.030** | | -0.043*** | -0.042*** | -0.029* | 0.028* | -0.027* | -0.0269* |
| **Summer** (1 yes, 0 no) | | -0.008 | | -0.0168 | -0.0166 | -0.003 | -0.002 | -0.002 | -0.0006 |
| **Fall** (1 yes, 0 no) | | -0.034** | | -0.0298 | -0.0299** | -0.017 | -0.017 | -0.016 | -0.015 |
| **Adjusted $R^2$ (weighted)** | | **0.012*** ** | | **0.16*** ** | | | | | |
| **America** (1 yes, 0 no) | | | -0.016 | | -0.013 | -0.015 | -0.014 | -0.016 | -0.068** |
| **Africa** (1 yes, 0 no) | | | N.A. | | N.A. | N.A. | N.A. | N.A. | N.A. |
| **Asia** (1 yes, 0 no) | | | 0.002 | | -0.0012 | -0.0006 | 0.0007 | 0.014 | 0.038 |
| **Oceania** (1 yes, 0 no) | | | 0.018 | | 0.026 | 0.019 | 0.018 | 0.011 | -0.055 |
| **Adjusted $R^2$ (weighted)** | | | **0.004*** ** | | **0.164*** ** | | | | |
| **Christmas** (1 yes, 0 no) | | | | | | 0.073*** | 0.073*** | 0.075*** | 0.074*** |
| **Adjusted $R^2$ (weighted)** | | | | | | **0.173*** ** | | | |
| **$log_{10}$AUTHORS** | | | | | | | 0.0059 | 0.0059 | 0.005 |
| **Adjusted $R^2$ (weighted)** | | | | | | | **0.174*** ** | | |
| **$log_{10}$HDI** | | | | | | | | 0.218 | 0.276* |

## 2008-2010 timespan

| | | | | | | | | | |
|---|---|---|---|---|---|---|---|---|---|
| **Adjusted $R^2$ (weighted)** | | | | | | | | **0.175*** ** | |
| **$log_{10}$LTO** | | | | | | | | | -0.141* |
| **Adjusted $R^2$** | | | | | | | | | **0.179*** ** |
| **E. Cell** | | | | | | | | | |
| *0* | *1* | *2* | *3* | *4* | *5* | *6* | *7* | *8* | *9* |
| **Intercept** | 0.2619*** | 0.296*** | 0.3076*** | 0.2477*** | 0.2419*** | 0.2445*** | 0.223*** | 0.2213*** | 0.2328*** |
| **Monday** (1 yes, 0 no) | 0.104*** | | | 0.1091*** | 0.1081*** | 0.1086*** | 0.1085*** | 0.1086*** | 0.1084*** |
| **Tuesday** (1 yes, 0 no) | 0.0508* | | | 0.0516* | 0.0501* | 0.0505* | 0.0505* | 0.0504* | 0.0503* |
| **Wednesday** (1 yes, 0 no) | 0.0504* | | | 0.053* | 0.0515* | 0.0514* | 0.0512* | 0.0512* | 0.0511* |
| **Thursday** (1 yes, 0 no) | 0.0912*** | | | 0.0887*** | 0.0879*** | 0.0882*** | 0.0873*** | 0.0856*** | 0.0861*** |
| **Weekend** (1 yes, 0 no) | -0.0604* | | | -0.0592* | -0.0616* | -0.0615* | -0.0622* | -0.0623* | -0.0626* |
| **Adjusted $R^2$ (weighted)** | **0.061*** ** | | | | | | | | |
| **Spring** (1 yes, 0 no) | | 0.0005 | | -0.0053 | -0.0058 | -0.0086 | -0.0095 | -0.0082 | -0.0086 |
| **Summer** (1 yes, 0 no) | | 0.019 | | 0.0177 | 0.0179 | 0.0152 | 0.0145 | 0.015 | 0.015 |
| **Fall** (1 yes, 0 no) | | 0.0434* | | 0.0438* | 0.0436* | 0.0409* | 0.0407 | 0.0418* | 0.0412* |
| **Adjusted $R^2$ (weighted)** | | **0.007*** ** | | **0.067*** ** | | | | | |
| **America** (1 yes, 0 no) | | | 0.0042 | | 0.0095 | 0.0094 | 0.0104 | 0.0105 | 0.0339 |
| **Africa** (1 yes, 0 no) | | | N.A. | | N.A. | N.A. | N.A. | N.A. | N.A. |
| **Asia** (1 yes, 0 no) | | | 0.0187 | | 0.0126 | 0.012 | 0.0126 | 0.009 | 0.0035 |
| **Oceania** (1 yes, 0 no) | | | -0.0381 | | -0.0322 | -0.0333 | -0.0357 | -0.036 | -0.0063 |
| **Adjusted $R^2$ (weighted)** | | | **0.0009*** ** | | **0.068*** ** | | | | |
| **Christmas** (1 yes, 0 no) | | | | | | -0.014 | -0.0145 | -0.013 | -0.0142 |
| **Adjusted $R^2$ (weighted)** | | | | | | **0.069*** ** | | | |
| **$log_{10}$AUTHORS** | | | | | | | 0.0201 | 0.0219 | 0.0205 |
| **Adjusted $R^2$ (weighted)** | | | | | | | **0.069*** ** | | |
| **$log_{10}$HDI** | | | | | | | | 0.0204 | -0.0283 |
| **Adjusted $R^2$ (weighted)** | | | | | | | | **0.068*** ** | |
| **$log_{10}$LTO** | | | | | | | | | 0.0599 |
| **Adjusted $R^2$** | | | | | | | | | **0.069*** ** |

**Note:** * Statistically significant at level 10%
** Statistically significant at level 5%
*** Statistically significant at level 1%
N.A. – Due to the lack of variability (no papers having PCA – Papers' Corresponding Authors – from Africa) this factor was automatically removed from the model.

## 2011-2013 timespan

| A. Consolidated data set† | | | | | | | | | |
|---|---|---|---|---|---|---|---|---|---|
| Models / Variables/ characteristics | **M1 (equation 3)** | **M2 (equation 4)** | **M3 (equation 5)** | **M4 (equation 6)** | **M5 (equation 7)** | **M6 (equation 8)** | **M7 (equation 9)** | **M8 (equation 10)** | **M9 (equation 11)** |
| *0* | *1* | *2* | *3* | *4* | *5* | *6* | *7* | *8* | *9* |
| **Intercept** | 0.276 | 0.284 | 0.286 | 0.279 | 0.278 | 0.269 | 0.27 | 0.271 | 0.266 |
| **Monday** (1 yes, 0 no) | -0.049 | | | -0.048 | -0.048 | -0.047 | -0.047 | -0.0472 | -0.047 |
| **Tuesday** (1 yes, 0 no) | 0.044 | | | 0.045 | 0.0453 | 0.0454 | 0.0454 | 0.0459 | 0.046 |
| **Wednesday** (1 yes, 0 no) | 0.068 | | | 0.069 | 0.069 | 0.07 | 0.07 | 0.07 | 0.07 |
| **Thursday** (1 yes, 0 no) | 0.067 | | | 0.067 | 0.068 | 0.0678 | 0.067 | 0.068 | 0.068 |
| **Weekend** (1 yes, 0 no) | -0.586 | | | -0.585 | -0.584 | -0.586 | -0.586 | -0.586 | -0.586 |
| **Adjusted $R^2$ (weighted)** | **0.673*** ** | | | | | | | | |
| **Spring** (1 yes, 0 no) | | 0.007 | | 0.0033 | 0.003 | 0.012 | 0.012 | 0.012 | 0.011 |
| **Summer** (1 yes, 0 no) | | -0.0066 | | -0.009 | -0.0092 | -0.0006 | -0.0006 | -0.0007 | -0.0009 |
| **Fall** (1 yes, 0 no) | | -0.0091 | | -0.004 | -0.004 | 0.0039 | 0.0039 | 0.0038 | 0.0036 |
| **Adjusted $R^2$ (weighted)** | | **0.001*** ** | | **0.672*** ** | | | | | |
| **America** (1 yes, 0 no) | | | 0.007 | | 0.0045 | 0.0045 | 0.0045 | 0.0044 | -0.0014 |
| **Africa** (1 yes, 0 no) | | | -0.021 | | -0.0123 | -0.0126 | -0.0126 | -0.012 | -0.016 |
| **Asia** (1 yes, 0 no) | | | -0.03 | | -0.003 | -0.0033 | -0.0033 | -0.0019 | 0.0009 |
| **Oceania** (1 yes, 0 no) | | | -0.024 | | -0.005 | -0.0052 | -0.0053 | -0.0054 | -0.013 |
| **Adjusted $R^2$ (weighted)** | | | **0.006*** ** | | **0.673*** ** | | | | |
| **Christmas** (1 yes, 0 no) | | | | | | 0.075 | 0.075 | 0.074 | 0.0744 |
| **Adjusted $R^2$ (weighted)** | | | | | | **0.676*** ** | | | |
| **$log_{10}$AUTHORS** | | | | | | | -0.0003 | -0.0004 | -0.00002 |
| **Adjusted $R^2$ (weighted)** | | | | | | | **0.676*** ** | | |
| **$log_{10}$HDI** | | | | | | | | 0.0061 | 0.0089 |
| **Adjusted $R^2$ (weighted)** | | | | | | | | **0.674*** ** | |
| **$log_{10}$LTO** | | | | | | | | | -0.0187 |
| **Adjusted $R^2$** | | | | | | | | | **0.675*** ** |
| **B. Physica A: Statistical mechanics and its applications** | | | | | | | | | |
| *0* | *1* | *2* | *3* | *4* | *5* | *6* | *7* | *8* | *9* |
| **Intercept** | 0.217*** | 0.169*** | 0.142*** | 0.230*** | 0.215*** | 0.194*** | 0.203*** | 0.211*** | 0.206*** |
| **Monday** (1 yes, 0 no) | -0.032* | | | -0.033* | -0.032* | -0.034* | -0.034* | -0.035* | -0.040** |
| **Tuesday** (1 yes, 0 no) | -0.059*** | | | -0.058*** | -0.058*** | -0.060*** | -0.061*** | -0.060*** | -0.063*** |

## 2011-2013 timespan

| | | | | | | | | | |
|---|---|---|---|---|---|---|---|---|---|
| **Wednesday** (1 yes, 0 no) | -0.010 | | | -0.008 | -0.008 | -0.008 | -0.008 | -0.009 | -0.014 |
| **Thursday** (1 yes, 0 no) | -0.063*** | | | -0.061*** | -0.061*** | -0.062*** | -0.063*** | -0.064*** | -0.070*** |
| **Weekend** (1 yes, 0 no) | -0.232*** | | | -0.230*** | -0.230*** | -0.233*** | -0.234*** | -0.235*** | -0.241*** |
| **Adjusted $R^2$ (weighted)** | **0.117*** ** | | | | | | | | |
| **Spring** (1 yes, 0 no) | | -0.023 | | -0.021 | -0.021 | 0.000 | 0.0001 | 0.0006 | -0.0044 |
| **Summer** (1 yes, 0 no) | | -0.010 | | -0.0047 | -0.0046 | 0.017 | 0.018 | 0.019 | 0.016 |
| **Fall** (1 yes, 0 no) | | -0.036** | | -0.0296 | -0.0303* | -0.007 | -0.007 | -0.007 | -0.008 |
| **Adjusted $R^2$ (weighted)** | | **0.003*** ** | | **0.118*** ** | | | | | |
| **America** (1 yes, 0 no) | | | 0.0257 | | 0.0262* | 0.025 | 0.026* | 0.031* | 0.023 |
| **Africa** (1 yes, 0 no) | | | 0.001 | | 0.021 | 0.011 | 0.014 | 0.032 | 0.048 |
| **Asia** (1 yes, 0 no) | | | 0.005 | | 0.0187 | 0.0184 | 0.0204 | 0.031* | 0.030* |
| **Oceania** (1 yes, 0 no) | | | 0.024 | | 0.0032 | 0.007 | 0.008 | 0.006 | -0.003 |
| **Adjusted $R^2$ (weighted)** | | | **0.001*** ** | | **0.119*** ** | | | | |
| **Christmas** (1 yes, 0 no) | | | | | | 0.137*** | 0.136*** | 0.136*** | 0.134*** |
| **Adjusted $R^2$ (weighted)** | | | | | | **0.133*** ** | | | |
| **$log_{10}$AUTHORS** | | | | | | | -0.023 | -0.020 | -0.009 |
| **Adjusted $R^2$ (weighted)** | | | | | | | **0.134*** ** | | |
| **$log_{10}$HDI** | | | | | | | | 0.166 | 0.138 |
| **Adjusted $R^2$ (weighted)** | | | | | | | | **0.136*** ** | |
| **$log_{10}$LTO** | | | | | | | | | -0.026 |
| **Adjusted $R^2$** | | | | | | | | | **0.139*** ** |
| **C. PLOS ONE** | | | | | | | | | |
| *0* | *1* | *2* | *3* | *4* | *5* | *6* | *7* | *8* | *9* |
| **Intercept** | 0.341*** | 0.340*** | 0.353*** | 0.342*** | 0.343*** | 0.334*** | 0.333*** | 0.334*** | 0.331*** |
| **Monday** (1 yes, 0 no) | -0065*** | | | -0.065*** | -0.065*** | -0.064*** | -0.064*** | -0.064*** | -0.064*** |
| **Tuesday** (1 yes, 0 no) | 0.066*** | | | 0.066*** | 0.066*** | 0.066*** | 0.0667*** | 0.067*** | 0.067*** |
| **Wednesday** (1 yes, 0 no) | 0.100*** | | | 0.1000* | 0.101*** | 0.102*** | 0.102*** | 0.103*** | 0.102*** |
| **Thursday** (1 yes, 0 no) | 0.098*** | | | 0.0978** | 0.098*** | 0.098*** | 0.098*** | 0.098*** | 0.098*** |
| **Weekend** (1 yes, 0 no) | -0.532*** | | | -0.531*** | -0.530*** | -0.532*** | -0.532*** | -0.531*** | -0.532*** |
| **Adjusted $R^2$** | **0.508*** ** | | | | | | | | |
| **Spring** (1 yes, 0 no) | | 0.0125*** | | 0.0082*** | 0.0082*** | 0.016*** | 0.016*** | 0.016*** | 0.016*** |
| **Summer** (1 yes, 0 no) | | -0.0045*** | | -0.0069*** | -0.0068*** | 0.0013 | 0.0113 | 0.0011 | 0.00099 |
| **Fall** (1 yes, 0 no) | | -0.0080*** | | -0.0028** | -0.0026** | 0.0056*** | 0.005*** | 0.0056*** | 0.0053*** |
| **Adjusted $R^2$** | | **0.0009*** ** | | **0.511*** ** | | | | | |
| **America** (1 yes, 0 no) | | | 0.0025* | | 0.0031*** | 0.0031*** | 0.0032*** | 0.0032*** | -0.0001 |

## 2011-2013 timespan

| | | | | | | | | | |
|---|---|---|---|---|---|---|---|---|---|
| **Africa** (1 yes, 0 no) | | | -0.0324*** | | -0.0186*** | -0.0189*** | -0.0189*** | -0.0155*** | -0.0207*** |
| **Asia** (1 yes, 0 no) | | | -0.0486*** | | -0.0077*** | -0.0073*** | -0.0073*** | -0.0058*** | -0.0041*** |
| **Oceania** (1 yes, 0 no) | | | -0.039*** | | -0.0069*** | -0.0065*** | -0.006*** | -0.00666*** | -0.011*** |
| **Adjusted $R^2$** | | | **0.0079*** ** | | **0.512*** ** | | | | |
| **Christmas** (1 yes, 0 no) | | | | | | 0.072*** | 0.072*** | 0.0722*** | 0.0721*** |
| **Adjusted $R^2$** | | | | | | **0.515*** ** | | | |
| **$log_{10}$AUTHORS** | | | | | | | 0.0016 | 0.0016 | 0.0023 |
| **Adjusted $R^2$** | | | | | | | **0.515*** ** | | |
| **$log_{10}$HDI** | | | | | | | | 0.1096 | 0.022* |
| **Adjusted $R^2$** | | | | | | | | **0.514*** ** | |
| **$log_{10}$LTO** | | | | | | | | | -0.0109*** |
| **Adjusted $R^2$** | | | | | | | | | **0.515*** ** |
| **D. Nature** | | | | | | | | | |
| *0* | *1* | *2* | *3* | *4* | *5* | *6* | *7* | *8* | *9* |
| **Intercept** | 0.3056*** | 0.284*** | 0.2704*** | 0.3096*** | 0.2994*** | 0.2895*** | 0.291*** | 0.281*** | 0.2710*** |
| **Monday** (1 yes, 0 no) | 0.0165* | | | 0.0165* | 0.016* | 0.0167* | 0.0168* | 0.017* | 0.0176** |
| **Tuesday** (1 yes, 0 no) | -0.0049 | | | -0.0051 | -0.0049 | -0.0052 | -0.0052 | -0.0045 | -0.0041 |
| **Wednesday** (1 yes, 0 no) | -0.0237** | | | -0.0238** | -0.023** | -0.0231** | -0.023** | -0.0229** | -0.023** |
| **Thursday** (1 yes, 0 no) | -0.0177* | | | -0.0176* | -0.0171* | -0.0156* | -0.0156* | -0.0153* | -0.0148 |
| **Weekend** (1 yes, 0 no) | -0.217*** | | | -0.2174*** | -0.2177*** | -0.2165*** | -0.2166*** | -0.2172*** | -0.216*** |
| **Adjusted $R^2$ (weighted)** | **0.247*** ** | | | | | | | | |
| **Spring** (1 yes, 0 no) | | -0.015* | | -0.0107 | -0.0109 | -0.0016 | -0.0016 | -0.0017 | -0.0019 |
| **Summer** (1 yes, 0 no) | | -0.006 | | 0.0019 | 0.0008 | 0.0101 | 0.0101 | 0.0104 | 0.0102 |
| **Fall** (1 yes, 0 no) | | -0.009 | | -0.0057 | -0.0066 | 0.0028 | 0.0028 | 0.0024 | 0.0023 |
| **Adjusted $R^2$ (weighted)** | | **0.001*** ** | | **0.251*** ** | | | | | |
| **America** (1 yes, 0 no) | | | 0.0109 | | 0.0167*** | 0.0163*** | 0.0163*** | 0.0171*** | -0.0017 |
| **Africa** (1 yes, 0 no) | | | -0.0864 | | -0.0032 | 0.0014 | 0.0019 | -0.030 | -0.042 |
| **Asia** (1 yes, 0 no) | | | 0.0034 | | 0.0108 | 0.012 | 0.012 | 0.0006 | 0.0063 |
| **Oceania** (1 yes, 0 no) | | | -0.0143 | | -0.0003 | -0.0008 | -0.0008 | 0.002 | -0.0210 |
| **Adjusted $R^2$ (weighted)** | | | **0.002*** ** | | **0.253*** ** | | | | |
| **Christmas** (1 yes, 0 no) | | | | | | 0.0729*** | 0.0724*** | 0.0725*** | 0.0733*** |
| **Adjusted $R^2$ (weighted)** | | | | | | **0.257*** ** | | | |
| **$log_{10}$AUTHORS** | | | | | | | -0.0007 | -0.0017 | -0.0017 |
| **Adjusted $R^2$ (weighted)** | | | | | | | **0.258*** ** | | |
| **$log_{10}$HDI** | | | | | | | | -0.245 | -0.234 |

## 2011-2013 timespan

| | | | | | | | | | |
|---|---|---|---|---|---|---|---|---|---|
| **Adjusted $R^2$ (weighted)** | | | | | | | | **0.259*** ** | |
| **$\log_{10}$LTO** | | | | | | | | | -0.052 |
| **Adjusted $R^2$** | | | | | | | | | **0.259*** ** |
| **E. Cell** | | | | | | | | | |
| *0* | *1* | *2* | *3* | *4* | *5* | *6* | *7* | *8* | *9* |
| **Intercept** | 0.3030*** | 0.28*** | 0.2453*** | 0.3334*** | 0.3137*** | 0.3075*** | 0.3222*** | 0.2907*** | 0.2535*** |
| **Monday** (1 yes, 0 no) | -0.0089 | | | -0.0146 | -0.0126 | -0.0128 | -0.0131 | -0.0108 | -0.117 |
| **Tuesday** (1 yes, 0 no) | -0.0169 | | | -0.0213 | -0.0203 | -0.0207 | -0.0213 | -0.0193 | -0.0198 |
| **Wednesday** (1 yes, 0 no) | -0.0322 | | | -0.0383 | -0.0442* | -0.0443* | -0.0447* | -0.0444* | -0.0442* |
| **Thursday** (1 yes, 0 no) | -0.0457* | | | -0.0454* | -0.0436 | -0.0447* | -0.045* | -0.0426 | -0.044 |
| **Weekend** (1 yes, 0 no) | -0.1971*** | | | -0.2033*** | -0.2041*** | -0.2022*** | -0.2021*** | -0.2006*** | -0.1996*** |
| **Adjusted $R^2$ (weighted)** | **0.07*** ** | | | | | | | | |
| **Spring** (1 yes, 0 no) | | -0.0517** | | -0.0602*** | -0.0585*** | -0.0514** | -0.0511** | -0.0513** | -0.045** |
| **Summer** (1 yes, 0 no) | | -0.0374 | | -0.0462** | -0.0464** | -0.0393* | -0.0395* | -0.0389* | -0.0369 |
| **Fall** (1 yes, 0 no) | | 0.0288 | | 0.0211 | 0.0179 | 0.0251 | 0.0259 | 0.0231 | 0.0261 |
| **Adjusted $R^2$ (weighted)** | | **0.018*** ** | | **0.091*** ** | | | | | |
| **America** (1 yes, 0 no) | | | 0.0141 | | 0.0213 | 0.0201 | 0.02 | 0.0228 | -0.0545 |
| **Africa** (1 yes, 0 no) | | | N.A. | | N.A. | N.A. | N.A. | N.A. | N.A. |
| **Asia** (1 yes, 0 no) | | | 0.0448 | | 0.0522 | 0.0529 | 0.0541 | 0.0187 | 0.0446 |
| **Oceania** (1 yes, 0 no) | | | 0.3769*** | | 0.3437*** | 0.3386*** | 0.342*** | 0.3555*** | 0.2555* |
| **Adjusted $R^2$ (weighted)** | | | **0.014*** ** | | **0.105*** ** | | | | |
| **Christmas** (1 yes, 0 no) | | | | | | 0.0596 | 0.0603 | 0.0605 | 0.0629 |
| **Adjusted $R^2$ (weighted)** | | | | | | **0.106*** ** | | | |
| **$\log_{10}$AUTHORS** | | | | | | | -0.0147 | -0.0171 | -0.0171 |
| **Adjusted $R^2$ (weighted)** | | | | | | | **0.106*** ** | | |
| **$\log_{10}$HDI** | | | | | | | | -0.8308 | -0.6873 |
| **Adjusted $R^2$ (weighted)** | | | | | | | | **0.108*** ** | |
| **$\log_{10}$LTO** | | | | | | | | | -0.2034 |
| **Adjusted $R^2$** | | | | | | | | | **0.11*** ** |

**Note:** * Statistically significant at level 10%

** Statistically significant at level 5%

*** Statistically significant at level 1%

† For the consolidated dataset for this timespan the regression coefficients' standard errors could not be computed. Subsequently, there is no information about their statistical significance.

N.A. – Due to the lack of variability (no papers having PCA – Papers' Corresponding Authors – from Africa) this factor was automatically removed from the model.

## 2011-2013 timespan

| A. Consolidated data set† | | | | | | | | | |
|---|---|---|---|---|---|---|---|---|---|
| Models / Variables/ characteristics | M1 (equation 3) | M2 (equation 4) | M3 (equation 5) | M4 (equation 6) | M5 (equation 7) | M6 (equation 8) | M7 (equation 9) | M8 (equation 10) | M9 (equation 11) |
| *0* | *1* | *2* | *3* | *4* | *5* | *6* | *7* | *8* | *9* |
| **Intercept** | 0.165 | 0.222 | 0.245 | 0.158 | 0.158 | 0.148 | 0.15 | 0.149 | 0.148 |
| **Monday** (1 yes, 0 no) | 0.087 | | | 0.087 | 0.088 | 0.09 | 0.09 | 0.091 | 0.090 |
| **Tuesday** (1 yes, 0 no) | 0.143 | | | 0.143 | 0.143 | 0.145 | 0.145 | 0.145 | 0.145 |
| **Wednesday** (1 yes, 0 no) | 0.143 | | | 0.143 | 0.143 | 0.143 | 0.143 | 0.143 | 0.142 |
| **Thursday** (1 yes, 0 no) | 0.119 | | | 0.119 | 0.1199 | 0.119 | 0.119 | 0.119 | 0.119 |
| **Weekend** (1 yes, 0 no) | -0.467 | | | -0.467 | -0.467 | -0.468 | -0.468 | -0.468 | -0.469 |
| **Adjusted $R^2$ (weighted)** | **0.72*** ** | | | | | | | | |
| **Spring** (1 yes, 0 no) | | 0.0115 | | 0.01 | 0.01 | 0.0199 | 0.019 | 0.019 | 0.019 |
| **Summer** (1 yes, 0 no) | | 0.022 | | 0.011 | 0.011 | 0.021 | 0.021 | 0.021 | 0.021 |
| **Fall** (1 yes, 0 no) | | 0.013 | | 0.006 | 0.0064 | 0.016 | 0.016 | 0.016 | 0.016 |
| **Adjusted $R^2$ (weighted)** | | **0.001*** ** | | **0.717*** ** | | | | | |
| **America** (1 yes, 0 no) | | | 0.001 | | 0.0006 | 0.0007 | 0.0006 | 0.0005 | -0.0011 |
| **Africa** (1 yes, 0 no) | | | -0.027 | | -0.005 | -0.0046 | -0.0047 | -0.0088 | -0.0103 |
| **Asia** (1 yes, 0 no) | | | -0.036 | | -0.0012 | -0.0009 | -0.0009 | -0.0018 | -0.00099 |
| **Oceania** (1 yes, 0 no) | | | -0.028 | | -0.008 | -0.0071 | -0.0073 | -0.0067 | -0.0095 |
| **Adjusted $R^2$ (weighted)** | | | **0.007*** ** | | **0.718*** ** | | | | |
| **Christmas** (1 yes, 0 no) | | | | | | 0.112 | 0.112 | 0.113 | 0.113 |
| **Adjusted $R^2$ (weighted)** | | | | | | **0.723*** ** | | | |
| **$log_{10}$AUTHORS** | | | | | | | -0.001 | -0.00102 | -0.0024 |
| **Adjusted $R^2$ (weighted)** | | | | | | | **0.723*** ** | | |
| **$log_{10}$HDI** | | | | | | | | -0.0087 | -0.0179 |
| **Adjusted $R^2$ (weighted)** | | | | | | | | **0.719*** ** | |
| **$log_{10}$LTO** | | | | | | | | | -0.0052 |
| **Adjusted $R^2$** | | | | | | | | | **0.719*** ** |
| **B. Physica A: Statistical mechanics and its applications** | | | | | | | | | |
| *0* | *1* | *2* | *3* | *4* | *5* | *6* | *7* | *8* | *9* |
| **Intercept** | 0.134*** | 0.150*** | 0.142*** | 0.137*** | 0.131*** | 0.130*** | 0.128*** | 0.129*** | 0.126*** |
| **Monday** (1 yes, 0 no) | 0.057*** | | | 0.055** | 0.055** | 0.055** | 0.055** | 0.058*** | 0.056** |
| **Tuesday** (1 yes, 0 no) | 0.077*** | | | 0.078*** | 0.077*** | 0.077*** | 0.077*** | 0.076*** | 0.074*** |

## 2011-2013 timespan

| | | | | | | | | | |
|---|---|---|---|---|---|---|---|---|---|
| **Wednesday** (1 yes, 0 no) | 0.052** | | | 0.052** | 0.052** | 0.051** | 0.051** | 0.053** | 0.048** |
| **Thursday** (1 yes, 0 no) | -0.016 | | | -0.014 | -0.013 | -0.013 | -0.013 | -0.010 | -0.011 |
| **Weekend** (1 yes, 0 no) | -0.151*** | | | -0.0149*** | -0.151*** | -0.151*** | -0.151*** | -0.152*** | -0.158*** |
| **Adjusted $R^2$ (weighted)** | **0.122*** ** | | | | | | | | |
| **Spring** (1 yes, 0 no) | | -0.024 | | -0.0141 | -0.0155 | -0.0149 | -0.014 | -0.015 | -0.014 |
| **Summer** (1 yes, 0 no) | | -0.015 | | -0.007 | -0.007 | -0.007 | -0.007 | -0.006 | -0.005 |
| **Fall** (1 yes, 0 no) | | 0.012 | | 0.016 | 0.0155 | 0.016 | 0.016 | 0.0159 | 0.020 |
| **Adjusted $R^2$ (weighted)** | | **0.003*** ** | | **0.121*** ** | | | | | |
| **America** (1 yes, 0 no) | | | 0.004 | | 0.005 | 0.005 | 0.005 | 0.006 | 0.002 |
| **Africa** (1 yes, 0 no) | | | -0.031 | | -0.023 | -0.023 | -0.023 | -0.019 | -0.046 |
| **Asia** (1 yes, 0 no) | | | -0.001 | | 0.012 | 0.012 | 0.012 | 0.016 | 0.012 |
| **Oceania** (1 yes, 0 no) | | | -0.027 | | -0.031 | -0.031 | -0.031 | -0.031 | -0.071 |
| **Adjusted $R^2$ (weighted)** | | | **0.0006*** ** | | **0.122*** ** | | | | |
| **Christmas** (1 yes, 0 no) | | | | | | 0.003 | 0.0037 | 0.0025 | 0.0088 |
| **Adjusted $R^2$ (weighted)** | | | | | | **0.122*** ** | | | |
| **$log_{10}$AUTHORS** | | | | | | | 0.0039 | 0.038 | 0.0001 |
| **Adjusted $R^2$ (weighted)** | | | | | | | **0.122*** ** | | |
| **$log_{10}$HDI** | | | | | | | | 0.031 | -0.043 |
| **Adjusted $R^2$ (weighted)** | | | | | | | | **0.123*** ** | |
| **$log_{10}$LTO** | | | | | | | | | -0.004 |
| **Adjusted $R^2$** | | | | | | | | | **0.126*** ** |
| **C. PLOS ONE†** | | | | | | | | | |
| *0* | *1* | *2* | *3* | *4* | *5* | *6* | *7* | *8* | *9* |
| **Intercept** | 0.194 | 0.271 | 0.3 | 0.182 | 0.183 | 0.172 | 0.172 | 0.171 | 0.171 |
| **Monday** (1 yes, 0 no) | 0.113 | | | 0.112 | 0.112 | 0.116 | 0.116 | 0.117 | 0.116 |
| **Tuesday** (1 yes, 0 no) | 0.194 | | | 0.194 | 0.194 | 0.197 | 0.197 | 0.197 | 0.197 |
| **Wednesday** (1 yes, 0 no) | 0.199 | | | 0.199 | 0.199 | 0.199 | 0.199 | 0.199 | 0.199 |
| **Thursday** (1 yes, 0 no) | 0.162 | | | 0.162 | 0.162 | 0.162 | 0.162 | 0.163 | 0.162 |
| **Weekend** (1 yes, 0 no) | -0.382 | | | -0.382 | -0.381 | -0.383 | -0.383 | -0.383 | -0.384 |
| **Adjusted $R^2$** | **0.570*** ** | | | | | | | | |
| **Spring** (1 yes, 0 no) | | 0.012 | | 0.0165 | 0.0165 | 0.026 | 0.026 | 0.026 | 0.0261 |
| **Summer** (1 yes, 0 no) | | 0.022 | | 0.0194 | 0.0194 | 0.029 | 0.029 | 0.029 | 0.0297 |
| **Fall** (1 yes, 0 no) | | 0.012 | | 0.012 | 0.0127 | 0.022 | 0.022 | 0.022 | 0.0225 |
| **Adjusted $R^2$** | | **0.001*** ** | | **0.575*** ** | | | | | |
| **America** (1 yes, 0 no) | | | -0.0033 | | -0.0011 | -0.0011 | -0.0011 | -0.0011 | -0.0023 |

## 2011-2013 timespan

| | | | | | | | | | |
|---|---|---|---|---|---|---|---|---|---|
| **Africa** (1 yes, 0 no) | | | -0.034 | | -0.007 | -0.0067 | -0.0067 | -0.011 | -0.011 |
| **Asia** (1 yes, 0 no) | | | -0.047 | | -0.0041 | -0.0038 | -0.0038 | -0.0048 | -0.0041 |
| **Oceania** (1 yes, 0 no) | | | -0.037 | | -0.0130 | -0.0121 | -0.0121 | -0.011 | -0.013 |
| **Adjusted $R^2$** | | | **0.007*** ** | | **0.576*** ** | | | | |
| **Christmas** (1 yes, 0 no) | | | | | | 0.1164 | 0.1164 | 0.1184 | 0.117 |
| **Adjusted $R^2$** | | | | | | **0.582*** ** | | | |
| **$log_{10}$AUTHORS** | | | | | | | 0.0003 | 0.0003 | -0.000004 |
| **Adjusted $R^2$** | | | | | | | **0.582*** ** | | |
| **$log_{10}$HDI** | | | | | | | | -0.023 | -0.0197 |
| **Adjusted $R^2$** | | | | | | | | **0.579*** ** | |
| **$log_{10}$LTO** | | | | | | | | | -0.0033 |
| **Adjusted $R^2$** | | | | | | | | | **0.580*** ** |

**D. Nature**

| *0* | *1* | *2* | *3* | *4* | *5* | *6* | *7* | *8* | *9* |
|---|---|---|---|---|---|---|---|---|---|
| **Intercept** | 0.3088*** | 0.2965*** | 0.2794*** | 0.3157*** | 0.3073*** | 0.3003*** | 0.3012*** | 0.3082*** | 0.314*** |
| **Monday** (1 yes, 0 no) | -0.0011 | | | -0.0021 | -0.0022 | -0.0005 | -0.0005 | -0.0003 | -0.0005 |
| **Tuesday** (1 yes, 0 no) | 0.0153 | | | 0.0147 | 0.0113 | 0.0138 | 0.0138 | 0.0139 | 0.0129 |
| **Wednesday** (1 yes, 0 no) | 0.0072 | | | 0.0064 | 0.006 | 0.0073 | 0.0073 | 0.0072 | 0.0075 |
| **Thursday** (1 yes, 0 no) | -0.0372* | | | -0.0369* | -0.0364* | -0.0361** | -0.0361** | -0.0364*** | -0.0365*** |
| **Weekend** (1 yes, 0 no) | -0.468*** | | | -0.1473*** | -0.1484*** | -0.1487*** | -0.1487*** | -0.149*** | -0.149*** |
| **Adjusted $R^2$ (weighted)** | **0.137*** ** | | | | | | | | |
| **Spring** (1 yes, 0 no) | | -0.0277** | | -0.0194* | -0.0198* | -0.0136 | -0.0135 | -0.0127 | -0.0127 |
| **Summer** (1 yes, 0 no) | | -0.0069 | | -0.0103 | -0.01 | -0.044 | -0.004 | -0.0034 | -0.0031 |
| **Fall** (1 yes, 0 no) | | 0.0087 | | 0.0133 | 0.0125 | 0.0185 | 0.0186 | 0.0192 | 0.020 |
| **Adjusted $R^2$ (weighted)** | | **0.01*** ** | | **0.145*** ** | | | | | |
| **America** (1 yes, 0 no) | | | 0.0141 | | 0.0143 | 0.0145 | 0.0144 | 0.0139 | 0.0259 |
| **Africa** (1 yes, 0 no) | | | 0.0345 | | 0.0105 | 0.0128 | 0.013 | 0.0542 | 0.126 |
| **Asia** (1 yes, 0 no) | | | 0.0063 | | 0.0098 | 0.0105 | 0.0105 | 0.02 | 0.0171 |
| **Oceania** (1 yes, 0 no) | | | -0.0006 | | 0.0178 | 0.0192 | 0.0193 | 0.0172 | 0.0324 |
| **Adjusted $R^2$ (weighted)** | | | **0.002*** ** | | **0.149*** ** | | | | |
| **Christmas** (1 yes, 0 no) | | | | | | 0.0529** | 0.0525** | 0.0523** | 0.053** |
| **Adjusted $R^2$ (weighted)** | | | | | | **0.151*** ** | | | |
| **$log_{10}$AUTHORS** | | | | | | | -0.0004 | -0.0002 | -0.0013 |
| **Adjusted $R^2$ (weighted)** | | | | | | | **0.151*** ** | | |
| **$log_{10}$HDI** | | | | | | | | 0.0864 | 0.165 |

## 2011-2013 timespan

| | | | | | | | | | |
|---|---|---|---|---|---|---|---|---|---|
| **Adjusted $R^2$ (weighted)** | | | | | | | | **0.152*** ** | |
| **$log_{10}$LTO** | | | | | | | | | 0.0328 |
| **Adjusted $R^2$** | | | | | | | | | **0.153*** ** |
| **E. Cell** | | | | | | | | | |
| *0* | *1* | *2* | *3* | *4* | *5* | *6* | *7* | *8* | *9* |
| **Intercept** | 0.3059*** | 0.2466*** | 0.2906*** | 0.2736*** | 0.2879*** | 0.2638*** | 0.2624*** | 0.2412*** | 0.2202*** |
| **Monday** (1 yes, 0 no) | -0.0112 | | | -0.0101 | -0.012 | -0.0146 | -0.0147 | -0.0137 | -0.0144 |
| **Tuesday** (1 yes, 0 no) | -0.0151 | | | -0.0167 | -0.0173 | -0.0144 | -0.0144 | -0.0136 | -0.0142 |
| **Wednesday** (1 yes, 0 no) | -0.0533* | | | -0.0504 | -0.0529 | -0.0605* | -0.0606* | -0.0602* | -0.0617* |
| **Thursday** (1 yes, 0 no) | -0.071 | | | -0.0112 | -0.0109 | -0.0083 | -0.0083 | -0.008 | -0.0093 |
| **Weekend** (1 yes, 0 no) | -0.1348*** | | | -0.1323*** | -0.1313 | -0.1395*** | -0.1395*** | -0.1375*** | -0.1399*** |
| **Adjusted $R^2$ (weighted)** | **0.039*** ** | | | | | | | | |
| **Spring** (1 yes, 0 no) | | 0.0713*** | | 0.0707*** | 0.0678** | 0.0964*** | 0.0964*** | 0.0954*** | 0.0956*** |
| **Summer** (1 yes, 0 no) | | 0.0067 | | 0.0163 | 0.0147 | 0.0444 | 0.0444 | 0.0427 | 0.0431 |
| **Fall** (1 yes, 0 no) | | 0.026 | | 0.0351 | 0.0344 | 0.0633** | 0.0634** | 0.0625** | 0.0627** |
| **Adjusted $R^2$ (weighted)** | | **0.017*** ** | | **0.056*** ** | | | | | |
| **America** (1 yes, 0 no) | | | -0.0258 | | -0.0183 | -0.0214 | -0.0214 | -0.0203 | -0.0641 |
| **Africa** (1 yes, 0 no) | | | N.A. | | N.A. | N.A. | N.A. | N.A. | N.A. |
| **Asia** (1 yes, 0 no) | | | -0.0022 | | 0.0092 | -0.0018 | -0.0018 | -0.0253 | -0.0208 |
| **Oceania** (1 yes, 0 no) | | | 0.1168 | | 0.0985 | 0.0964 | 0.0962 | 0.1021 | 0.0474 |
| **Adjusted $R^2$ (weighted)** | | | **0.006*** ** | | **0.059*** ** | | | | |
| **Christmas** (1 yes, 0 no) | | | | | | 0.2111*** | 0.2109*** | 0.2101*** | 0.2123*** |
| **Adjusted $R^2$ (weighted)** | | | | | | **0.09*** ** | | | |
| **$log_{10}$AUTHORS** | | | | | | | 0.0014 | 0.0023 | 0.0021 |
| **Adjusted $R^2$ (weighted)** | | | | | | | **0.09*** ** | | |
| **$log_{10}$HDI** | | | | | | | | -0.531 | -0.5196 |
| **Adjusted $R^2$ (weighted)** | | | | | | | | **0.091*** ** | |
| **$log_{10}$LTO** | | | | | | | | | -0.1139 |
| **Adjusted $R^2$** | | | | | | | | | **0.094*** ** |

**Note:** * Statistically significant at level 10%

** Statistically significant at level 5%

*** Statistically significant at level 1%

† For the consolidated and PLOS ONE datasets for this timespan the regression coefficients' standard errors could not be computed. Subsequently, there is no information about their statistical significance.

N.A. – Due to the lack of variability (no papers having PCA – Papers' Corresponding Authors – from Africa) this factor was automatically removed from the model.

**SI. Kurtosis formula and its components**

$$Kurtosis = \frac{N(N+1)}{(N-1)(N-2)(N-3)}\left[\frac{S_4}{V(x)^2}\right]\text{-}3\frac{(N-1)^2}{(N-2)(N-3)}$$

where

N is number of cases

$S_4$ is sum of deviations to the mean raised to the fourth power

V(x) is the population variance in its unbiased version

$$S_4 = \sum(y - \bar{y})^4$$

$$V(x) = \frac{S_2}{n-1}$$

where $S_2$ is sum of squared deviation to the mean

$$S_2 = \sum(y - \bar{y})^2$$

For normal distributions the value of kurtosis indicator should be 0. A positive value indicates a Leptokurtic distribution while a negative one appears in case of a Platykurtic distribution.

**SI. Geographic descriptive information**

All maps, which are designed within our study, rely on information for the whole timespan covered (2001-2016). At a glance, in figure SI.3, it is very easy to see that United States of America with a little bit more than 46 thousand papers (25.8% from the total) is at the top followed by China with more than 26 thousand papers (14.6% from the total), Germany with 10 600 papers (6% from the total), and United Kingdom with around 10 100 papers (5.7% from total). At the bottom of the ranking there are more than a dozen of important (from territorial view point) countries/territories which record no paper in the dataset: Afghanistan, Antarctica (as expected, since it is not a country per se), Democratic Republic of Congo, Chad, Guyana, Kirghizstan, Papua New Guinea, Paraguay, Somalia, South Sudan, Surinam, Turkmenistan, Tajikistan, Western Sahara etc. We have to admit that there are two potential immediate explanations. The first one is regarding the lack of development of the countries enlisted in this group, while the second one could be the fact that in our dataset, Nature and Cell are elite journals while PLOS ONE and Physica A could be considered as being in the higher segment of the journals rankings. Is not our purpose here to rank the journals but the last statement rely for scientometrics indicators like Impact Factor (IF) or Article Influence Score (AIS).

The use of gross figures (number of papers) when one compares countries could be misleading due to the size effect (i.e. big – from the population viewpoint – countries tend to rank up and vice versa). Therefore, papers population ratio (PPR) is used (figure SI.4) for a better emphasize of the papers' distribution per countries. The leading group comprise (alphabetically): Australia, Denmark, Iceland, Israel, Netherlands, Norway, Singapore, Switzerland, and Sweden. The second group is mainly formed by two compact territories: North America (Canada and United States of America) and Western Europe – with some exceptions which belong to the first group (Austria, Belgium, Finland, France, Germany, Italy, Luxembourg, Portugal, Spain, and United Kingdom). In addition, there are Estonia and Slovenia – two Eastern and Central European countries – and two insular ones from Pacific: New Zeeland and Taiwan.

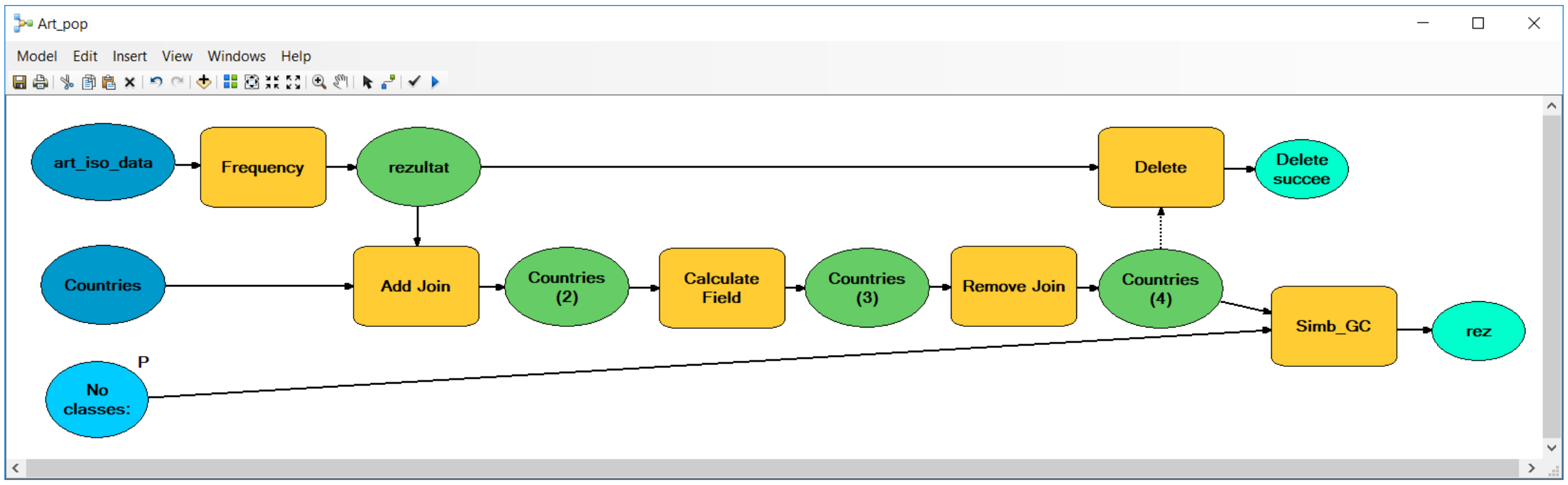

**Figure SI.1.** Geoprocessing model for compute and display the total number of papers published by corresponding authors per population for each country

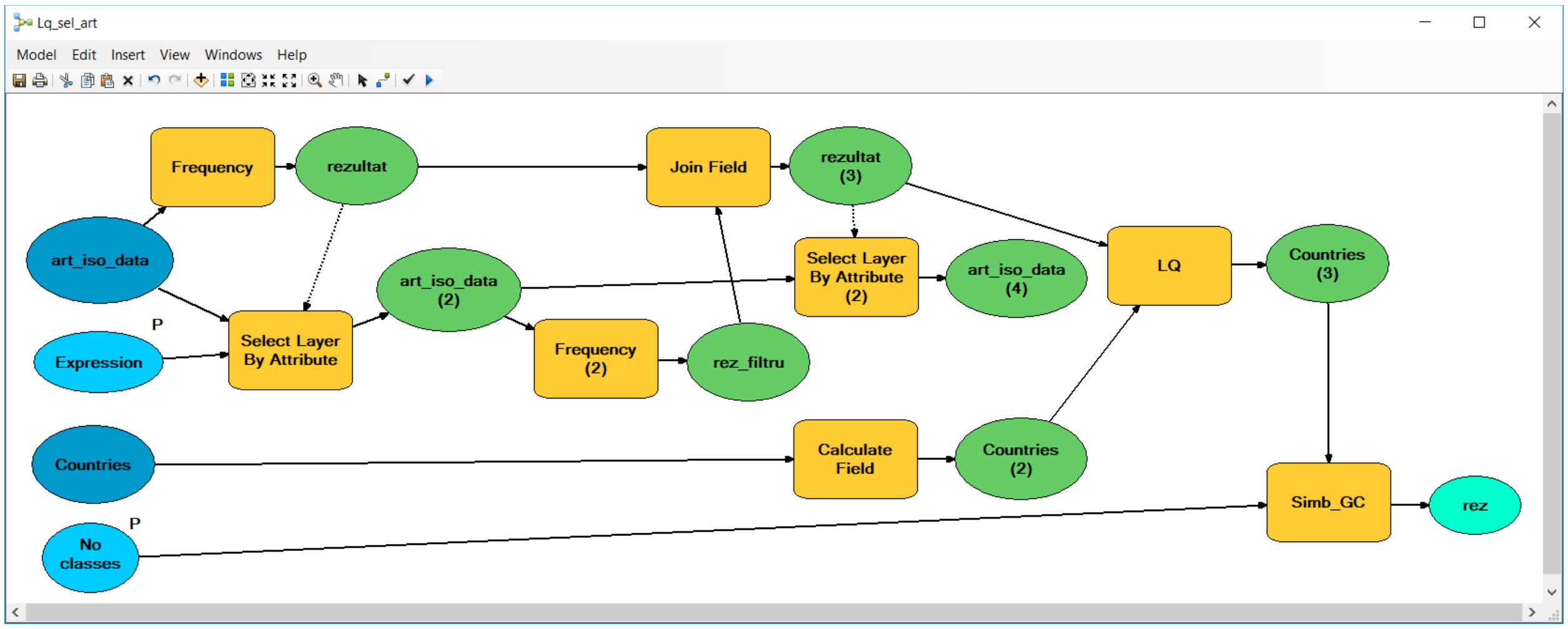

**Figure SI.2.** Geoprocessing model for Localization Quotient indicator (LQ)

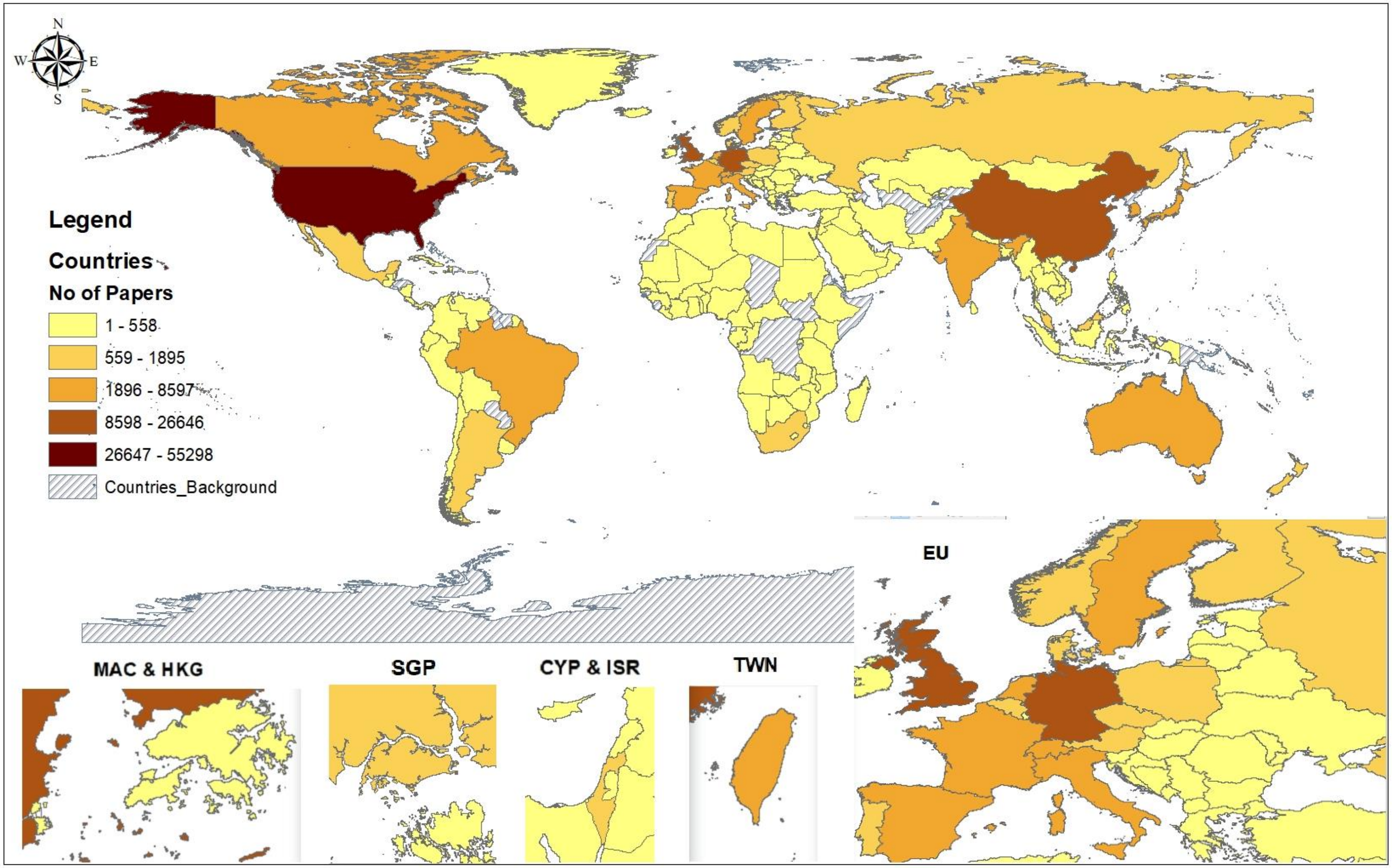

**Figure SI.3.** Distribution of the papers by PCA's country of origin within the consolidate dataset (2001-2016)

**Note:** The consolidated dataset rely on papers from: Physica A, PLOS ONE, Nature and Cell. PCA= Papers' Corresponding Authors; MAC=Macao (China); HKG=Hong Kong (China); SGP=Singapore; CYP=Cyprus; ISR=Israel; TWN=Taiwan and EU=Europe. Countries marked with shading lines record no information in our dataset.

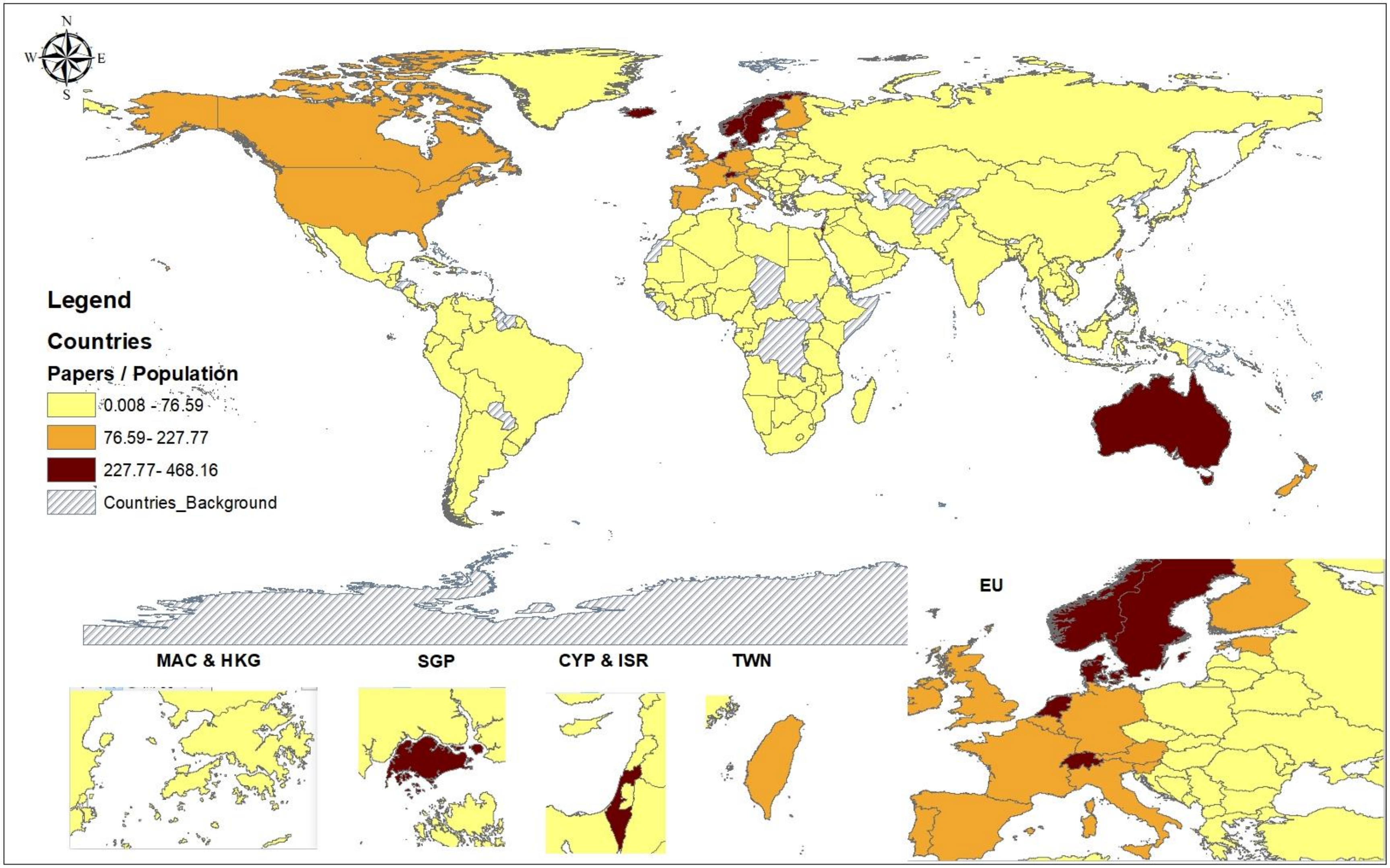

**Figure SI.4.** Distribution of the PPR by PCA's country of origin within the consolidate dataset (2001-2016)

**Note:** The consolidated dataset rely on papers from: Physica A, PLOS ONE, Nature and Cell. PPR=Papers Population Ratio (population in million inhabitants); PCA= Papers' Corresponding Authors; MAC=Macao (China); HKG=Hong Kong (China); SGP=Singapore; CYP=Cyprus; ISR=Israel; TWN=Taiwan and EU=Europe. Countries marked with shading lines record no information in our dataset.